\documentclass[]{bytedance_seed}

\usepackage[toc,page,header]{appendix}

\usepackage[utf8]{inputenc}
\usepackage{newunicodechar}
\newunicodechar{≈}{\approx}
\usepackage{multicol}
\usepackage{tikz}
\usepackage{booktabs}
\usepackage{graphicx}
\usepackage{xcolor}
\usepackage{hyperref}
\usepackage{listings}
\usepackage{multirow}
\usepackage{array}
\usepackage{tabularx}
\usepackage{enumitem}
\usepackage{float}
\usepackage[ruled,vlined]{algorithm2e}
\usepackage{natbib}
\PassOptionsToPackage{font=small,labelfont=bf}{subcaption}
\usepackage{subcaption}  % 提供 subfigure 环境
\RestyleAlgo{ruled}

\usetikzlibrary{arrows,shapes,positioning,shadows,trees,calc,fit,backgrounds,decorations.pathreplacing}

\SetAlgorithmName{Algorithm}{Algorithm}{List of Algorithms}
\SetAlgoLined
\SetAlgoVlined
\SetAlgoNoLine
\SetAlgoNoEnd

\setlist[itemize]{leftmargin=*}

\newcommand{\CompanyA}{ByteDance\xspace}
\newcommand{\PlatformB}{the Seed Serving Platform}
\newcommand{\ModelC}{Doubao-Seed-1.6-thinking\xspace}

\title{Taming the Chaos: Coordinated Autoscaling for Heterogeneous and Disaggregated LLM Inference}

\author[1,2,*]{Rongzhi Li}
\author[1,*]{Ruogu Du}
\author[1,*]{Zefang Chu}
\author[1]{Sida Zhao}
\author[1]{Chunlei Han}
\author[1]{Zuocheng Shi}
\author[1]{Yiwen Shao}
\author[1]{Huanle Han}
\author[1]{Long Huang}
\author[1]{Zherui Liu}
\author[1]{Shufan Liu}

\contribution[1]{ByteDance Seed}
\contribution[2]{National University of Singapore}
\contribution[*]{Equal contribution}

\abstract{
Serving Large Language Models (LLMs) is a GPU-intensive task where traditional autoscalers fall short, particularly for modern Prefill-Decode (P/D) disaggregated architectures. This architectural shift, while powerful, introduces significant operational challenges, including inefficient use of heterogeneous hardware, network bottlenecks, and critical imbalances between prefill and decode stages. We introduce HeteroScale, a coordinated autoscaling framework that addresses the core challenges of P/D disaggregated serving. HeteroScale combines a topology-aware scheduler that adapts to heterogeneous hardware and network constraints with a novel metric-driven policy derived from the first large-scale empirical study of autoscaling signals in production. By leveraging a single, robust metric to jointly scale prefill and decode pools, HeteroScale maintains architectural balance while ensuring efficient, adaptive resource management. Deployed in a massive production environment on tens of thousands of GPUs, HeteroScale has proven its effectiveness, increasing average GPU utilization by a significant 26.6 percentage points and saving hundreds of thousands of GPU-hours daily, all while upholding stringent service level objectives.
}

\date{\today}
\correspondence{Shufan Liu at \email{liushufan.amos@bytedance.com}}

\begin{document}
\maketitle

%不需要目录就注释掉 注意目录不要和第一页放在一块 要有\newpage
%\newpage
%\tableofcontents
%\newpage

% \input{sections/introduction}
% \input{sections/relatedwork}
% \input{sections/approach}
% \input{sections/experiments}

\twocolumn

\section{Introduction}
Large Language Models (LLMs) are revolutionizing applications from chatbots~\cite{chatgpt2022, dam2024complete} to search engines~\cite{xiong2024searchengineservicesmeet}, but serving them is extremely GPU-intensive~\cite{nvidia2023llminference}. State-of-the-art models have limited throughput, making resource efficiency critical~\cite{guo2025deepseek, yang2025qwen3, deepseek_r1_bench, qwen3_baseten}. LLM traffic's strong diurnal pattern~\cite{xia2025skylb, jaiswal2025servingmodelsfastslowoptimizing, stojkovic2024dynamollmdesigningllminference} necessitates effective autoscaling to prevent massive resource waste. However, traditional autoscalers like Kubernetes's Horizontal Pod Autoscaler (HPA)~\cite{kubernetes,k8shpa} struggle to meet the demands of modern LLM serving architectures, particularly in Prefill–Decode (P/D) disaggregated settings~\cite{zhong2024distserve,patel2024splitwiseefficientgenerativellm,strati2024dejavukvcachestreamingfast,zhu2025megascale,rajbhandari2022deepspeedmoeadvancingmixtureofexpertsinference}.

P/D disaggregation separates the compute-intensive prefill phase from the memory-bound decode phase, allowing for independent optimization. While this improves efficiency, it introduces a trio of interconnected scaling and scheduling challenges at large scale that naive solutions cannot address:

\begin{enumerate}[leftmargin=*, wide, labelwidth=!, labelindent=0pt]
 
    \item \textbf{Heterogeneous Hardware Inefficiency:} Prefill and decode phases have different computational demands (compute-intensive vs.\ memory-bandwidth-bound, respectively)~\cite{kamath2025pod}. A naive remedy is to run on a homogeneous pool of general-purpose GPUs, but this over-provisions compute for decode or HBM bandwidth for prefill.  Recent measurements show that such one-size-fits-all provisioning inflates cost per generated token by 41\% compared with a phase-aware heterogeneous deployment~\cite{zhu2025megascale}. Homogeneous pools therefore increase cost and aggravate fragmentation in large heterogeneous clusters.
    
    \item \textbf{Network Bottlenecks:} Transferring the large Key-Value (KV) cache~\cite{vaswani2017attention} is a bandwidth-intensive operation requiring high-throughput interconnects~\cite{zhong2024distserve,qin2024mooncake}. A baseline, topology-agnostic scheduler that treats all GPUs as a flat resource pool places instances wherever capacity is available. Our empirical observations show that such placements across different network switches can reduce the available bandwidth for KV cache transfer by approximately 20\%. This bandwidth reduction directly leads to higher transfer latency, creating a severe performance bottleneck.
   
    \item \textbf{Architectural Imbalance:} Maintaining an optimal ratio of prefill-to-decode (P/D) instances is crucial to prevent one phase from bottlenecking the other, which cripples overall throughput~\cite{qin2024mooncake, wang2025burstgptrealworldworkloaddataset}. A naive approach would be to apply a standard HPA to each pool independently based on GPU utilization. This fails because decode-phase GPU utilization is a misleading metric, often staying high due to KV cache memory pressure even at low loads. Independent scaling thus leads to architectural imbalance, starving one phase while the other sits idle and crippling overall throughput.

\end{enumerate}

To address these challenges, we present \textbf{HeteroScale}, an autoscaling system for P/D disaggregated LLM services. It delivers a coordinated, topology-aware, and resource-efficient scaling strategy through three key innovations:

\begin{itemize}
    \item \textbf{A Framework for Heterogeneous Resource Management.} Our system elevates the P/D ratio and hardware requirements to first-class scheduling constraints. The scheduler intelligently places different service roles on the most suitable hardware types, honoring network affinity and P/D balance simultaneously. This maximizes performance and eliminates the waste inherent in homogeneous allocation strategies.
    
    \item \textbf{Novel Abstractions for Network-Aware Scheduling.} We introduce the \textit{Deployment Group} to enforce network affinity constraints and the \textit{RDMA Subgroup} to manage resource priority based on network topology. These abstractions ensure that prefill and decode instances are co-located for low-latency KV cache transfer while optimizing the use of scarce, high-performance hardware.
    
    \item \textbf{A Comprehensive Analysis of Scaling Policies with Production Data.} We present the first large-scale empirical analysis of autoscaling metrics for P/D disaggregated serving, using massive production datasets. Our study establishes decode Tokens-Per-Second (TPS) as the most robust signal, unlike conventional hardware metrics which we found to be misleading. This core data-driven insight enables our coordinated scaling policy, which uses this single signal to scale both prefill and decode pool together, thus maintaining architectural balance.

\end{itemize}

HeteroScale's capabilities are proven in one of the world's largest production environments at \CompanyA. Managing tens of thousands of GPUs, it consistently delivers substantial performance benefits, saving hundreds of thousands of GPU-hours daily while boosting average GPU utilization by \textbf{26.6} percentage points and SM activity by \textbf{9.2} percentage points. Crucially, these efficiency improvements are achieved while upholding all Service Level Objectives (SLOs), establishing its architecture as a new benchmark for robust, large-scale LLM serving.

\section{Background and Motivation}
\label{sec:background}
% \subsection{LLM Inference Basics}
\subsection{P/D Disaggregated Serving}
LLM inference consists of two distinct phases—prefill and decode—characterized by its own computational demands and operational behaviors. In the prefill phase, the model processes the entire input prompt in parallel, building KV cache for all input tokens in one go. In contrast, the decode phase operates autoregressively, generating one token at a time while relying on previously created KV cache. These phases exhibit fundamentally different computational characteristics. From a computation perspective, prefill is highly compute-intensive and benefits from substantial parallelism, whereas decode is inherently sequential and less demanding in raw computation power. Their memory usage also diverges: prefill’s memory footprint grows with the length of the input prompt, while decode’s memory demand increases with the size of the accumulated KV cache. In terms of batching efficiency, prefill can fully exploit large, homogeneous batches to maximize throughput, but decode, restricted to producing one token per step, gains far less from batching and quickly becomes memory-bound. Moreover, the two phases affect latency differently: prefill determines the initial response latency (Time-To-First-Token, TTFT), while decode dictates the token generation speed (Time-Between-Tokens, TBT)~\cite{databricks_ttft,agrawal2024taming}.

Traditional LLM serving executes both the prefill and decode phases on the same instance. In contrast, P/D disaggregated serving~\cite{zhong2024distserve,patel2024splitwiseefficientgenerativellm,strati2024dejavukvcachestreamingfast,qin2024mooncake,jin2024p} separates these phases onto distinct sets of instances, enabling each to be optimized independently. Such separation improves resource utilization by allowing prefill and decode nodes to be provisioned individually, even on heterogeneous hardware, so that their differing compute and memory demands can be better matched. It also enhances batching efficiency, since the prefill and decode nodes have different constraints on the batch size. Together, these benefits lead to lower latency, higher throughput, and more cost-effective LLM services.

% \subsection{Autoscaling in Conventional Cloud Environments}
% In traditional cloud environments, mature autoscaling solutions, such as the Kubernetes Horizontal Pod Autoscaler (HPA)~\cite{k8shpa}, are widely used to adjust resources dynamically. However, while effective for conventional workloads, these systems fall short when applied to the unique demands and architectural patterns of LLM services. First, the choice of metric is often too narrow: relying on a single hardware metric cannot capture the diverse and dynamic resource demands of LLM inference, especially in P/D disaggregated settings. Second, these systems lack the coordination mechanisms and flexibility required to manage interdependent components in large-scale, heterogeneous deployments. In a P/D disaggregated architecture, the prefill and decode pools must be scaled in a coordinated manner to maintain performance balance, but traditional autoscalers treat the service as a monolithic deployment. This coordination gap creates substantial operational overhead when scheduling at scale, and the challenge becomes even more severe in scenarios involving disaggregated Mixture-of-Experts (MoE) serving~\cite{zhu2025megascale}.

\subsection{Motivation and Challenges}
\label{sec:seedserving}
% HeteroScale is deployed on \CompanyA's production LLM \PlatformB, a large-scale Kubernetes-based system managing thousands of GPUs across multiple data centers. In this environment, conventional autoscalers fall short, creating substantial operational overhead---a challenge that becomes even more severe for disaggregated Mixture-of-Experts (MoE) serving~\cite{zhu2025megascale}. HeteroScale's design is tailored to the specific challenges of this environment, which is characterized by three key features:

HeteroScale is deployed on \PlatformB, \CompanyA's production LLM serving environment, a large-scale Kubernetes-based system managing thousands of GPUs across multiple data centers. In this environment, conventional autoscalers like the Kubernetes HPA fall short for the unique demands of LLM services. Their reliance on narrow hardware metrics cannot capture the dynamic resource demands in P/D disaggregated settings , and they lack the coordination mechanisms to manage interdependent components like prefill and decode pools, which must be scaled in a balanced manner. This coordination gap creates substantial operational overhead, a challenge that becomes even more severe for disaggregated Mixture-of-Experts (MoE) serving~\cite{zhu2025megascale}. HeteroScale's design is tailored to the specific challenges of this environment, which is characterized by three key features:

\textbf{Hardware Heterogeneity}: The platform manages a deep hierarchy of heterogeneous hardware. Resources are organized into logical clusters within actual Kubernetes clusters (also known as physical clusters), which contain service deployments running on diverse machine configurations, including various GPU types (e.g. NVIDIA H20 and L20) with high-speed RDMA interconnects.

\textbf{Hierarchical Network Topology}: The network architecture is explicitly hierarchical, comprising racks of machines under local switches (S0), which are aggregated into minipods (S1) and bigpods (S2). This topology is critical for P/D disaggregated serving, where the transfer of large KV caches necessitates network-aware placement to minimize latency.

\textbf{Co-location and Grouping Abstractions}: Services are managed through logical abstractions, most notably the Deployment Group. This composite object bundles the different roles of a service (e.g., prefill and decode) that must share a common scheduling domain, such as being co-located under the same network switch. While roles within a Deployment Group can be scaled independently, this abstraction creates complex scheduling interdependencies. These characteristics, such as hardware heterogeneity, a strict network topology, and the need for coordinated scaling of disaggregated components, created practical challenges in metric selection and resource management that motivated the design of HeteroScale.

When deploying and operating P/D-disaggregated LLM serving at scale on \PlatformB, we encountered several significant operational challenges.

\textbf{Challenge 1: Scheduling with constraints.} Prefill and decode workers may operate on heterogeneous hardware and carry different SLO priorities, since they predominantly influence distinct performance metrics—TTFT for prefill and TBT for decode. Shifts in workload characteristics, including model choice, prompt length, and response length, can alter the demand distribution between the two phases. Such imbalances risk leaving resources underutilized in one pool while creating bottlenecks in the other, making coordinated scaling between them a non-trivial task. Additionally, P/D disaggregation introduces strong network affinity constraints: prefill and decode nodes must exchange large KV cache data efficiently, which directly contributes to latency. Scheduling must account for multiple constraints, maintaining balanced capacity between pools while reducing the overhead of inter-node communication.

\textbf{Challenge 2: Heterogeneous resource management.} Our clusters are composed of highly heterogeneous resources, including GPUs of different generations, memory capacities, compute capabilities, and interconnect bandwidths. The services running on these clusters are equally diverse—some rely on homogeneous hardware pools, while others require heterogeneous configurations. Coordinating allocation in this mixed environment is challenging. First, scaling conflicts can arise, such as resource contention between heterogeneous and homogeneous services when they share the same hardware pool. Second, mismatched placements—where services are not aligned with their optimal hardware—can lead to performance degradation, wasted capacity, and increased fragmentation, ultimately reducing overall efficiency.

\textbf{Challenge 3: Metric selection.} In traditional autoscaling systems such as Kubernetes's Horizontal Pod Autoscaler (HPA)~\cite{k8shpa}, decisions are often driven by coarse-grained metrics like GPU utilization. However, in LLM serving with P/D disaggregated, these metrics can be misleading. Decode nodes, for instance, may show high GPU utilization despite operating at a low workload due to KV cache memory pressure. This calls for deeper investigation into metrics that better capture the true workload and performance characteristics of each pool.

\textbf{Challenge 4: Large-scale production environment.} Given the nature of our production environment, spanning tens of thousands of GPUs and processing trillions of tokens daily, a practical solution should be highly scalable and capable of handling diverse operational demands. It must be compatible with various types of services and support different deployment modes—including common deployments, P/D disaggregation, and disaggregated MoE serving—while accommodating heterogeneous resources. Additionally, the design should keep maintenance costs low, ensure high reliability under dynamic workloads.

These challenges motivat the development of HeteroScale, our comprehensive autoscaling system designed specifically for P/D disaggregated LLM services.

\section{System Design}
\label{sec:sysdesign}
HeteroScale is designed to address the unique challenges of autoscaling P/D disaggregated LLM services. The system consists of three main layers: autoscaling layer with policy engine, federated pre-scheduling layer and sub-cluster scheduling layer. This section details the design of each layer and the key mechanisms that enable efficient autoscaling.

\subsection{System Architecture}

\begin{figure}
    \centering
    \includegraphics[width=1\linewidth]{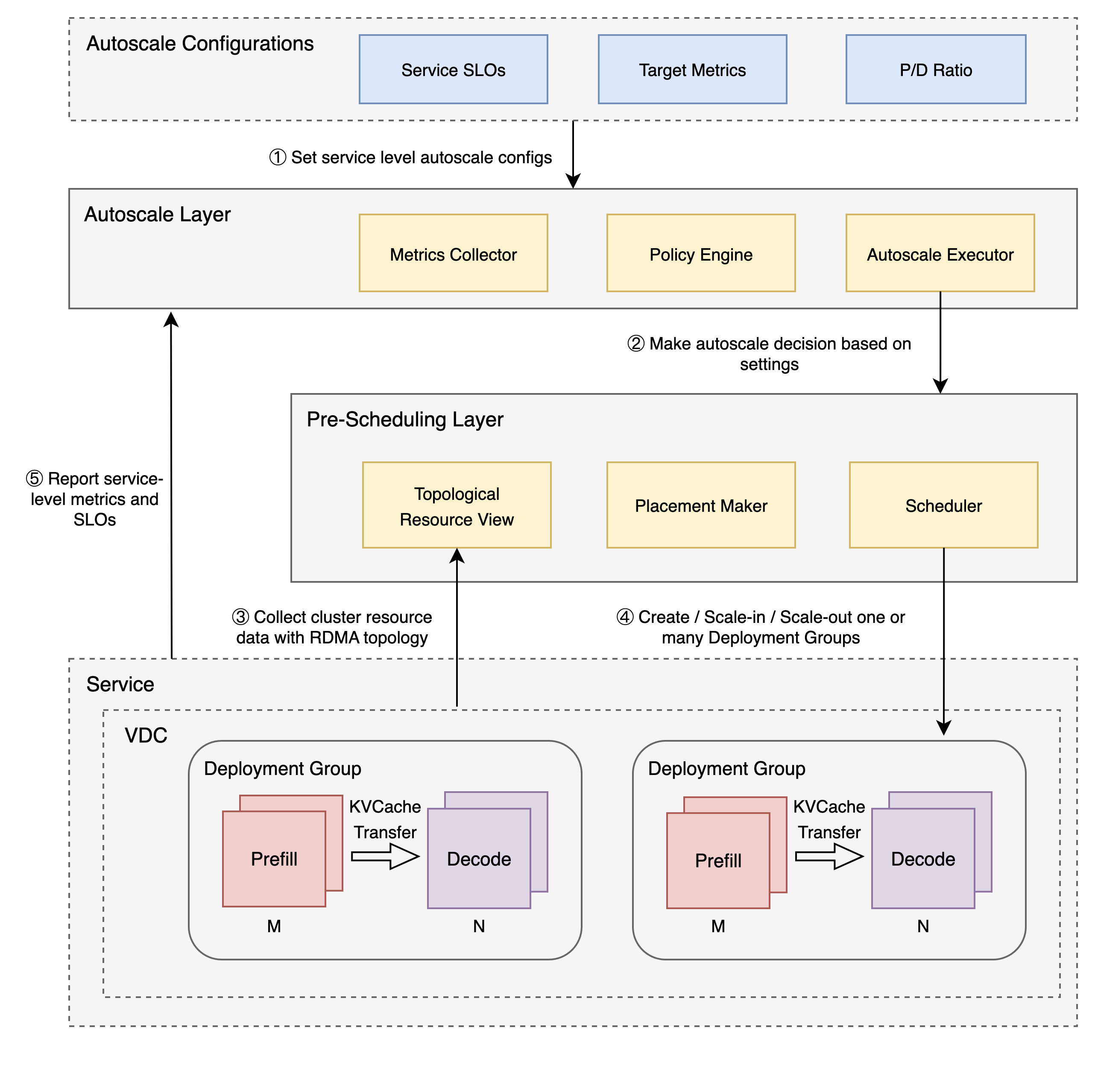}
    \caption{HeteroScale System Architecture}
    \label{fig:system-architecture}
\end{figure}

Figure~\ref{fig:system-architecture} illustrates the overall architecture of HeteroScale. The system is designed as a layered architecture with clear separation of concerns:

\begin{itemize}
    \item \textbf{Autoscaling Layer with Policy Engine}: Provides configuration validation, storage, APIs, metrics collection, and various policies for making scale-in and scale-out decisions.
    
    \item \textbf{Federated Pre-Scheduling Layer}: Executes scaling decisions, which is in charge of managing resource allocation, assembling topological resource view, and scheduling  \hyperref[subsubsec:deploymentgroup]{Deployment Groups}. This layer interfaces with the underlying infrastructure to implement scaling actions, including scaling in/out existing resources or creating batches of Geployment Groups with specified P/D ratios.

    \item \textbf{Sub-cluster Scheduling Layer}: All scaling, creation, and deletion of Deployment Groups originating in the pre-scheduling layer are delegated through this layer, which propagates the calls down to the underlying Kubernetes API server, where the corresponding CRDs are created or updated. At the same time, this layer exposes the node API and supplies it upward for topology assembly.
    
\end{itemize}

The layers communicate through well-defined interfaces, allowing for modular development and testing. The monitoring component provides feedback to the policy engine, creating a closed-loop control system that continuously adapts to changing workload conditions.

\subsection{Autoscaling Layer with Policy Engine}

The auto-scaling framework integrates configuration management and policy engine capabilities to deliver a holistic solution for automated scaling decisions. The configuration management module empowers backend engineers and operators to define production service parameters aligned with desired SLOs and performance targets. Concurrently, the policy engine periodically evaluates these configurations across services, leveraging real-time metric observations to execute scaling actions.

\subsection{Scaling Policies}
One of the main challenges in autoscaling P/D disaggregated services lies in maintaining architectural balance under dynamic loads. Scaling one component in isolation risks bottlenecks that undermine disaggregation. Therefore, HeteroScale's policy engine is designed not merely to adjust capacity, but to act as a coordinated orchestrator for the interdependent prefill and decode stages. We implement two policy paradigms that provide complementary control strategies: a periodic policy for predictable, coarse-grained adjustments, and a metrics-driven policy for fine-grained, real-time optimization. These policies translate high-level service objectives into concrete scaling actions that preserve the crucial P/D ratio across heterogeneous hardware. In addition, we develop a workload-aware framework for policy curation.

\subsubsection{Periodic Scaling Policy}

The periodic scaling policy adjusts resources based on time-of-day patterns, enabling proactive scaling based on expected workload patterns. This policy is beneficial for services with predictable traffic patterns, such as day-night cycles or weekly patterns. By defining scaling schedules according to periods with static target instance and P/D ratios, expected resources could be automatically allocated to target services. In production environments, periodic scaling is employed for services that operate under specific constraints or involve experimental configurations, which are not amenable to metrics-driven scaling policies.

\subsubsection{Metrics-driven Scaling Policy}\label{sec:metric-driven-scaling-policy}

\begin{figure*}[t!]
    \centering % Center the entire figure group

    \begin{subfigure}[b]{0.24\textwidth}
        \centering
        \includegraphics[width=\linewidth]{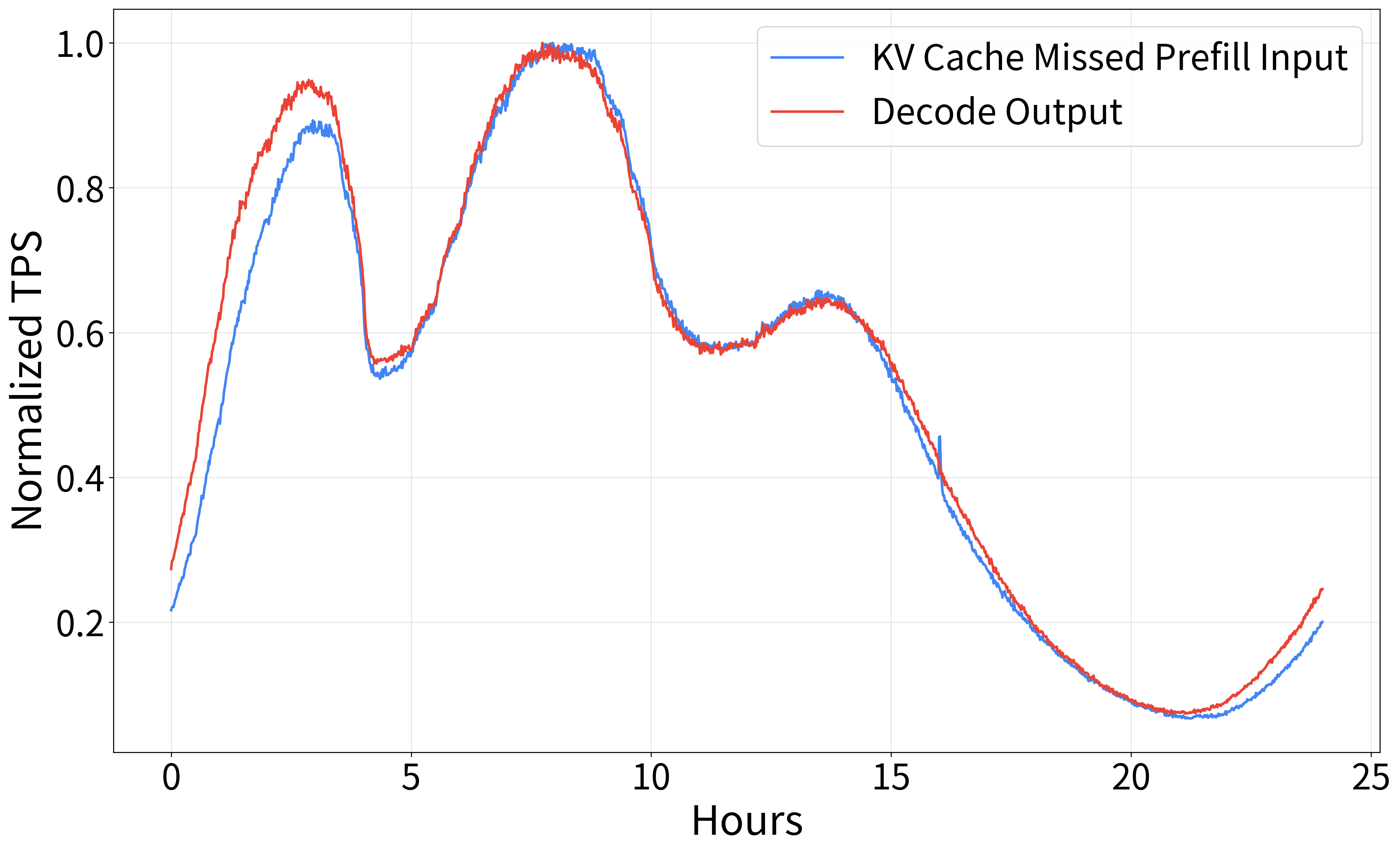}
        % \subcaption{TPS metrics showing significant differences between peak and off-peak periods, demonstrating good load reflection capability.}
        \subcaption{Normalized TPS}
        \label{fig:tps-metric}
    \end{subfigure}
    % \hfill % Adds flexible space between the two subfigures
    \begin{subfigure}[b]{0.24\textwidth}
        \centering
        \includegraphics[width=\linewidth]{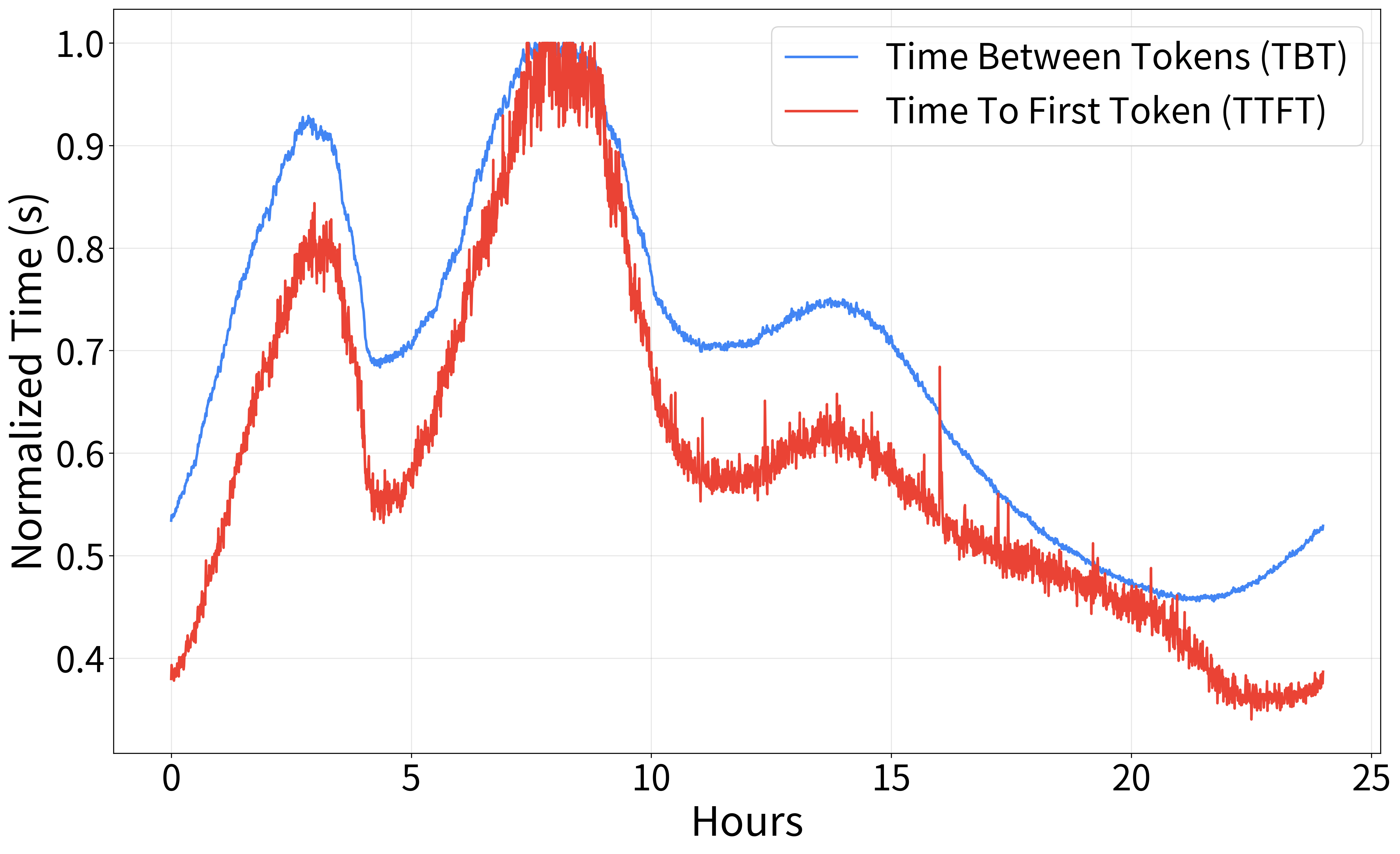}
        % \subcaption{Latency metrics \placeholder{update caption and diagram}}
        \subcaption{Normalized latencies}
        \label{fig:latency-metric}
    \end{subfigure}
    % \vspace{1em} % Add some vertical space between the rows
    \begin{subfigure}[b]{0.24\textwidth}
        \centering
        \includegraphics[width=\linewidth]{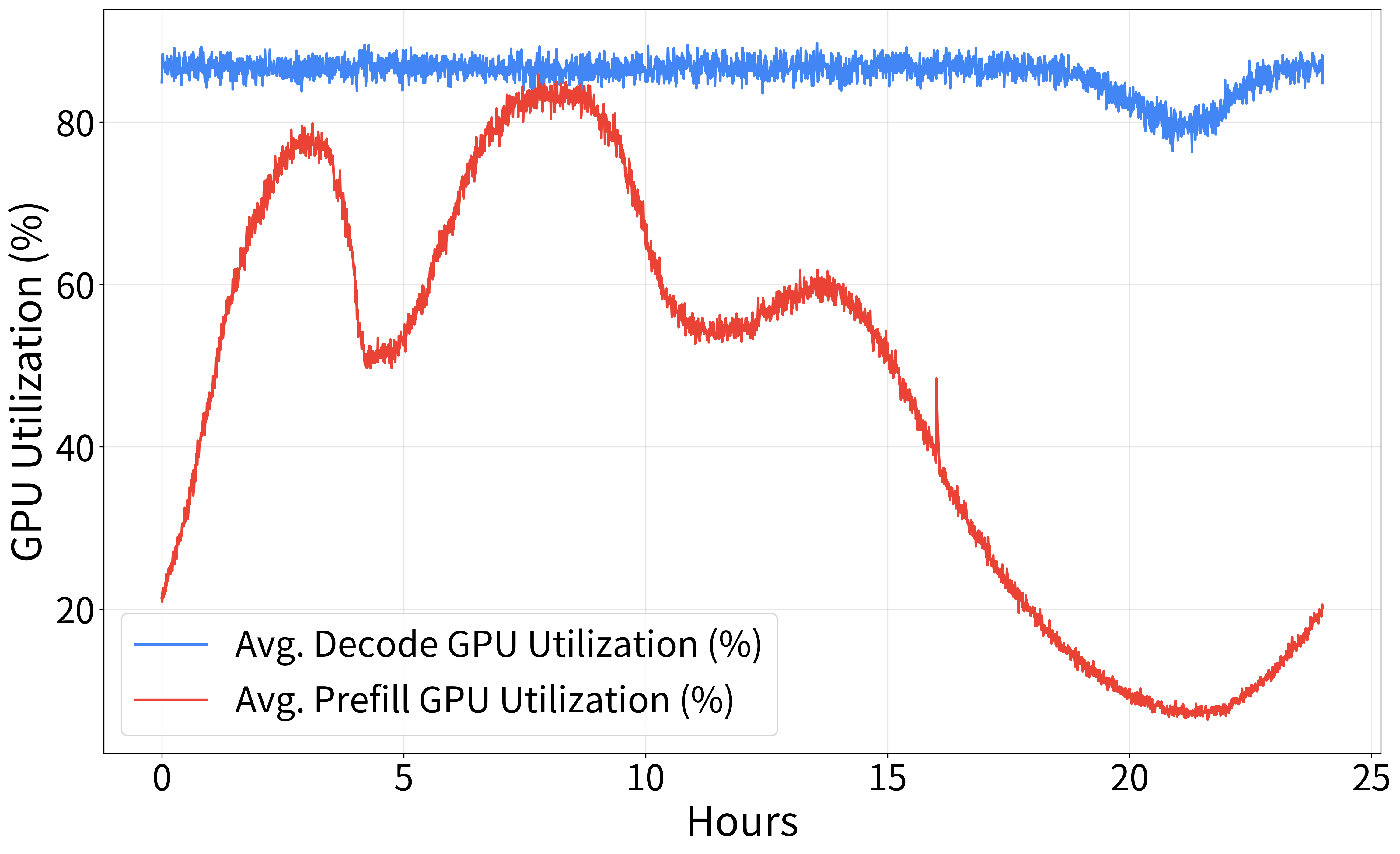}
        % \subcaption{GPU Utilization \placeholder{update caption}}
        \subcaption{GPU utilization}
        \label{fig:gpu-util} % Corrected label
    \end{subfigure}
    % \hfill
    \begin{subfigure}[b]{0.24\textwidth}
        \centering
        \includegraphics[width=\linewidth]{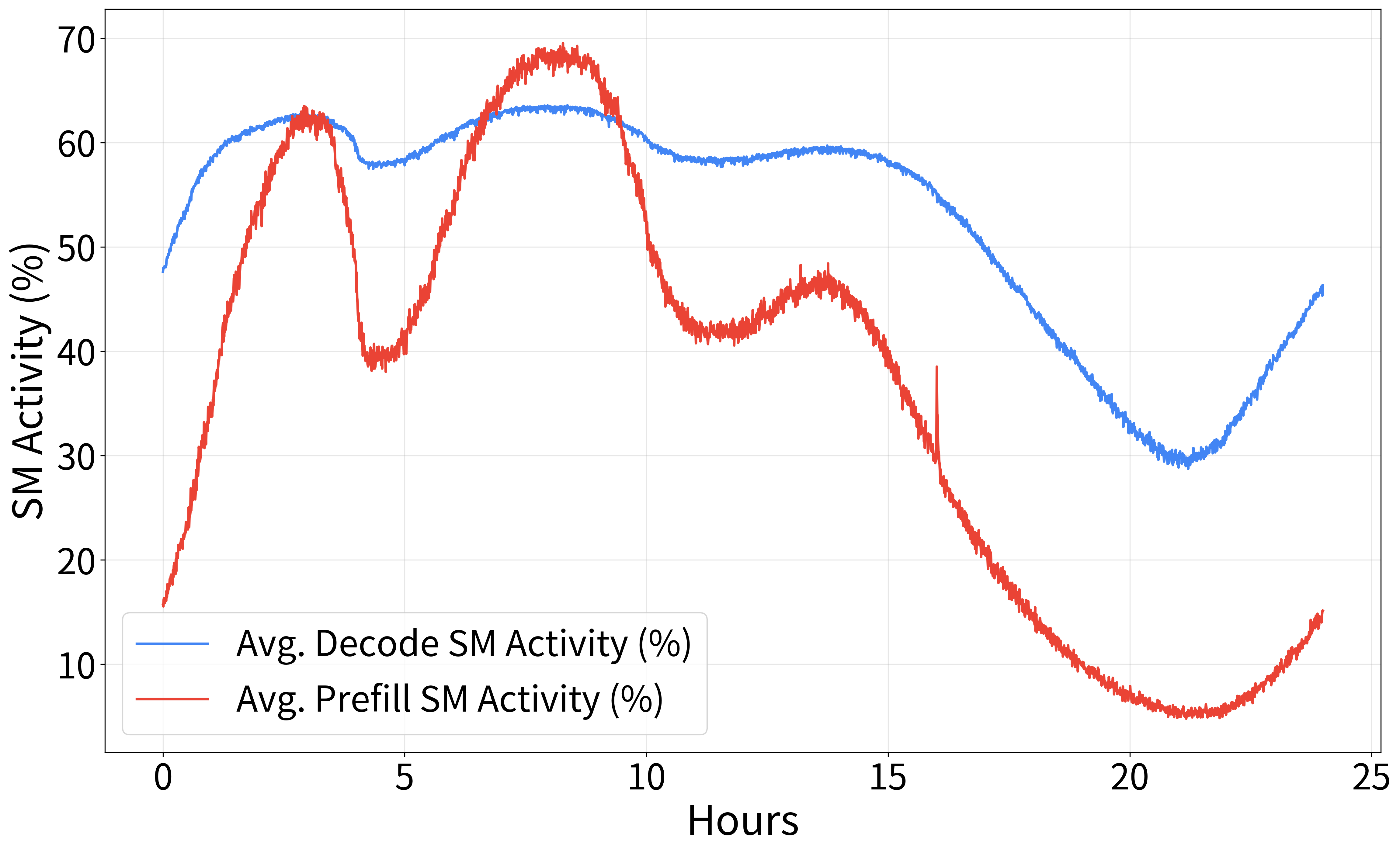}
        % \subcaption{SM Activity \placeholder{update caption}}
        \subcaption{SM activity}
        \label{fig:sm-activity} % Corrected label
    \end{subfigure}
    
    \caption{Performance metrics of an \textbf{open-domain dialogue} service with \ModelC without autoscaling.}
    \label{fig:open-dialogue-combined}
\end{figure*}

While traditional autoscalers often default to hardware utilization, our investigation revealed that such metrics could be misleading in a P/D disaggregated environment. This necessitates a principled empirical search for a metric that could serve as a robust proxy for the overall health and load of the system. A viable metric must not only track load pressure with a high signal-to-noise ratio but also be universally applicable across diverse models and hardware. Failing to identify the correct signal would lead to instability and resource waste. To characterize a P/D-disaggregated serving stack, we group candidate metrics into three classes:
\begin{itemize}
  \item \textbf{Throughput}: Prefill and decode TPS;
  \item \textbf{Hardware}: Prefill/decode GPU SM activity and GPU utilization~\cite{nvidia_sm_activity,nvidia_gpu_util,trainy_sm_eff};
  \item \textbf{Latency}: TTFT and TBT~\cite{databricks_ttft,agrawal2024taming}.
\end{itemize}
We used metric traces from an open-domain dialogue service (Figure~\ref{fig:open-dialogue-combined}) with hundreds of GPUs for preliminary verification. To demonstrate that these insights are modality-agnostic, we repeated the study on a vision–language search service; the results appear in the appendix (Figure~\ref{fig:vlm-combined}). The same qualitative patterns hold for other workloads such as web search, long-form content understanding, real-time audio conversation, real-time video processing, and code generation. 

\noindent{\textbf{Throughput Metrics Analysis.}}
We found that throughput metrics show significant differences between peak and off-peak periods, with high signal-to-noise ratios, accurately reflecting service load conditions (as shown in Figure~\ref{fig:tps-metric}). These metrics respond quickly to traffic changes, enabling timely scaling operations. Due to the interference of the KV cache hit rate, prefill TPS measurements under caching are unreliable for autoscaling, thus only "KV cache missed prefill TPS" and decode TPS are considered.

\noindent{\textbf{Hardware-level Metrics Analysis.}}
Hardware-level metrics show a clear polarization: the hardware metrics of prefill stage instances (such as prefill GPU utilization and SM activity) respond sensitively to load changes with high signal-to-noise ratios, while the decode stage hardware metrics (such as decode GPU utilization and SM activity) maintain high values even under slight pressure, showing low sensitivity to load changes (Figure~\ref{fig:gpu-util},~\ref{fig:sm-activity}). This difference stems from the distinct characteristics of the two stages. The prefill stage is compute-intensive, so its hardware utilization is linearly correlated to load. In contrast, the decode stage is memory-bound, with a large portion of its utilization coming from KV cache storage and data transfer operations. These memory operations keep hardware-level metrics at consistently high levels regardless of modest drops in workload, making them far less sensitive to real load changes.

\noindent{\textbf{Latency Metrics Analysis.}}
Latency metrics—TTFT and TBT—capture workload pressure but react in a distinctly non-linear fashion (Figure~\ref{fig:latency-metric}). When the workload is on the lower end, both curves remain nearly flat; once the load moves to the higher end, they shoot upward abruptly. Such a cliff-like transition makes it impossible to scale resources proportionally from latency alone and instead demands a negative-feedback controller. Among the two, TBT provides a clearer signal, whereas TTFT suffers from a lower signal-to-noise ratio, limiting its effectiveness as a primary autoscaling indicator.

From the above analysis, a clear pattern emerges: throughput and hardware-level metrics vary proportionally and predictably with changes in resource allocation, whereas latency metrics exhibit a non-linear, threshold-driven behavior. This dichotomy demands a hybrid policy approach rather than a one-size-fits-all solution. Instead, a hybrid approach is required—one that leverages proportional control for linear metrics and negative-feedback mechanisms for non-linear ones. Guided by this insight, we design two complementary scaling algorithms that together form the basis of HeteroScale’s stable and efficient autoscaling framework. The full algorithms are detailed in Appendix~\ref{sec:scaling_algorithms}.

\paragraph{\textbf{Proportional Control for Linear Metrics}} The key innovation here is its application in a coordinated fashion: the scaling signal from one component (e.g., decode TPS) is used to calculate the required capacity for \textit{both} the prefill and decode pools, strictly enforcing the target P/D ratio. This transforms a simple algorithm into a powerful mechanism for maintaining architectural integrity. The system calculates the desired total capacity based on the target-per-instance metric. Then it applies the P/D ratio to determine the final instance counts for each role, preventing the instabilities that would arise from scaling them independently.

\paragraph{\textbf{Negative Feedback for Non-Linear Metrics}} For highly non-linear, cliff-like metrics such as latency (TTFT and TBT), a proportional response would be dangerously unstable, leading to severe over-provisioning and oscillation. Here, we apply a more conservative \textbf{negative feedback strategy}. This approach functions as a safety mechanism rather than a primary scaling driver. It uses a multi-tier system of thresholds to trigger fixed, incremental adjustments only when SLOs are at risk of being breached. For example, a moderate latency increase triggers a small, fixed-percentage scaling event (e.g., 10\%), while a severe breach triggers a larger, more urgent response (e.g., 20\%). This cautious, step-based approach prevents the system from overreacting to latency's volatile behavior, providing a crucial stability layer that complements the primary proportional strategy.

\subsubsection{Workload-centric Policy Curation}
In HeteroScale, autoscaling follows a workload-centric pipeline (Algorithm~\ref{alg:policy-pipeline}). It begins with a pressure test that, given a service and its workload profile, identifies the optimal P/D ratio and estimates the expected performance metric under load. Next, each candidate scaling policy is simulated under these baseline conditions to assess its effectiveness. Finally, the system selects the policy that maximizes the chosen objective (e.g., throughput while maintaining SLO compliance).

On our production platform, decode TPS serves as the primary scaling metric on our production platform. A comprehensive evaluation presented in Section~\ref{sec:eval} highlights the effectiveness and limitations of the aforementioned real-time metrics for scaling P/D-disaggregated services.

\begin{algorithm}[h]
\caption{Workload-centric Policy Curation}
\label{alg:policy-pipeline}
\SetAlgoLined
\SetKwFor{ForEach}{for each}{do}{end}

\KwIn{Service $\mathcal{S}$; workload $\mathcal{W}$; candidate policies $\mathcal{P}$}
\KwOut{Optimal policy $p_{\mathrm{opt}}$; optimal P/D ratio $r_{\mathrm{opt}}$; expected metric $\hat{m}$}

$(r_{\mathrm{opt}}, \hat{m}) \gets \textsc{PressureTest}(\mathcal{S}, \mathcal{W})$\;

\ForEach{$p \in \mathcal{P}$}{
  $score[p] \gets \textsc{Simulate}(\mathcal{S}, \mathcal{W}, p, r_{\mathrm{opt}}, \hat{m})$ }

$p_{\mathrm{opt}} \gets \arg\max\limits_{p \in \mathcal{P}} \, score[p]$\;

\Return{$(p_{\mathrm{opt}}, r_{\mathrm{opt}}, \hat{m})$}\;
\end{algorithm}
\subsection{Federated Pre-Scheduling Layer}
The federated pre-scheduling layer provides a higher-level view of GPU resources and makes scheduling decisions. It is responsible for translating the "what to scale" decisions from the policy engine into "where to place" decisions on the physical infrastructure. This is achieved through a heterogeneous resource management framework and a network affinity-aware scheduling algorithm that considers both service requirements and global cluster efficiency. The detailed logic for this scheduling process is available in Appendix~\ref{sec:scaling_algorithms}.

\subsubsection*{Heterogeneous Resource Management Framework} Managing heterogeneous GPU resources efficiently is essential to maximize resource utilization and service performance. HeteroScale implements a heterogeneous resource management framework that efficiently allocates and manages diverse GPU resources.

Algorithm \ref{alg:hetero-resource-allocation} in appendix outlines the heterogeneous resource allocation process. The algorithm takes into account resource requests, available resources by type, service priorities, and resource constraints. It sorts requests by priority and attempts to allocate preferred resources for each request. If preferred resources are not available, it tries alternative compatible resources.

% In the future, we plan to add capabilities such as proactive spot provisioning (purposely creating temporary instances based on predicted demand for other services to leverage) and more sophisticated resource modeling.

\subsubsection*{Deployment Group Abstraction for Network Affinity}\label{subsubsec:deploymentgroup}
In P/D disaggregated services, performance is critically dependent on the latency and bandwidth of the network link connecting prefill and decode instances, as this link carries the KV cache~\cite{chen2024kvdirect}. To minimize this communication overhead, instances must often be co-located within a network domain that supports high-speed interconnects like RDMA, such as under a common aggregation switch (e.g., S2).

To manage this co-location requirement while allowing for flexible scaling, HeteroScale introduces the \textbf{Deployment Group} abstraction. A Deployment Group is a logical container for the prefill and decode roles of a single service. Key characteristics include:
\begin{itemize}
    \item \textbf{Shared Scheduling Domain:} All instances within a group are bound by a common \textbf{network affinity constraint}. For high-performance services, this may mandate placement under the same S2 switch. For services with less stringent needs, this constraint can be relaxed to the physical cluster level.
    \item \textbf{Independent Scaling Roles:} Within a group, the prefill and decode roles function as independent deployment units that can be scaled out or in separately, subject to the system's P/D ratio maintenance logic.
\end{itemize}

This abstraction enables a more sophisticated, priority-aware scaling strategy. When a service requests a scale-out, the scheduler evaluates whether to expand an existing Deployment Group or provision a \textit{new one} in a different network domain. This decision is not merely based on available capacity but is guided by the priority of the underlying hardware resources. For instance, the system may choose to create a new Deployment Group in a lower-priority resource pool to conserve high-priority resources within an existing group's domain for more critical workloads. This intelligent placement prevents a service from being bottlenecked by local resource exhaustion while also enabling global, priority-based optimization of valuable, network-proximate hardware.

\subsubsection*{Prioritizing Resources with RDMA Subgroups} 
\begin{figure}[t]
    \includegraphics[width=\linewidth]{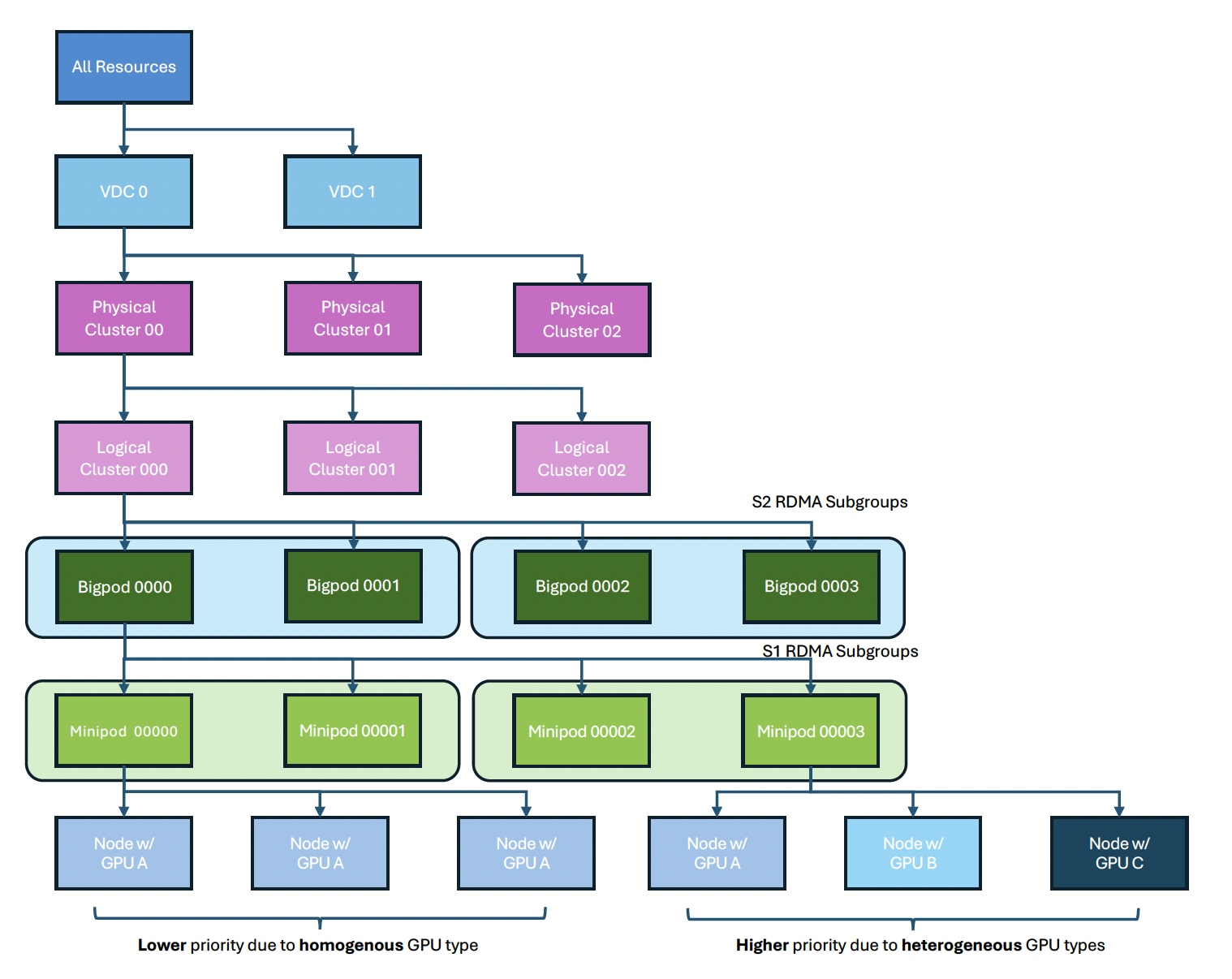}
    \caption{Topological resource tree illustrating the hierarchical organization of resources from data centers (VDCs) down to individual nodes. This view, which includes network domains like S1/S2 RDMA Subgroups, is used by the federated pre-scheduler to make network affinity-aware placement decisions.}
    \label{fig:resource-tree}
\end{figure}

The strategic placement of Deployment Groups requires a method for valuing different resource pools. In a large-scale cluster, not all hardware is equal. A rack containing multiple GPU types under a single S1 switch is a scarce, high-value asset, as it allows prefill and decode roles to be placed on different, specialized hardware while maintaining minimal network latency. To prevent services with loose affinity requirements from consuming these premium resources, HeteroScale introduces a priority system, which are logical collections of S1 or S2 switches as illustrated in Figure~\ref{fig:resource-tree}.

This system is implemented through \textbf{RDMA Subgroups}. An RDMA Subgroup is a logical collection of one or more S1/S2 switches, classified into a distinct priority tier based on the hardware they contain. By assigning priorities to these hardware pools, the scheduler can make more intelligent resource allocation decisions. We define three primary priority tiers, ranked from lowest to highest:

\begin{itemize}
    \item \textbf{Low Priority: S2 Homogeneous GPU Subgroups.} These contain S2 switches where all underlying GPUs are of a single type. They are the most common resource configuration and are suitable for the widest range of services.

    \item \textbf{Medium Priority: S2 Heterogeneous GPU Subgroups.} In these subgroups, an S2 switch manages a mix of GPU types, but each of its underlying S1 switches remains homogeneous. This allows for hardware specialization within the broader S2 domain.

    \item \textbf{High Priority: S1 Heterogeneous GPU Subgroups.} These are the most valuable pools, containing S1 switches that directly connect machines with different GPU types. They enable the most demanding heterogeneous P/D configurations with the tightest possible network affinity.
\end{itemize}

When making a placement decision, the scheduler uses this hierarchy to guide its choice. For a service with low-affinity requirements (e.g., only needing cluster-level co-location), the scheduler will strongly prefer to create or expand Deployment Groups within low-priority subgroups. Conversely, when a service explicitly requires a high-affinity, heterogeneous setup (e.g., different GPUs under one S1 switch), the scheduler filters for compatible high-priority subgroups. This ensures that scarce, high-performance resource pools are reserved for the workloads that need them most, optimizing global cluster efficiency.

\subsubsection*{P/D Ratio Maintenance} One of the key challenges in P/D disaggregated serving is maintaining an optimal ratio between prefill and decode instances. This ratio depends on various factors, including model architecture, prompt length distribution, and generation length distribution. HeteroScale implements a P/D ratio maintenance mechanism that maintains the preset ratio during scaling operations. In the online environment, we use a fixed P/D ratio for scaling. This P/D ratio is derived from service pressure tests and historical empirical data.

% In future work, we will continue to explore the optimal service P/D ratio configuration under heterogeneous hardware combinations in a more dynamic manner based on the various metrics of services.
The P/D ratio calculation and maintenance process takes into account the current number of prefill and decode instances, the target P/D ratio, a scaling threshold, and historical workload data. It first calculates the current P/D ratio and checks if an adjustment is needed based on the threshold. If an adjustment is required, it calculates the optimal instance counts based on workload data and applies a smooth transition to avoid abrupt changes. The prefill and decode instances are always scaled in or out in simultaneously. This approach is intended to prevent a scenario where, after either prefill or decode instances are successfully scaled out individually, the other one fails to scale due to insufficient resources, thereby avoiding the issue of an imbalanced P/D ratio. 

However, under certain circumstances, scaling prefill and decode instances simutaneously may also lead to P/D ratio imbalance. For instance, after we create a new Deployment Group with a fixed P/D ratio, prefill and decode instances may start out of order due to differences in configurations and startup strategies. This can result in a temporary imbalance of the P/D ratio, which affects SLOs such as TTFT. To address this this issue, we have added corresponding support at service framework level: if the ratio of prefill to decode instances in ready state deviates significantly from the configured ratio, service discovery for the role with a larger quantity will be suspended, and registration will proceed only after the other instances from the other role complete their service discovery registration and instances in ready state recover to a tolerable P/D ratio.

\subsubsection*{The Affinity-Aware Scheduling Algorithm}
The core of HeteroScale's scheduling algorithm lies in its network affinity-aware scheduling loop, which translates scaling policies into concrete pod placements. This process, outlined in Algorithm \ref{alg:hetero-resource-allocation}, ensures that every placement decision honors both the service's specific network requirements and the global resource priorities of the cluster. The scheduling cycle proceeds as follows:

1. \textbf{Topology Discovery:} At the start of each cycle, the controller builds a fresh \textit{topological resource tree} (Figure~\ref{fig:resource-tree}). This provides a live, hierarchical view of all available GPUs and their precise locations within the network fabric (node, S1/S2 switch).

2. \textbf{Request Sorting:} All pending scaling requests generated by the autoscaling policy engine are sorted, primarily by service priority, to ensure that critical workloads are allocated resources first.

3. \textbf{Candidate Evaluation (for Scale-Out):} For each high-priority scale-out request, the scheduler identifies all valid placement options. This involves evaluating both expanding the service's \textit{existing} Deployment Groups and creating \textit{new} ones in different, compatible network domains.

4. \textbf{Priority-Based Selection:} Each candidate placement is scored based on the priority of its associated RDMA Subgroup. The scheduler then selects the optimal placement that satisfies the service's affinity constraints while minimizing the consumption of high-priority resources. For instance, it will prefer to place a low-affinity service in a low-priority subgroup, even if this requires creating a new Deployment Group, thereby preserving high-value hardware for workloads that require it.

5. \textbf{Virtual Allocation:} Once a placement decision is made, the chosen resources are virtually deducted from the topological tree for the remainder of the cycle. This atomic reservation prevents them from being considered for lower-priority requests within the same loop.

Handling scale-in requests is more straightforward. The scheduler simply selects one or more of the service's Deployment Groups to scale in, typically targeting those occupying high-priority resource pools to free them up. To maintain a consistent state, the released resources are not immediately added back to the available pool; rather, the entire resource view is rebuilt from the underlying cluster state at the beginning of the next scheduling cycle.

\subsubsection*{Extending to Disaggregated MoE}
HeteroScale extends to disaggregated MoE by adapting its Deployment Group abstraction for services with distinct prefill and decode components. The prefill stage, itself comprising attention (attn) and feed-forward (ffn) instances, is co-located by the scheduler within a high-affinity S1 switch, while the entire prefill-decode pair is placed under a common S2 switch. This hierarchical scheduling enables dual-ratio control: maintaining a strict attn-to-ffn ratio within prefill replicas and a proportional balance between prefill and decode components, which ensures architectural stability and performance.

\subsection{Sub-cluster Scheduling Layer}
All operations on Deployment Groups initiated in the pre-scheduling layer are routed through this component, which delegates the requests to the Kubernetes API server for the creation or update of the corresponding CRDs. At the same time, this layer exposes the node API upward, enabling topology construction. Detailed discussion of this layer is outside the scope of this paper.

\subsection{System Stability Mechanisms}
Ensuring system stability is critical for autoscaling systems, particularly in production environments. HeteroScale implements several mechanisms to maintain system stability:

\textbf{Anti-flapping Mechanisms.}
Rapid oscillations between scaling in and scaling out, commonly referred to as flapping, can lead to resource waste and system instability. HeteroScale mitigates this problem through a combination of complementary mechanisms. First, it enforces cooling periods, ensuring a minimum interval between scaling actions to prevent rapid oscillations. This is reinforced by hysteresis thresholds, which use different trigger points for scaling out and scaling in, creating a buffer zone that promotes stability. In addition, dampening factors are applied to moderate the scale of adjustments, further smoothing the system’s response to changing conditions.

\textbf{Disaster Recovery Measures}
System failures or unexpected events can interrupt autoscaling operations and threaten service stability. \PlatformB  employs a set of disaster recovery measures. One key mechanism is \textbf{soft scaling in}, in which instances identified for removal are withdrawn from service discovery but kept running. During this observation period, the system monitors SLOs such as latency in real time. If performance remains within targets, the instances are then terminated; however, if degradation is detected, they are reinstated immediately, avoiding the startup delay associated with provisioning new instances. The platform also preserves critical state information to enable fast resumption of normal operations after a failure. Finally, when resources become constrained, it applies graceful degradation strategies, maintaining essential functionality while temporarily reducing non-critical services.

\section{Evaluation} \label{sec:eval}
To validate the effectiveness of HeteroScale's design, we conducted a series of targeted experiments in both controlled environments and production settings with P/D disaggregated LLM services powered by \ModelC at 
\CompanyA.

HeteroScale has been successfully deployed across multiple services on tens of thousands of GPUs, demonstrating significant and consistent improvements in resource utilization in a large-scale, real-world environment.

\subsection{P/D Ratio}
Given the nature of heterogeneous hardware and the distinct bottlenecks in the prefill (P) and decode (D) stages, achieving maximum performance often requires an asymmetric allocation of prefill and decode instances \cite{qin2024mooncake, wang2025burstgptrealworldworkloaddataset}. We conducted two P/D ratio investigation experiments on different services, each with unique input-output length distributions and SLO constraints, running on 16 nodes, each equipped with eight H20 GPUs. Service~A had an average input length of roughly 3k tokens and an output length of about 350 tokens, giving an input-output (I/O) ratio of 8.5. Its SLOs were set to TTFT $\leq$ 1 s and TBT $\leq$ 40 ms. Service~B had a longer input length of around 7.8k tokens and an output length of about 700 tokens, corresponding to an I/O ratio of 11, with SLOs of TTFT $\leq$ 1 s and TBT $\leq$ 20 ms. 

\begin{figure*}[t]
    \centering

    % ----- Left: one figure with two subfigures (using the subfigure environment) -----
    \begin{minipage}[t]{0.7\textwidth}
        \centering
        % Use subfigure for the left chart
        \begin{subfigure}[t]{0.5\linewidth}
            \centering
            \includegraphics[width=\linewidth]{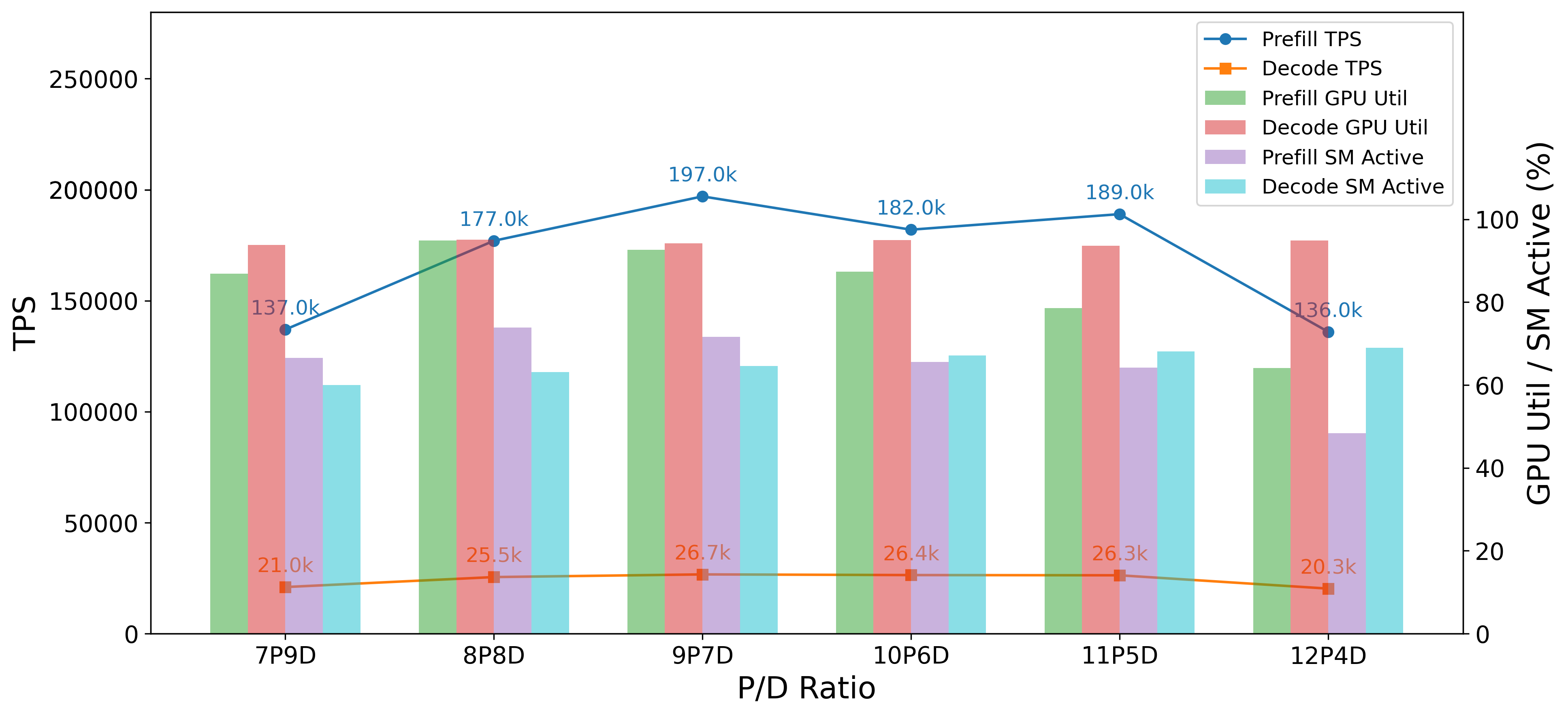}
            \caption{Service A} % Use \caption inside subfigure
            \label{fig:pd-servicea}
        \end{subfigure}%
        % Use subfigure for the right chart
        \begin{subfigure}[t]{0.5\linewidth}
            \centering
            \includegraphics[width=\linewidth]{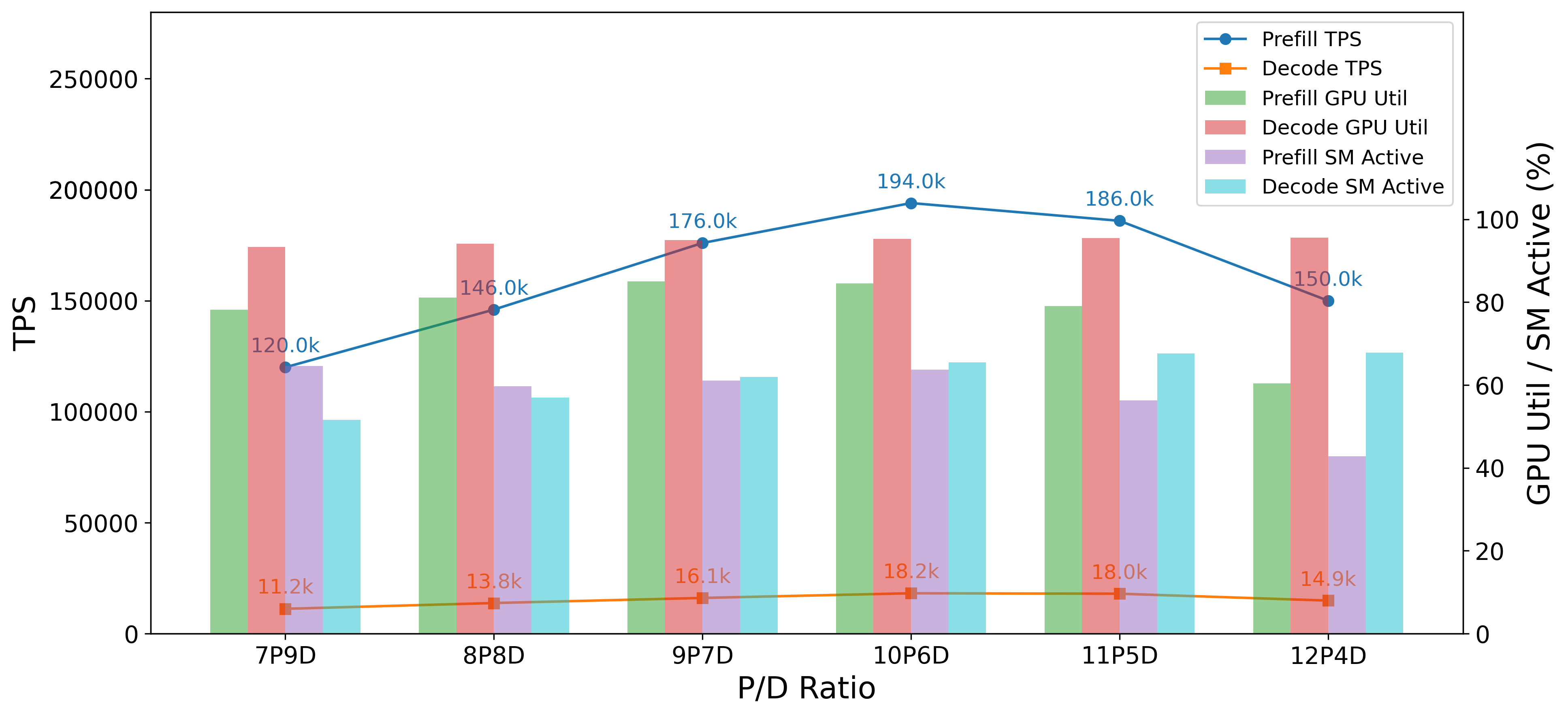}
            \caption{Service B} % Use \caption inside subfigure
            \label{fig:pd-serviceb}
        \end{subfigure}

        % This now serves as the main caption for the two subfigures above
        \caption{Maximum TPS with different P/D ratios for two services with distinct input–output length distributions and SLO constraints}
        \label{fig:pd-ratio-tps}
    \end{minipage}%
    % ----- Right: a separate figure with its own caption (this part is unchanged) -----
    \begin{minipage}[t]{0.30\textwidth}
        \centering
        \includegraphics[width=\linewidth]{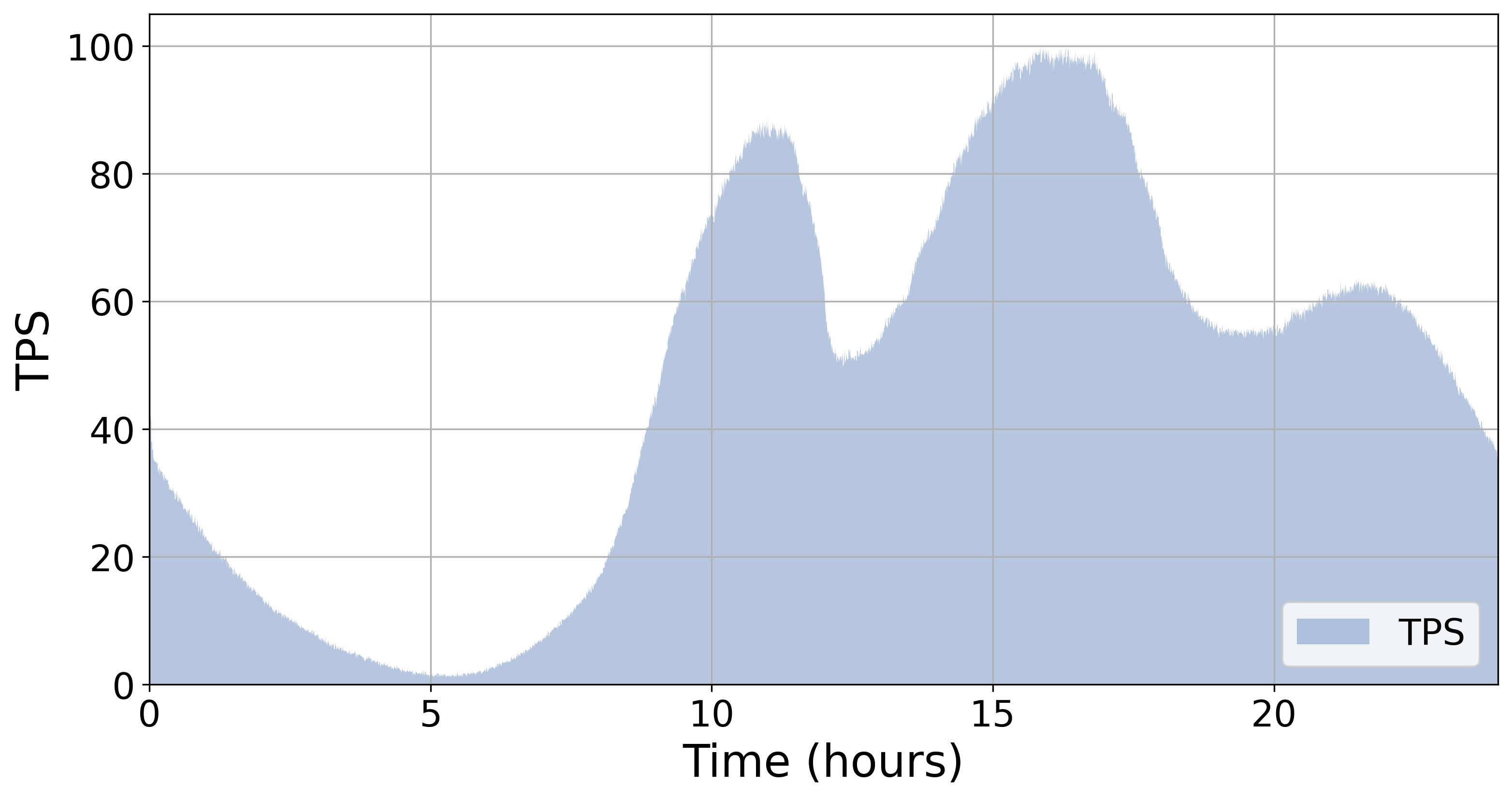}
        \captionof{figure}{Daily Workload (Norm.)}
        \label{fig:recorded-workload}
    \end{minipage}
\end{figure*}

As shown in Figure~\ref{fig:pd-ratio-tps}, both services exhibit a clear midrange peak in throughput as the P/D ratio varies, with underperformance on either side due to SLO violations. At lower ratios, scarce prefill instances cause TTFT to exceed preset threshold, capping throughput despite idle decode capacity. In contrast, higher ratios lead to excess prefill instances that overwhelm decode resources, pushing TBT beyond its limit, and dampening maximum throughput. 

This optimal P/D ratio is not fixed; in our production experience, it spans a considerable range—from $1P/5D$ to $9P/1D$—depending on the input-output length distribution, hardware configurations, and SLO priorities. This variability underscores the need for a flexible scheduling algorithm capable of adapting to diverse P/D ratios.

% Notably, this optimal P/D ratio is not fixed and can shift depending on multiple factors. For example, the optimal P/D ratio for Service~A is observed at $9P/7D$, while Service~B peaks at $10P/6D$. This difference is primarily explained by their I/O length ratios: Service~B’s larger ratio imposes greater load on the prefill stage, shifting the optimal allocation toward more prefill instances. In our production experience, we observe the optimal ratio span a considerable range—from $1P/5D$ to $9P/1D$—depending on input-output length distribution, hardware configurations, and SLO priorities.  These dynamics highlight the importance of a flexible scheduling algorithm capable of adapting to diverse P/D ratios.

% Notably, this optimal P/D ratio is not fixed; it shifts with hardware configurations, workload patterns, SLO priorities, and other contextual factors. This variability underscores the need for a flexible scheduling algorithm capable of adapting to diverse P/D ratios.

% Notably, this optimal P/D ratio is not fixed and can shift depending on multiple factors. In our production experience, we observe the optimal ratio span a considerable range—from 1:5 to 9:1—depending on input-output length distribution, hardware configurations, and SLO priorities.  These dynamics highlight the importance of a flexible scheduling algorithm capable of adapting to diverse P/D ratios.

\subsection{Metrics-based Autoscaling Policy}
As discussed in Section \ref{sec:metric-driven-scaling-policy}, we selected eight candidate metrics for evaluation: TPS, SM activity, and GPU utilization—each measured independently for prefill and decode instances (specifically, prefill SM activity, decode SM activity, prefill GPU utilization, decode GPU utilization, prefill TPS, and decode TPS)—in addition to two latency-based metrics (TTFT and TBT). To systematically evaluate these metrics, we conducted a replay of a trace collected from production workloads, enabling comparison across key performance dimensions.

\subsubsection{Workload}
To ground the experiments in realistic conditions, we sampled workload traces from an open-domain dialogue service. As one of the most prevalent applications of LLMs, this type of workloads serve as a representative benchmark for evaluating serving performance.

The daily workload reveals distinct diurnal patterns (see Figure~\ref{fig:recorded-workload}). User activity remains low during late-night and early-morning hours, followed by a sharp increase in the morning.  After a midday dip, activity rises again toward a secondary peak in the afternoon, then gradually declines and stabilizes. These fluctuations present significant challenges for autoscaling, particularly in detecting and reacting to rapid shifts in demand. To validate the selected metrics, we extracted an eight-hour workload segment spanning from morning to mid-afternoon. This segment includes two prominent peaks and valleys, offering a balanced and challenging testbed for assessing each metric’s responsiveness to real-world workload variability.

\begin{figure*}[t!]
    \centering
    \begin{subfigure}[b]{0.24\textwidth}
        \centering
        \includegraphics[width=\linewidth]{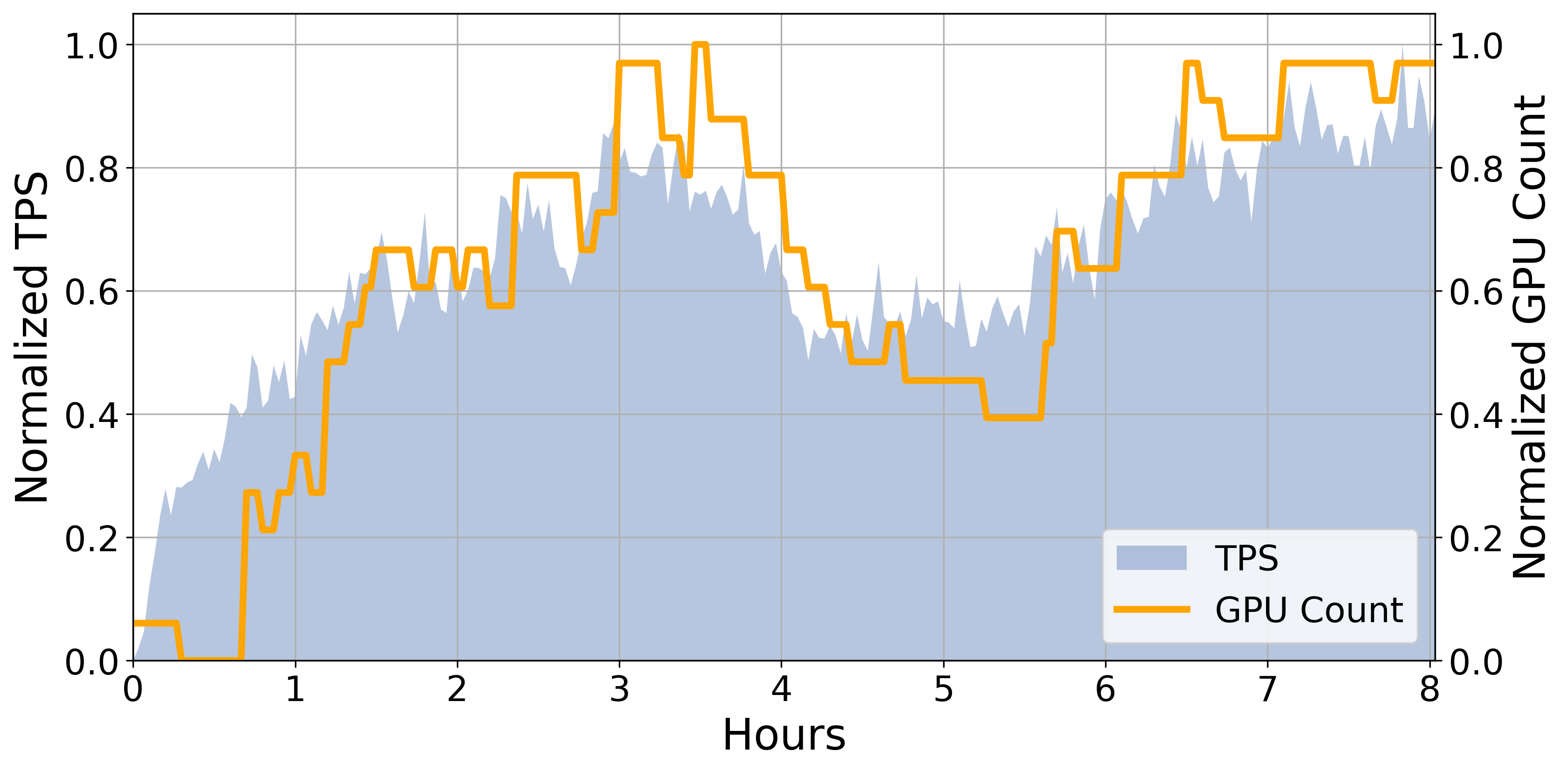}
        \subcaption{Prefill TPS}
        \label{fig:prefill-tps-autoscaling}
    \end{subfigure}
    \begin{subfigure}[b]{0.24\textwidth}
        \centering
        \includegraphics[width=\linewidth]{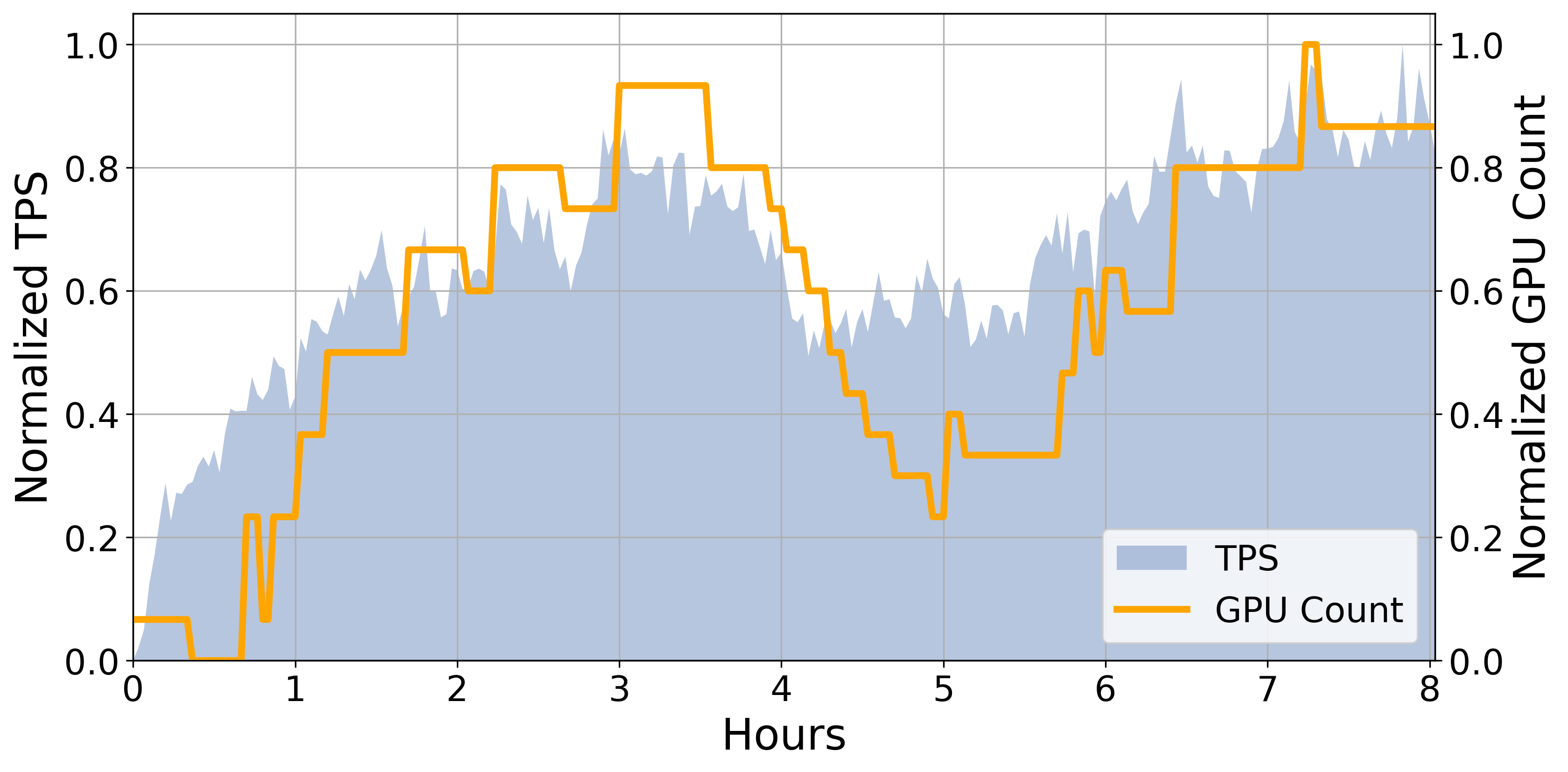}
        \subcaption{Decode TPS}
        \label{fig:decode-tps-autoscaling}
    \end{subfigure}
    \begin{subfigure}[b]{0.24\textwidth}
        \centering
        \includegraphics[width=\linewidth]{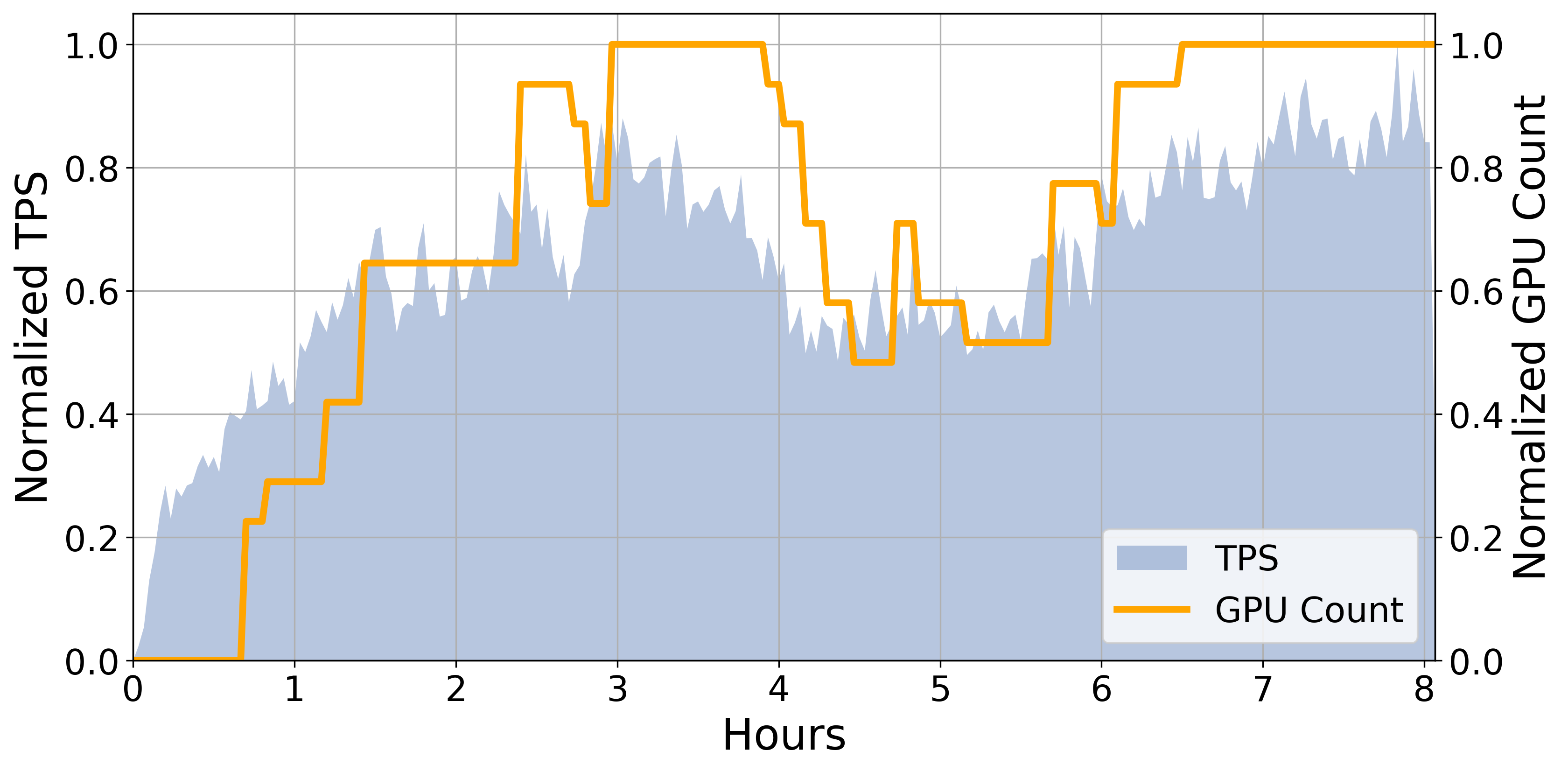}
        \subcaption{Prefill GPU Util}
        \label{fig:prefill-gpu-autoscaling}
    \end{subfigure}
    \begin{subfigure}[b]{0.24\textwidth}
        \centering
        \includegraphics[width=\linewidth]{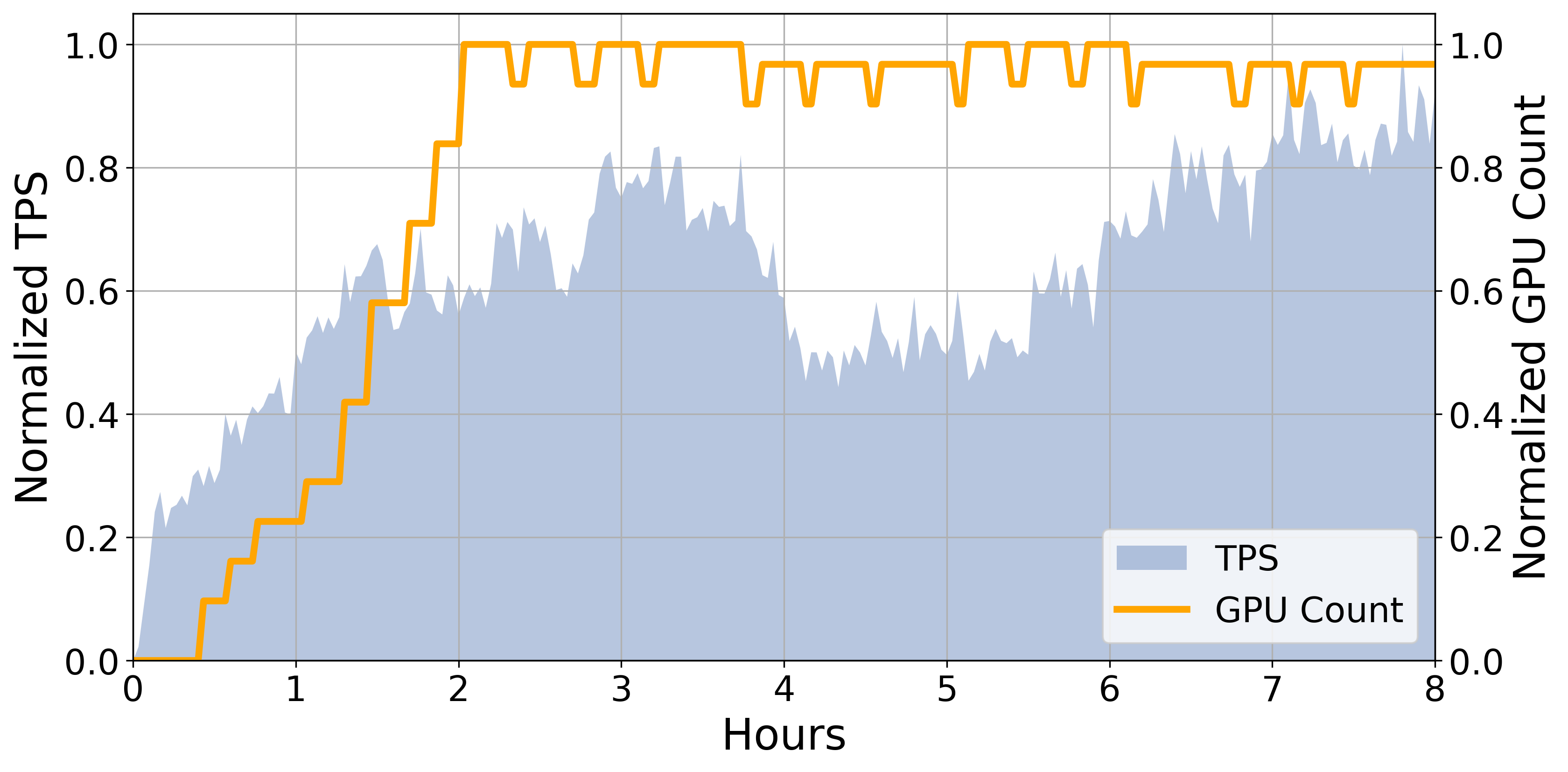}
        \subcaption{Decode GPU Util}
        \label{fig:decode-gpu-autoscaling}
    \end{subfigure}

    \begin{subfigure}[b]{0.24\textwidth}
        \centering
        \includegraphics[width=\linewidth]{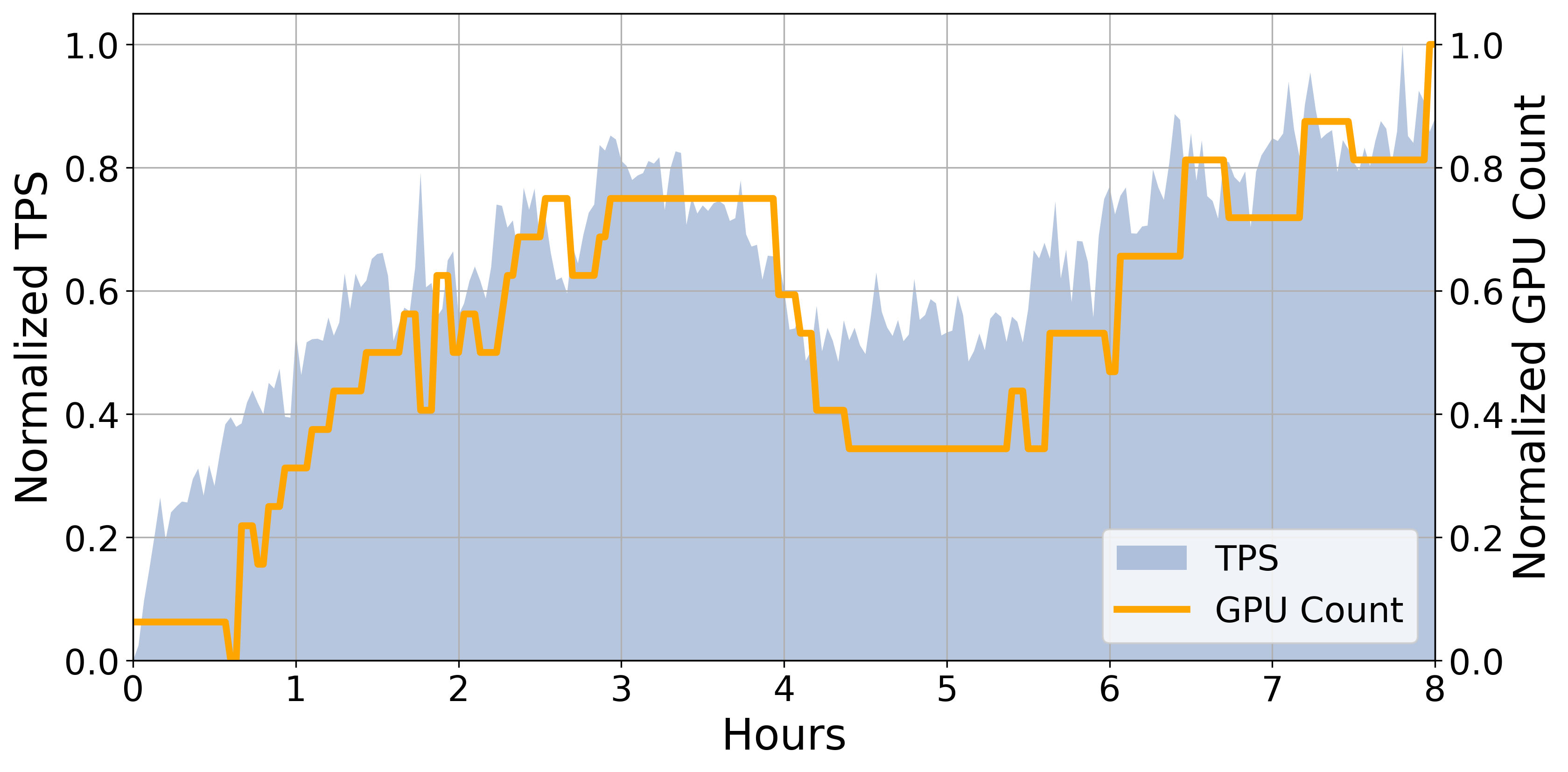}
        \subcaption{Prefill SM Activity}
        \label{fig:prefill-sm-autoscaling}
    \end{subfigure}
    \begin{subfigure}[b]{0.24\textwidth}
        \centering
        \includegraphics[width=\linewidth]{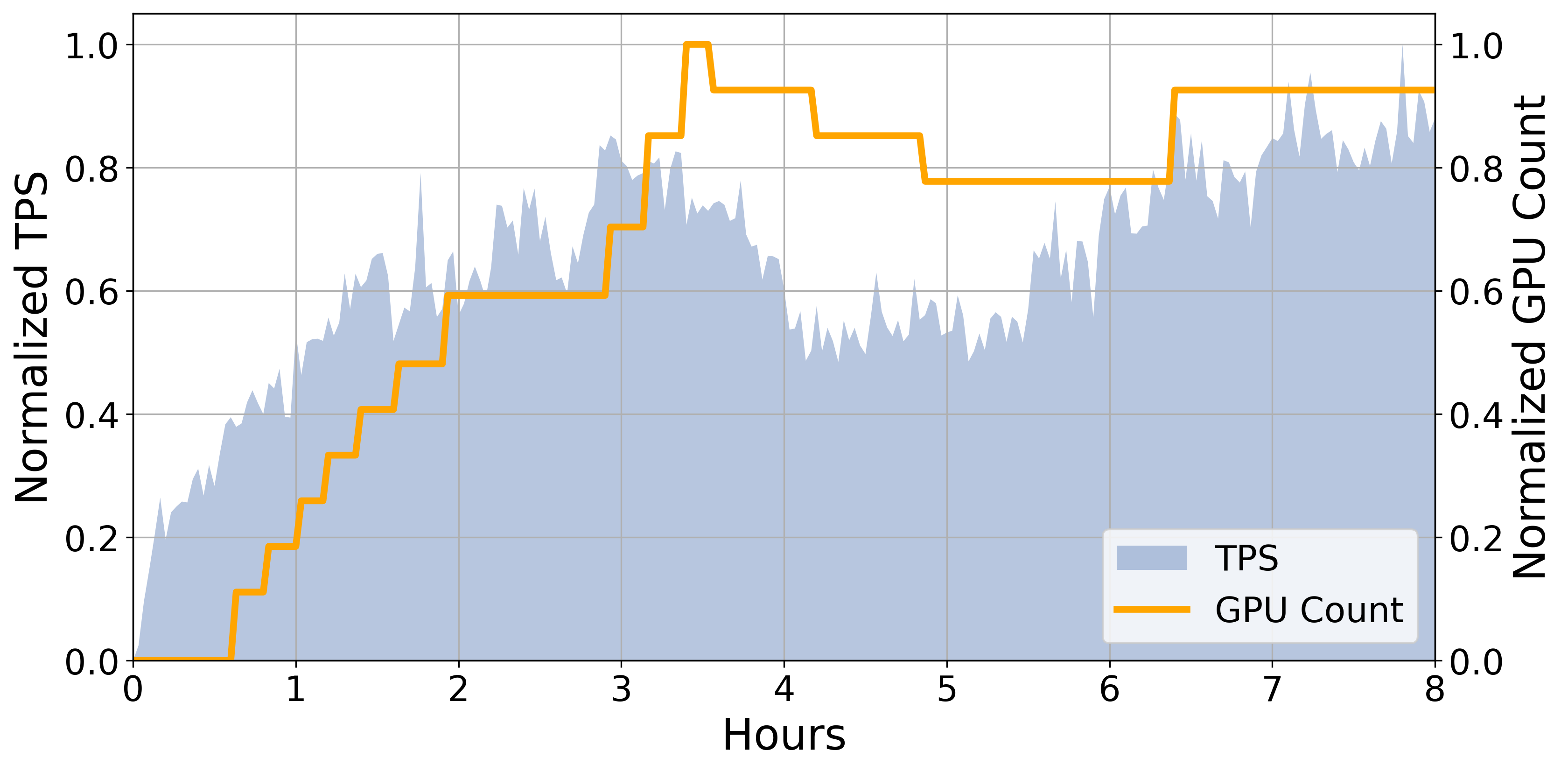}
        \subcaption{Decode SM Activity}
        \label{fig:decode-sm-autoscaling}
    \end{subfigure}
    \begin{subfigure}[b]{0.24\textwidth}
        \centering
        \includegraphics[width=\linewidth]{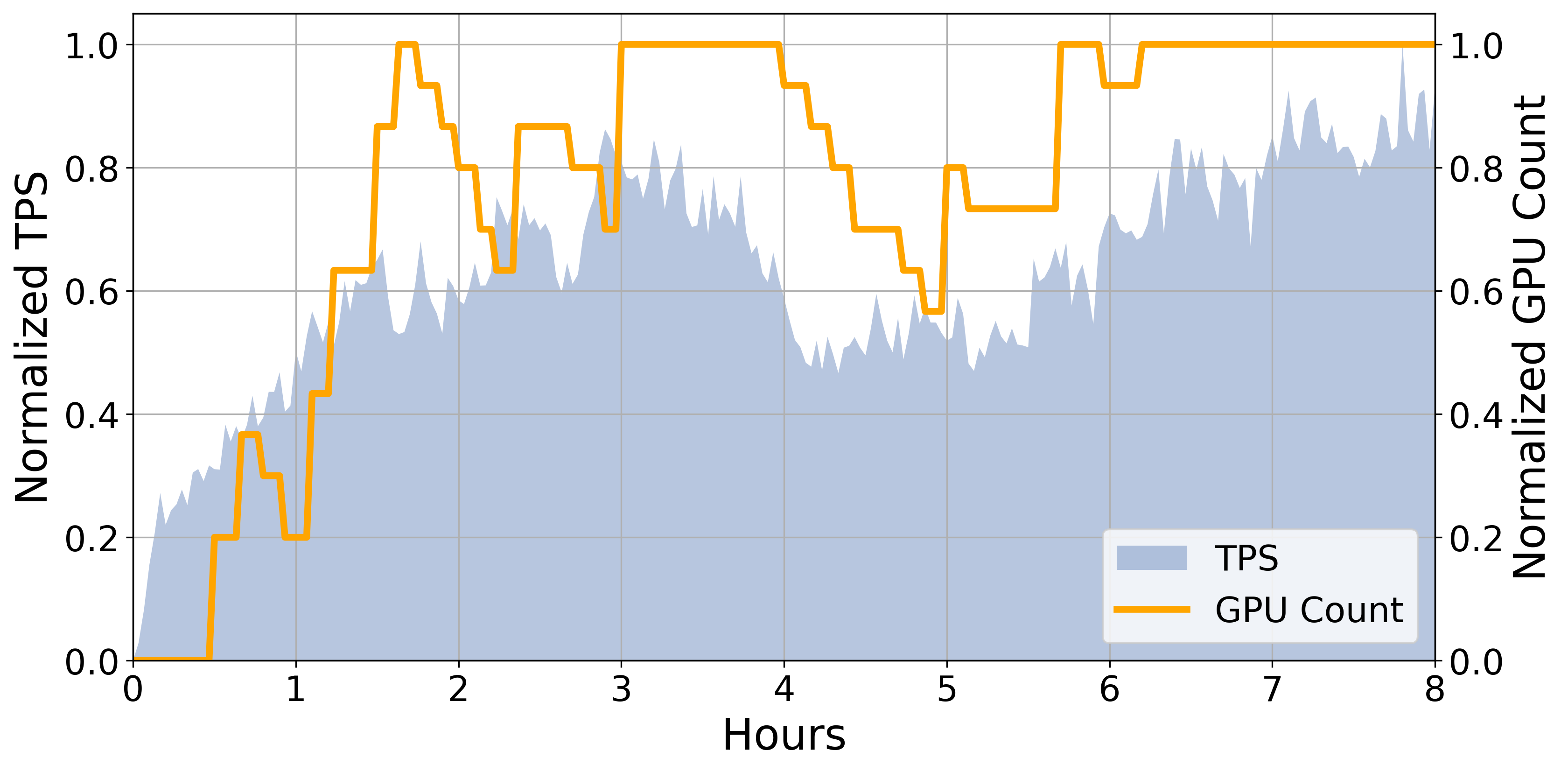}
        \subcaption{Time To First Token (TTFT)}
        \label{fig:ttft-autoscaling}
    \end{subfigure}
    \begin{subfigure}[b]{0.24\textwidth}
        \centering
        \includegraphics[width=\linewidth]{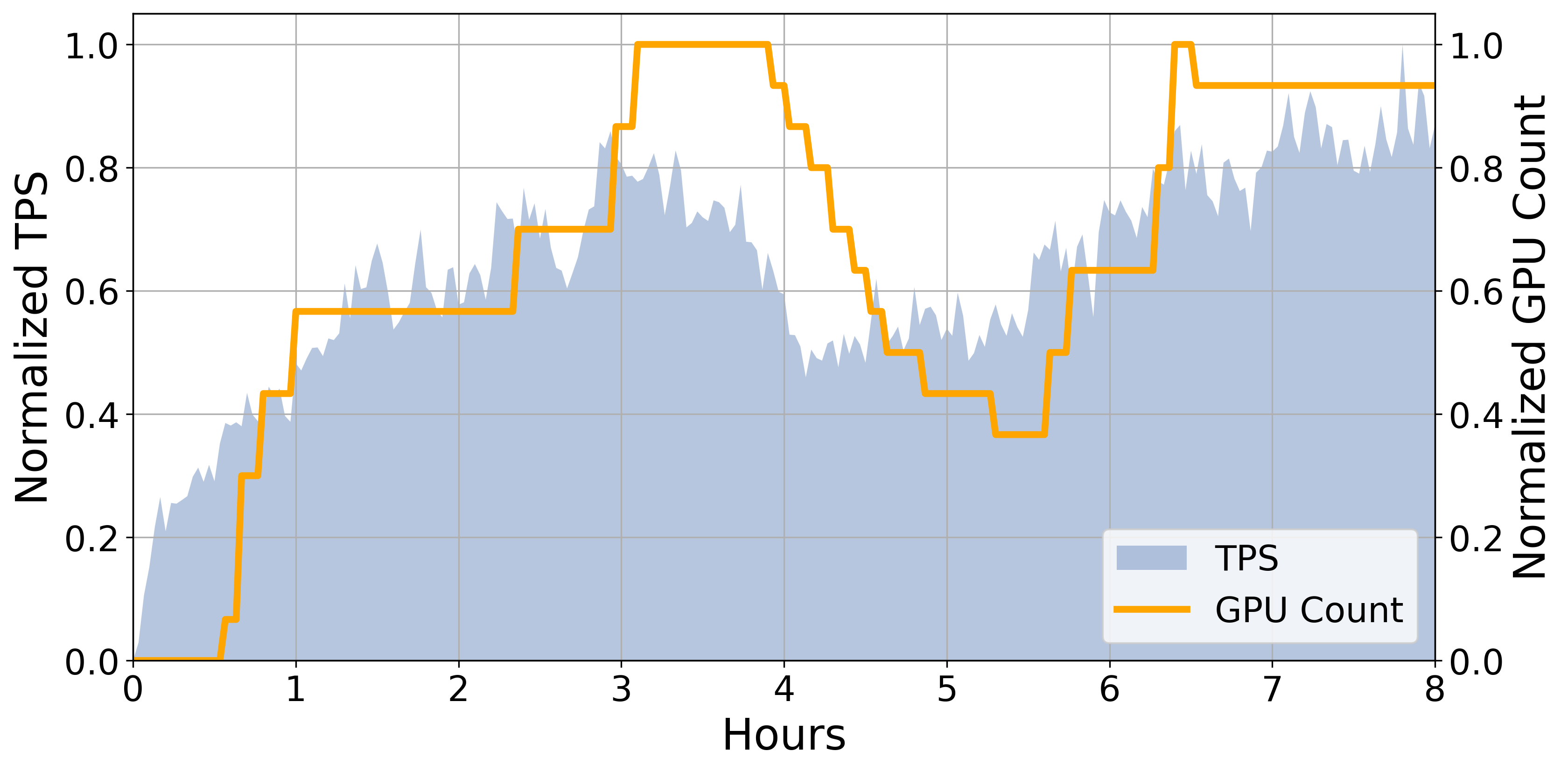}
        \subcaption{Time Between Tokens (TBT)}
        \label{fig:tpot-autoscaling}
    \end{subfigure}
    \caption{\textbf{Testing environment}: Autoscaling results with different metrics}
    \label{fig:combined-metrics}
\end{figure*}

\begin{figure*}[t!]
    \centering
    \begin{subfigure}[b]{0.24\textwidth}
        \centering
        \includegraphics[width=\linewidth]{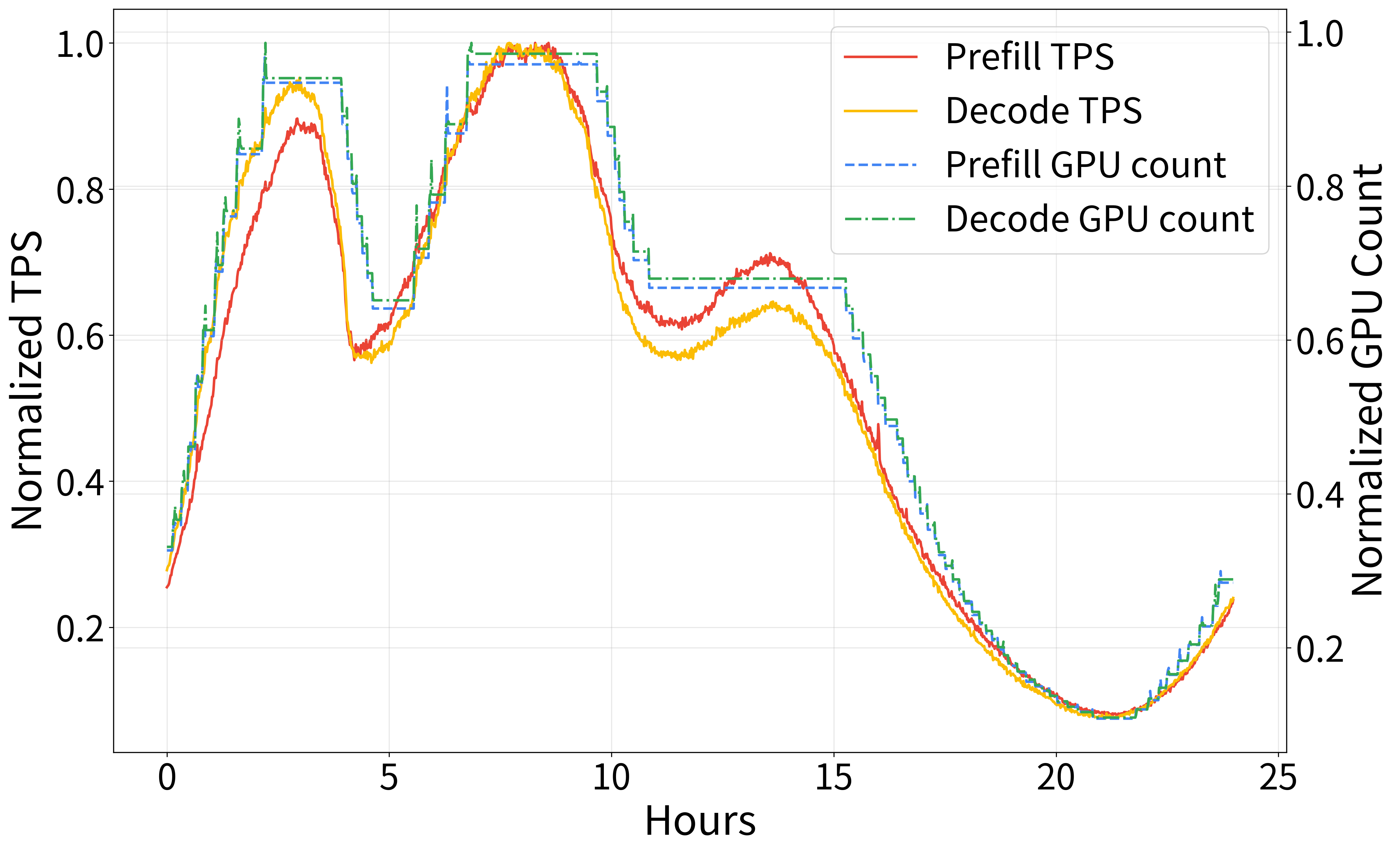}
        \subcaption{TPS/GPU (Norm.)}
        \label{fig:prod-workload-pod-num}
    \end{subfigure}
    \begin{subfigure}[b]{0.24\textwidth}
        \centering
        \includegraphics[width=\linewidth]{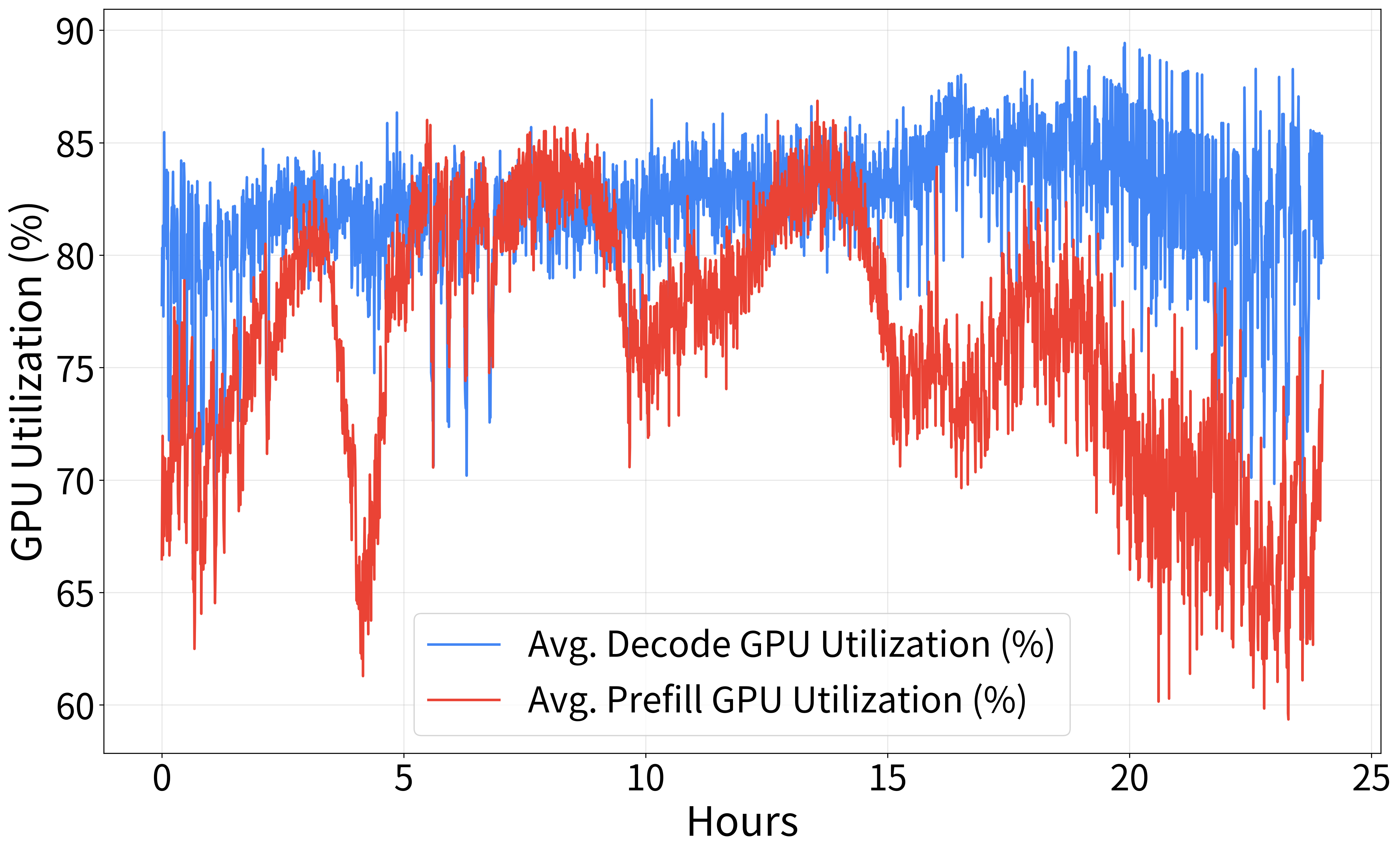}
        \subcaption{GPU Utilization}
        \label{fig:prod-gpu-util}
    \end{subfigure}
    \begin{subfigure}[b]{0.24\textwidth}
        \centering
        \includegraphics[width=\linewidth]{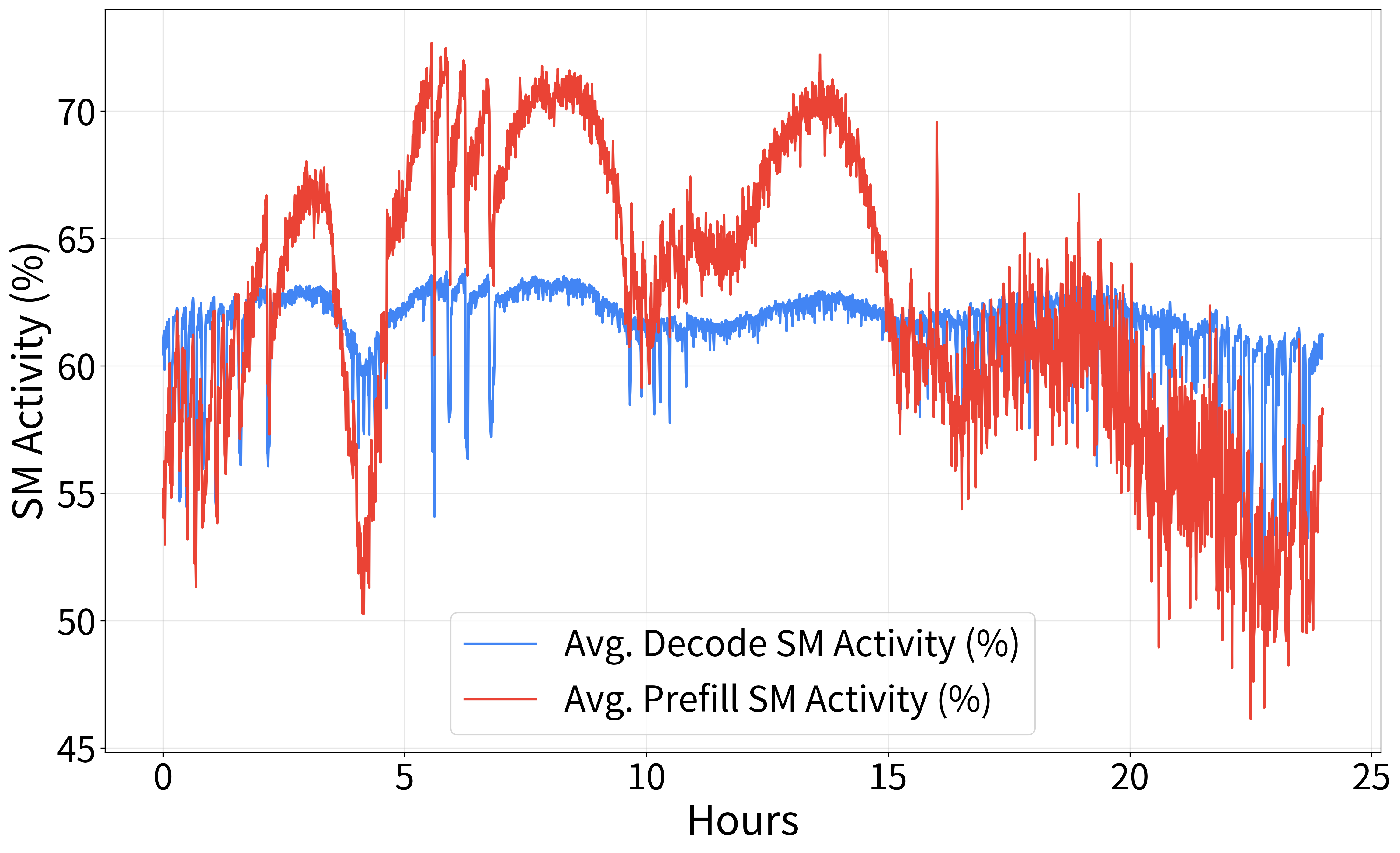}
        \subcaption{SM Activity}
        \label{fig:prod-sm-activity}
    \end{subfigure}
    \begin{subfigure}[b]{0.24\textwidth}
        \centering
        \includegraphics[width=\linewidth]{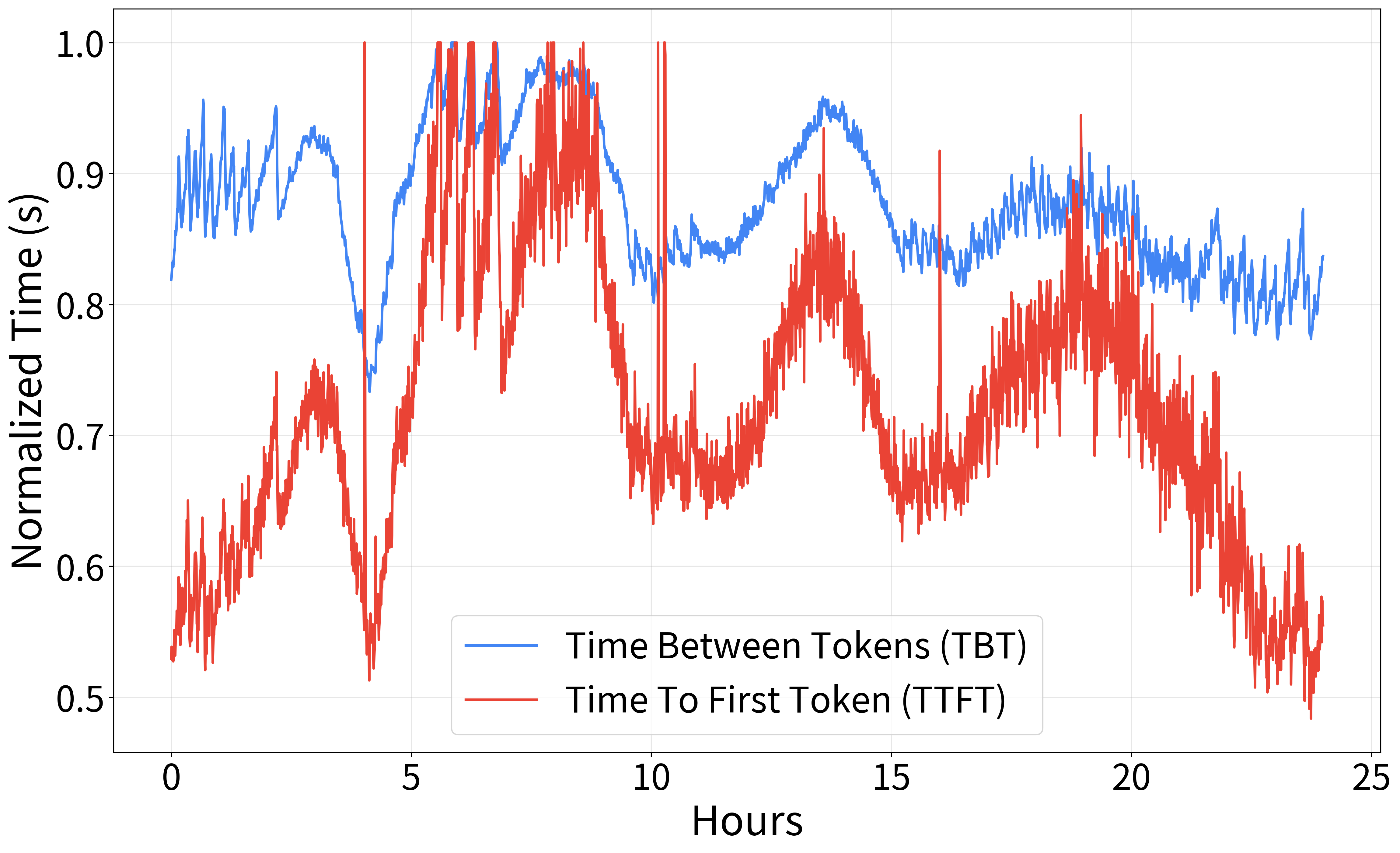}
        \subcaption{Normalized Latencies}
        \label{fig:prod-latency}
    \end{subfigure}
    \caption{\textbf{Production environment}: Performance metrics of an open-domain dialogue service with TPS-based autoscaling}
    \label{fig:prod-metrics}
\end{figure*}

\subsubsection{Experiment Results}
The evaluation was performed under standardized conditions: all experiments began with identical numbers of prefill and decode instances, shared the same resource quotas, and applied uniform scaling thresholds calibrated to induce scaling behavior under equivalent load conditions. Scaling events—including the timing and quantity of instance additions or removals—were recorded and visualized in Figure~\ref{fig:combined-metrics} to assess each metric’s responsiveness and suitability. More service metrics could be found in Appendix~\ref{sec:experiment-appendix}.

\textbf{TPS-based Autoscaling.} Experimental results (Figures~\ref{fig:prefill-tps-autoscaling} and~\ref{fig:decode-tps-autoscaling}) show that TPS-based autoscaling responds effectively to workload dynamics: during peak periods, the system promptly scales out resource instances to match the surge, ensuring that request demands are met without performance degradation; conversely, during valley periods, it efficiently scales in to avoid resource waste, maintaining cost-effectiveness. Typically, within a given service, the input-output length distribution remains relatively stable, leading to consistent behavior in both decode TPS and prefill TPS. This correlation allows the two metrics to be used interchangeably in principle. As a result, both forms of TPS metrics demonstrate sufficient reliability and responsiveness to be considered viable options for autoscaling.

\textbf{Utilization-based Autoscaling.} The effectiveness of this strategy differs significantly between the prefill and decode phases. The GPU utilization of prefill instances exhibits reasonable validity as a scaling signal (see Figure~\ref{fig:prefill-gpu-autoscaling}), though it is less sensitive to workload changes compared with TPS-based metrics. In contrast, GPU utilization for decode instances is ineffective for guiding scaling decisions. Figure~\ref{fig:decode-gpu-autoscaling} shows that decode GPU utilization remains at a high level even when the workload decreases, which aligns with our observation in Figure~\ref{fig:gpu-util}. Similar behaviors are observed with SM activity metrics. Prefill SM activity correlates well with workload fluctuations and may be a strong candidate for further investigation. Decode SM activity, however, exhibits the same limitation as decode GPU utilization—persistently high values even during low request volumes—making it unsuitable as a reliable autoscaling indicator.
    
\textbf{Latency-based Autoscaling.} We evaluated two latency-related metrics—TTFT and TBT—as potential autoscaling signals. As discussed in Section~\ref{sec:metric-driven-scaling-policy}, these metrics rely on a negative feedback strategy, which inherently leads to delayed reactions, overshooting, and frequent oscillations in resource allocation. This pattern is especially evident in our experiments with TTFT (see Figure~\ref{fig:ttft-autoscaling}), where scale-out actions tend to overshoot the required capacity, followed by frequent corrective adjustments. TBT demonstrates relatively smoother responses and fewer fluctuations compared to TTFT, reacting reasonably to workload variations (see Figure~\ref{fig:tpot-autoscaling}). However, tuning the associated hyperparameters poses a significant challenge. The nonlinear relationship between latency and resource allocation necessitates a multi-tier feedback-driven scaling mechanism. This design inevitably introduces numerous hyperparameters. For example, one must decide how wide each adjustment interval should be and how much additional capacity to provision at each step. achieving a balance between meeting SLOs and maximizing throughput demands that the system operates within a narrow and highly sensitive configuration range, which further complicates the parameter space. In practice, tuning so many interdependent parameters is prohibitively difficult in large-scale production environments, making this approach challenging to deploy reliably.

Based on the comparative analysis, TPS-based and prefill SM activity-based autoscaling strategies exhibit the most reliable responsiveness to workload variations. To simplify large-scale deployment in our production environment, we select the TPS-based metric due to its strong alignment with business objectives and straightforward configurability. Among TPS options, decode TPS is preferred, as it can be uniformly distributed across instances. In contrast, prefill instances often differ in hardware configurations, introducing additional complexity in metric normalization. To ensure consistent scaling behavior and reduce operational overhead, decode TPS is adopted as the primary autoscaling signal in our production system.

\subsection{Production Deployment Analysis}
To validate its effectiveness in a real-world setting, HeteroScale has been deployed in production at \CompanyA, where it now manages tens of thousands of GPUs across numerous services. On a daily basis, these services collectively process trillions of prefill tokens and generate hundreds of billions of decode tokens. A comparative analysis on a representative day revealed substantial utilization gains. Services with HeteroScale enabled showed a 26.6 percentage points increase in GPU utilization and a 9.2 percentage point increase in SM activity compared to services without autoscaling. Furthermore, an analysis comparing performance on a recent date with a date prior to the scaled deployment showed that overall GPU utilization increased by 8.6 percentage points and SM activity rose by 6.5 percentage points. Hundreds of thousands of GPU-hours are saved each day. These all underscores the system's broad impact on cluster efficiency.

The TPS-based policy is the most widely adopted, managing 64\% of the total GPU fleet under HeteroScale's control.  A detailed comparison of the two main policies showed that the dynamic, metrics-driven approach yields higher efficiency. The TPS-based policy delivered a 10.0 percentage points higher GPU utilization and an 11.1 percentage points higher SM activity compared to the periodic policy. This suggests that its ability to react to real-time workload fluctuations allows it to maintain higher resource pressure more consistently than the static, time-based schedules of the periodic policy.

For a more granular view, we collected and analyzed performance data from the same open-domain dialogue service shown in Figure~\ref{fig:open-dialogue-combined}. To demonstrate that our approach is effective across different modalities, we also gathered data for a vision-language search service, with the results detailed in Appendix~\ref{appendix:servicemetrics}.

As illustrated in Figure~\ref{fig:prod-workload-pod-num}, the number of both prefill and decode instances closely tracks the TPS variations, confirming that the essential P/D ratio is well-maintained throughout the workload fluctuations. With HeteroScale, the overall GPU usage is reduced by 41.3\%. This efficiency gain is reflected in the average prefill GPU utilization, which increased from 46.8\% to 76.2\%, and prefill SM activity, which rose from 36.6\% to 62.5\%. Concurrently, decode GPU utilization remained high (86.0\% $\rightarrow$ 82.2\%) and decode SM activity increased from 53.0\% to 61.6\%—a characteristic of the memory-bound decode stage—ensuring high performance while the system dynamically adjusts the number of active instances to match the load.

The latency metrics, TTFT and TBT, both vary within a smaller, more stable range compared to the non-autoscaled service. The apparent large fluctuations in Figure~\ref{fig:prod-latency} are an artifact of normalization. The occasional spikes observed in TTFT are caused by temporary P/D ratio imbalances that can occur during scaling operations. We have addressed this by implementing a soft P/D ratio maintenance mechanism at the service discovery level, which controls the registration of new instances to prevent such imbalances.

The data from the vision-language search service, presented in Figure~\ref{fig:vlm-combined-with-autoscale} in Appendix~\ref{appendix:servicemetrics}, shows a similar pattern of improved performance and stability, underscoring the general applicability across modalities of HeteroScale. The metrics for the autoscaled service may appear similar to the non-autoscaled service at times because the policy's minimum instance count is set to a relatively high threshold. This configuration is a deliberate choice to maintain a higher number of instances during off-peak hours, prioritizing service stability and readiness for sudden traffic surges.

\section{Related Work}
\label{sec:relatedwork}

\subsection{LLM Serving Systems}
Many systems optimize LLM serving.
Traditional systems like vLLM~\cite{kwon2023efficient}, TensorRT‑LLM~\cite{tensorrtllm}, SGLang~\cite{zheng2024sglang} and DeepSpeed‑Inference~\cite{aminabadi2022deepspeed} focus on single‑ or multi‑GPU inference via continuous batching, kernel fusion, and memory optimizations.
More recent systems address prefill/decode disaggregation, including DistServe~\cite{zhong2024distserve}, P/D‑Serve~\cite{jin2024p}, SplitWise~\cite{patel2024splitwise}, TetriInfer~\cite{hu2024inference}, MemServe~\cite{hu2024memservecontextcachingdisaggregated}, and Mooncake~\cite{qin2024mooncake}.
Earlier systems such as Orca~\cite{yu2022orca}, FastServe~\cite{wu2023fast}, and AlpaServe~\cite{li2023alpaserve} optimize batching, scheduling, or multiplexing without separating prefill and decode.
Orca proposes iteration‑level scheduling and selective batching; FastServe uses pre‑emptive scheduling to cut latency;
and AlpaServe targets statistical multiplexing to handle bursty workloads.

\subsection{Autoscaling Technologies}
Cloud autoscaling includes rule-based approaches like Kubernetes HPA~\cite{k8shpa}, VPA~\cite{k8svpa}, and KEDA~\cite{keda}, and learning-based ones like AutoScale~\cite{gandhi2012autoscale}, DeepScaling~\cite{wang2022deepscaling}, and Resource Central~\cite{cortez2017resource}.
However, these systems are not designed for P/D disaggregated LLM services.

\subsection{Heterogeneous Resource Management}
Systems like Heterogeneity-aware Scheduler~\cite{narayanan2020heterogeneity}, Tiresias~\cite{gu2019tiresias}, and Gandiva~\cite{xiao2018gandiva} optimize allocation for mixed workloads.
Preemption mechanisms such as Morpheus~\cite{jyothi2016morpheus}, Themis~\cite{mahajan2020themis}, and AntMan~\cite{xiao2020antman} improve multi-tenant sharing.
GPU-aware frameworks, including HexGen~\cite{jiang2023hexgen} and Mélange~\cite{griggs2024m}, leverage hardware heterogeneity by partitioning pipelines and solving cost-aware scheduling across GPU pools.
\subsection{Network-aware Scheduling}
Network-aware scheduling has been explored through systems like Sinbad~\cite{chowdhury2013leveraging}, Varys~\cite{chowdhury2014efficient}, and NetCache~\cite{jin2017netcache}, which optimize data placement to reduce network congestion.
Topology-aware systems like Firmament~\cite{gog2016firmament}, Paragon~\cite{delimitrou2013paragon}, and Quincy~\cite{isard2009quincy} consider network topology in placement decisions, informing our approach.

\section{Future Work}
\label{sec:futurework}

While HeteroScale provides a comprehensive solution for autoscaling P/D disaggregated LLM services, the field is rapidly evolving. We plan to extend this work in several key directions:

\begin{itemize}
    \item \textbf{Exploration of Advanced and Agnostic Metrics:} To refine scaling decisions, we will investigate more granular metrics, including internal statistics from inference engines like vLLM, TensorRT-LLM, or SGLang, which can offer deeper insights into system bottlenecks. A key goal is to identify and validate metrics that are model-agnostic, hardware-agnostic, and workload-agnostic. Such universal metrics would significantly simplify the configuration and deployment of autoscaling across a diverse and growing landscape of services.

    \item \textbf{Dynamic P/D Ratio Adaptation:} Building on our current fixed-ratio mechanism, we will explore methods for making minor, dynamic adjustments to the P/D ratio in real-time. This would allow the system to automatically compensate for ``workload drift''---subtle changes in user behavior, prompt complexity, or generation length---thereby continuously optimizing the balance between prefill and decode resources to maximize efficiency.

    \item \textbf{KV Cache-Aware Autoscaling:} We plan to develop a KV cache-aware autoscaling paradigm. By directly incorporating metrics such as cache hit rates, eviction statistics, and memory pressure into the policy engine, HeteroScale could make more intelligent scaling decisions.
\end{itemize}

\section{Conclusion}
\label{sec:conclusion}
This paper introduced HeteroScale, a system that successfully addresses the primary challenges of scaling Prefill-Decode (P/D) disaggregated LLM services. By tackling hardware inefficiency, network latency, and architectural imbalance head-on, our work represents a significant step forward for large-scale AI infrastructure. Key innovations, including network-aware scheduling abstractions and a coordinated, metrics-driven scaling policy, have proven highly effective. The system's value is not just theoretical; its successful deployment in a demanding, large-scale production environment has yielded substantial improvements in resource efficiency and major operational savings, without compromising performance. The design and proven results of HeteroScale establish a new and effective benchmark for building the robust, efficient, and scalable LLM serving platforms of the future.

\clearpage

\bibliographystyle{plainnat}
\bibliography{references/references}

\clearpage

\beginappendix

\twocolumn[ % open up both columns temporarily
\section{Extra Analysis of Service Metrics}
\label{appendix:servicemetrics}
\vspace{1em} % optional spacing
]
\begin{figure}[H] % Using [H] from the 'float' package to place it HERE
    \centering

    % --- First Row ---
    \begin{subfigure}[b]{0.4\textwidth}
        \centering
        \includegraphics[width=\linewidth]{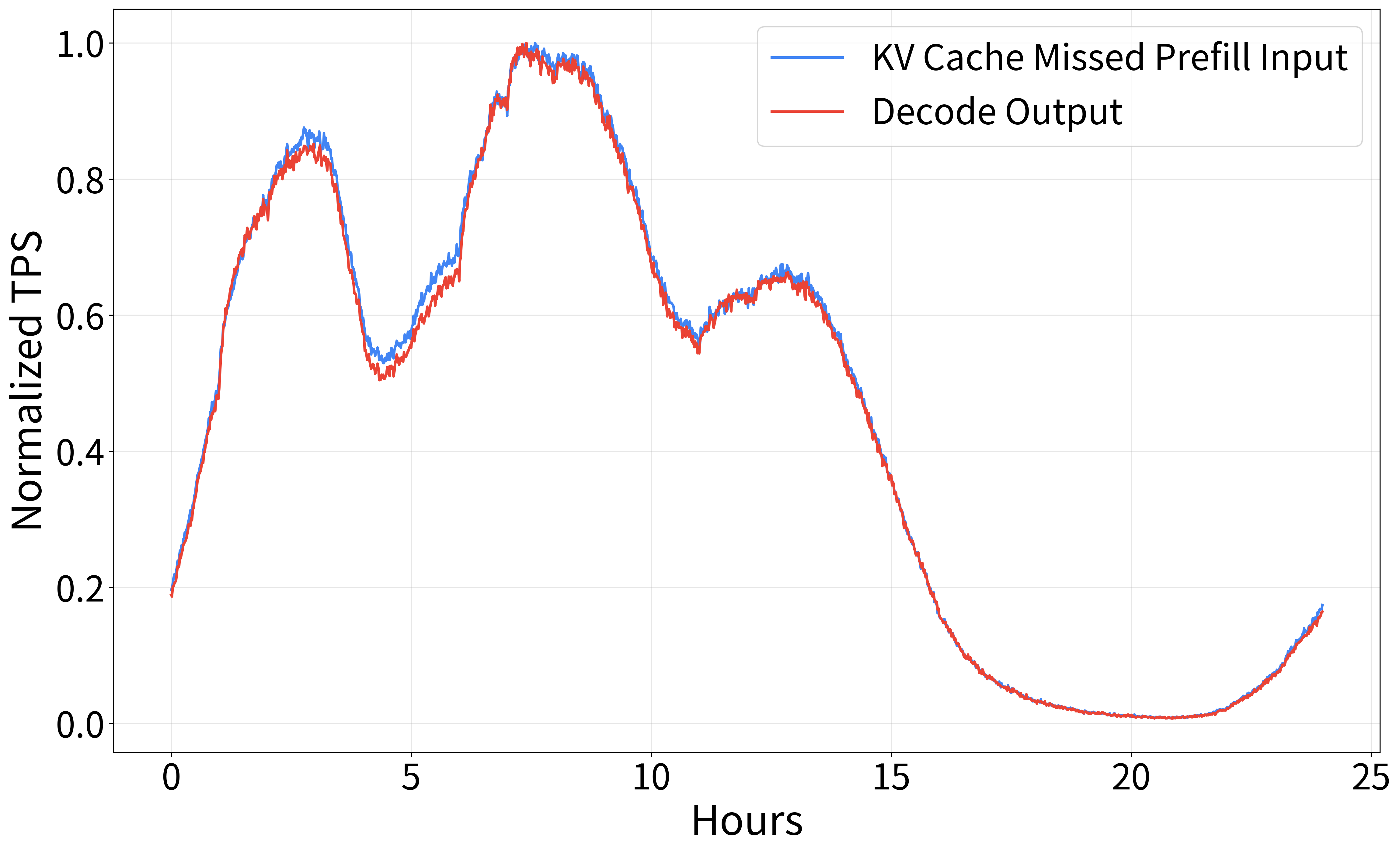}
        \caption{Normalized TPS}
        \label{fig:tps-metric-vlm}
    \end{subfigure}
    \hfill % Pushes the first and second images apart
    \begin{subfigure}[b]{0.4\textwidth}
        \centering
        \includegraphics[width=\linewidth]{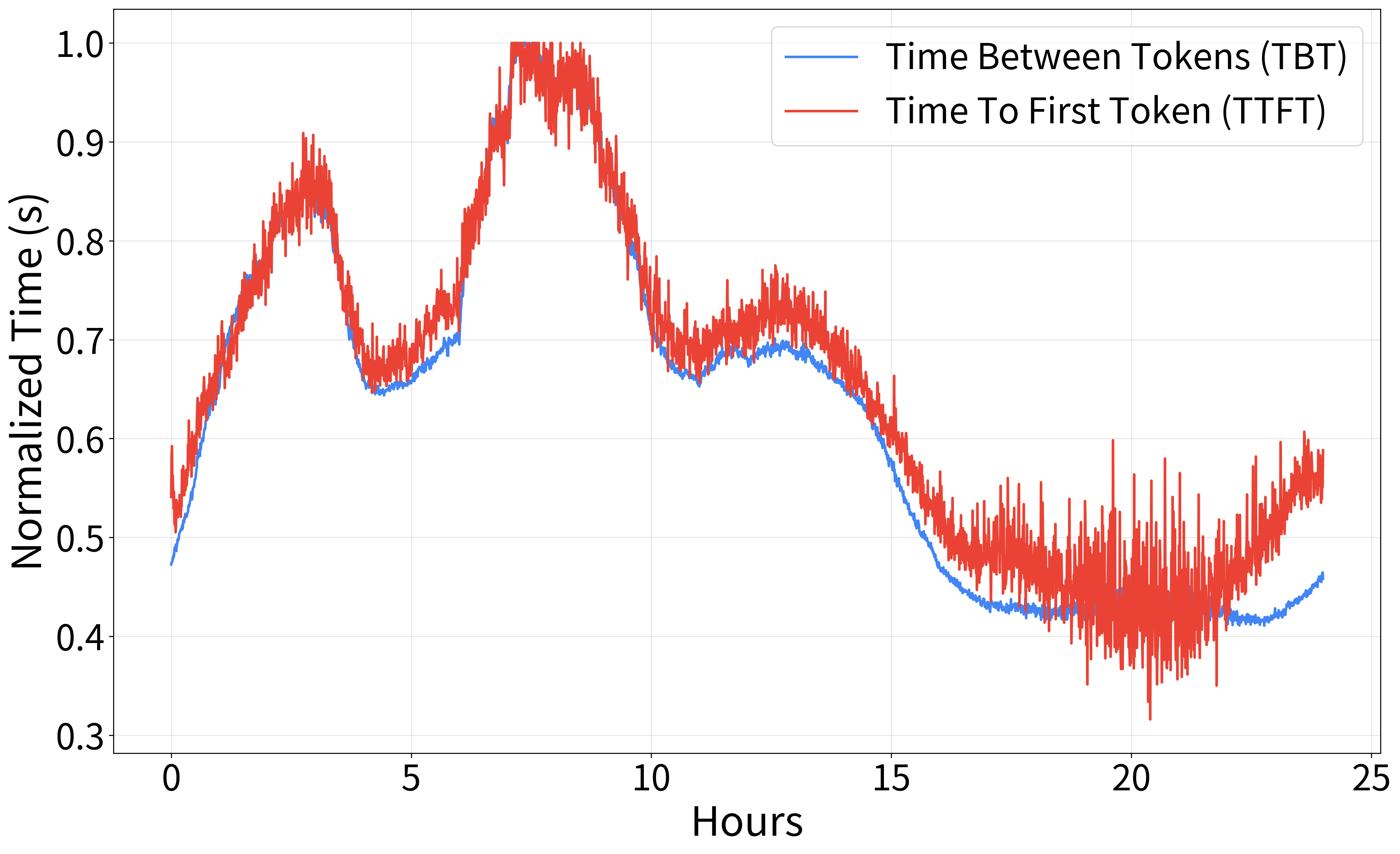}
        \caption{Normalized latencies}
        \label{fig:latency-metric-vlm}
    \end{subfigure}
    
    \vspace{1em} % Adds some vertical space between the two rows

    % --- Second Row ---
    \begin{subfigure}[b]{0.4\textwidth}
        \centering
        \includegraphics[width=\linewidth]{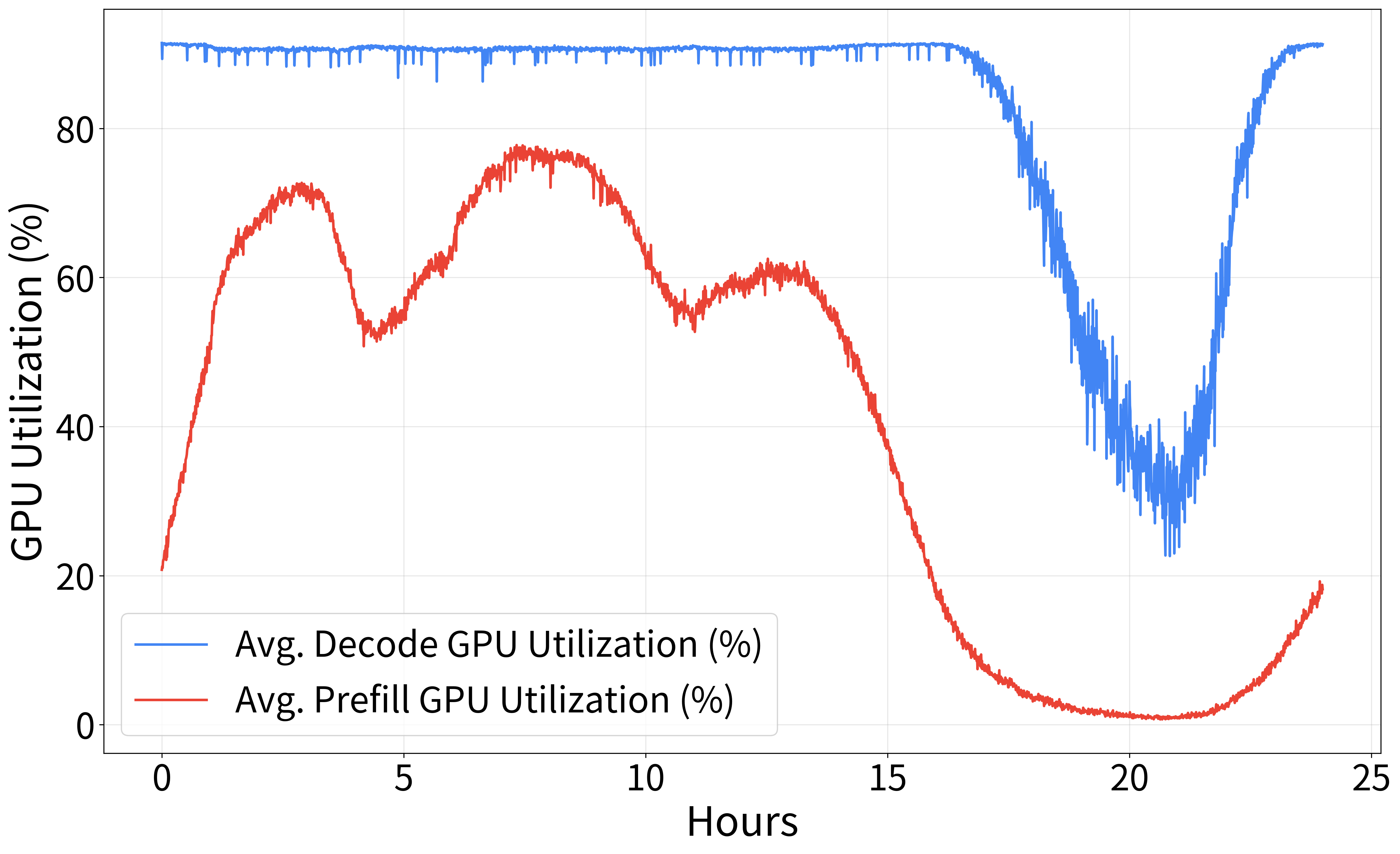}
        \caption{GPU utilization}
        \label{fig:gpu-util-vlm}
    \end{subfigure}
    \hfill % Pushes the third and fourth images apart
    \begin{subfigure}[b]{0.4\textwidth}
        \centering
        \includegraphics[width=\linewidth]{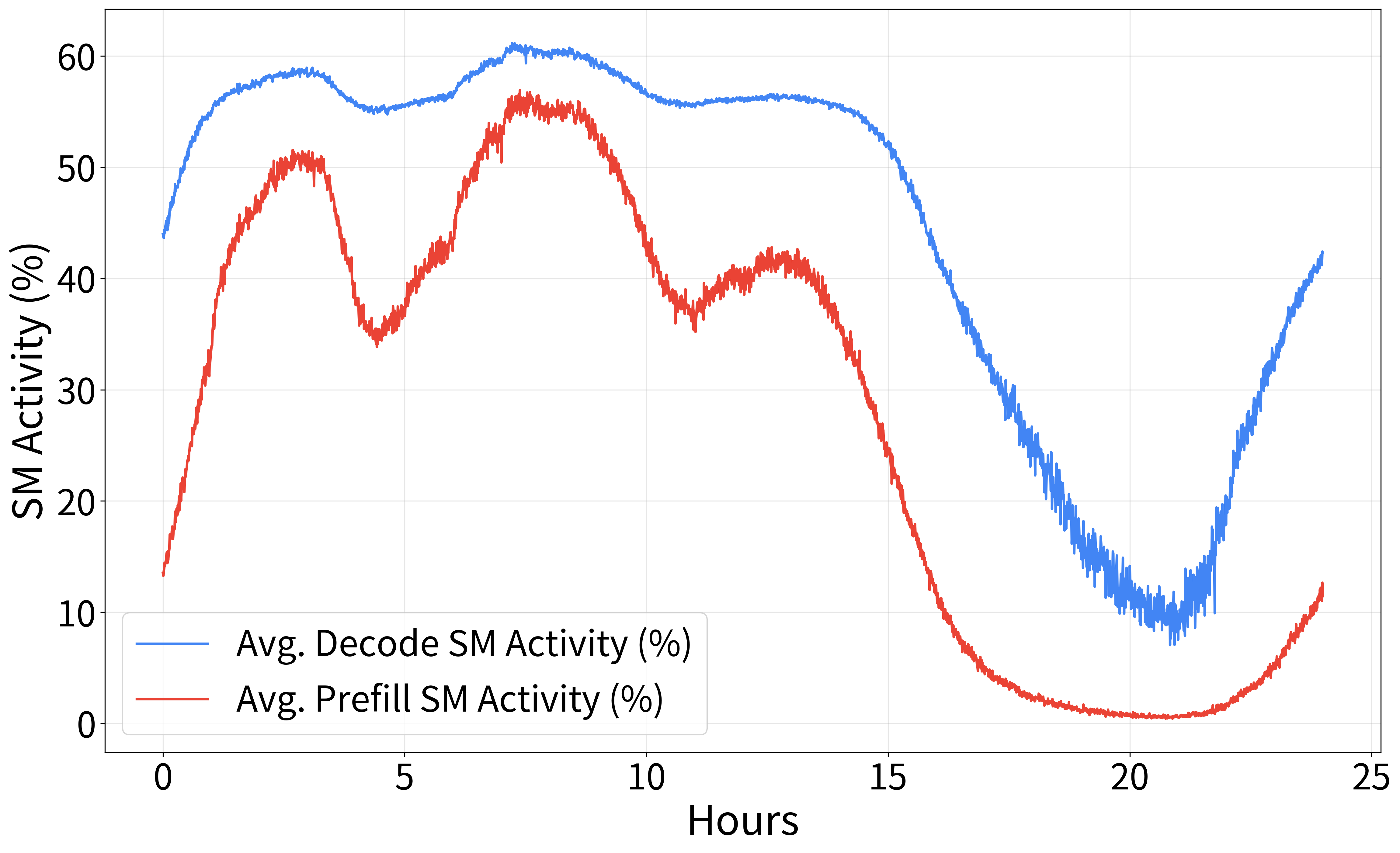}
        \caption{SM activity}
        \label{fig:sm-activity-vlm}
    \end{subfigure}

    % The MAIN caption for the entire figure, placed inside the environment
    \caption{Performance metrics of a \textbf{vision-language search} service with \ModelC \textbf{without} autoscaling.}
    \label{fig:vlm-combined}
\end{figure}

\begin{figure}[H] % Using [H] from the 'float' package to place it HERE
    \centering

    % --- First Row ---
    \begin{subfigure}[b]{0.4\textwidth}
        \centering
        \includegraphics[width=\linewidth]{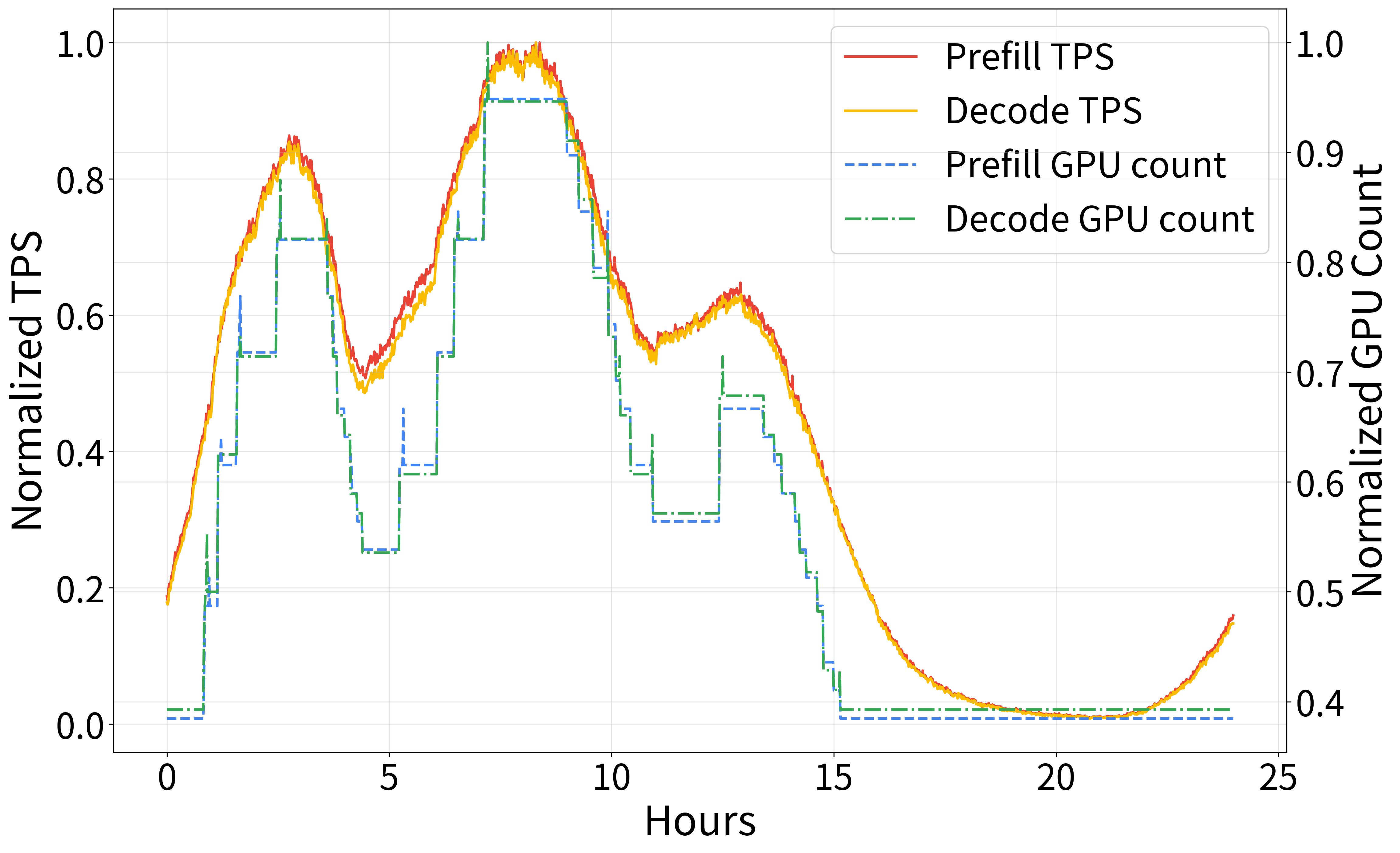}
        \caption{TPS and GPU Count (Normalized)}
        \label{fig:tps-metric-vlm}
    \end{subfigure}
    \hfill % Pushes the first and second images apart
    \begin{subfigure}[b]{0.4\textwidth}
        \centering
        \includegraphics[width=\linewidth]{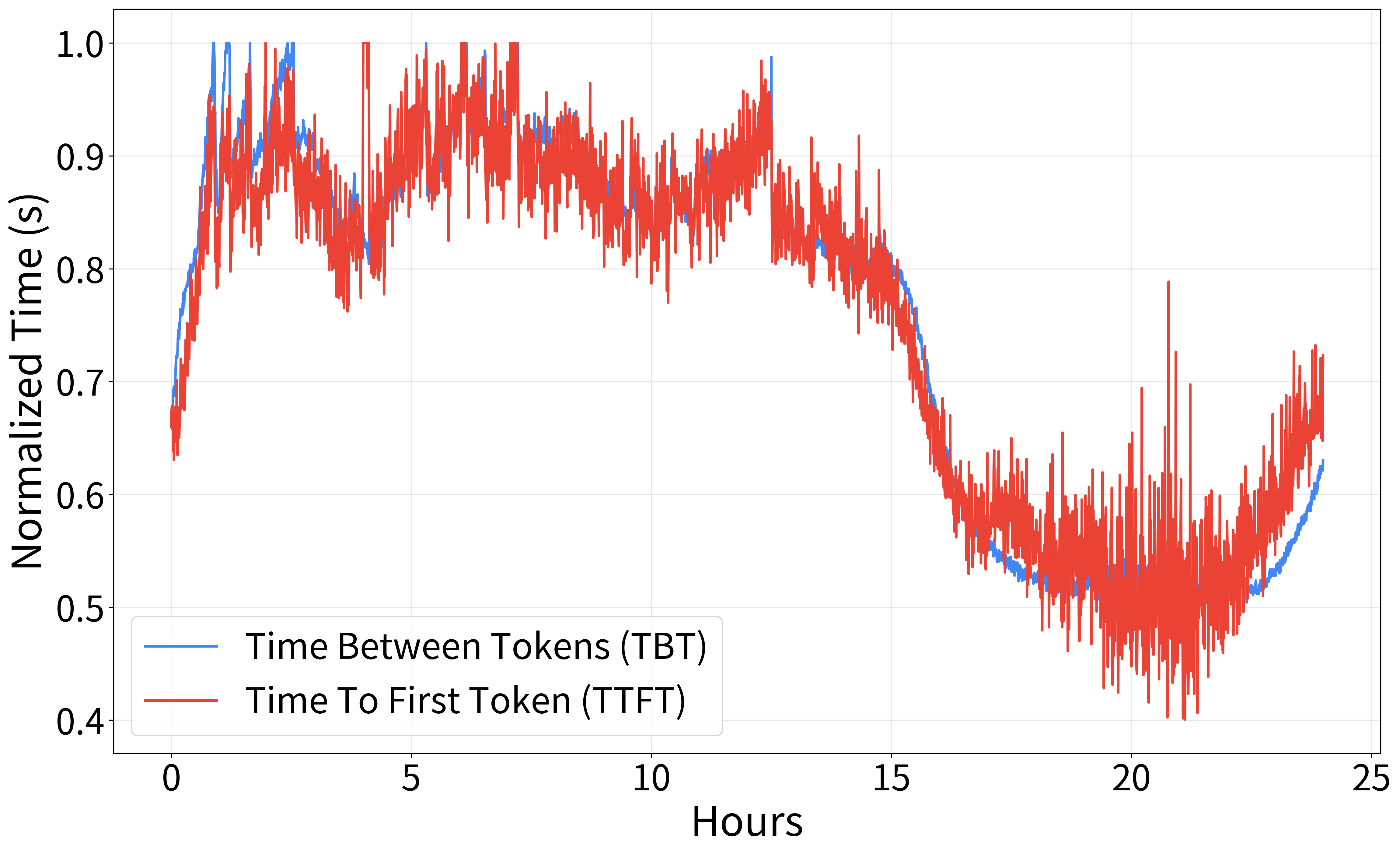}
        \caption{Normalized latencies}
        \label{fig:latency-metric-vlm}
    \end{subfigure}
    
    \vspace{1em} % Adds some vertical space between the two rows

    % --- Second Row ---
    \begin{subfigure}[b]{0.4\textwidth}
        \centering
        \includegraphics[width=\linewidth]{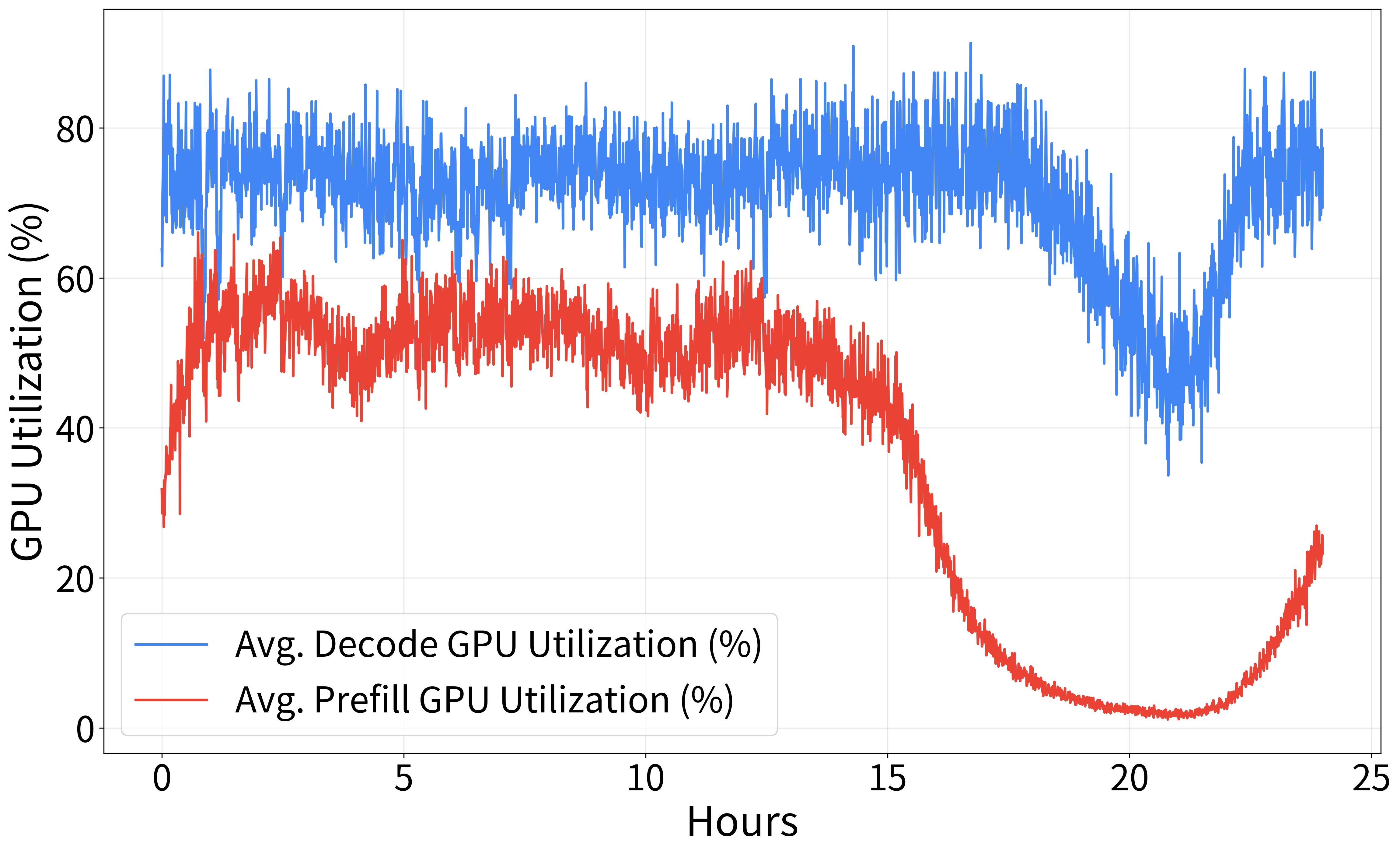}
        \caption{GPU utilization}
        \label{fig:gpu-util-vlm}
    \end{subfigure}
    \hfill % Pushes the third and fourth images apart
    \begin{subfigure}[b]{0.4\textwidth}
        \centering
        \includegraphics[width=\linewidth]{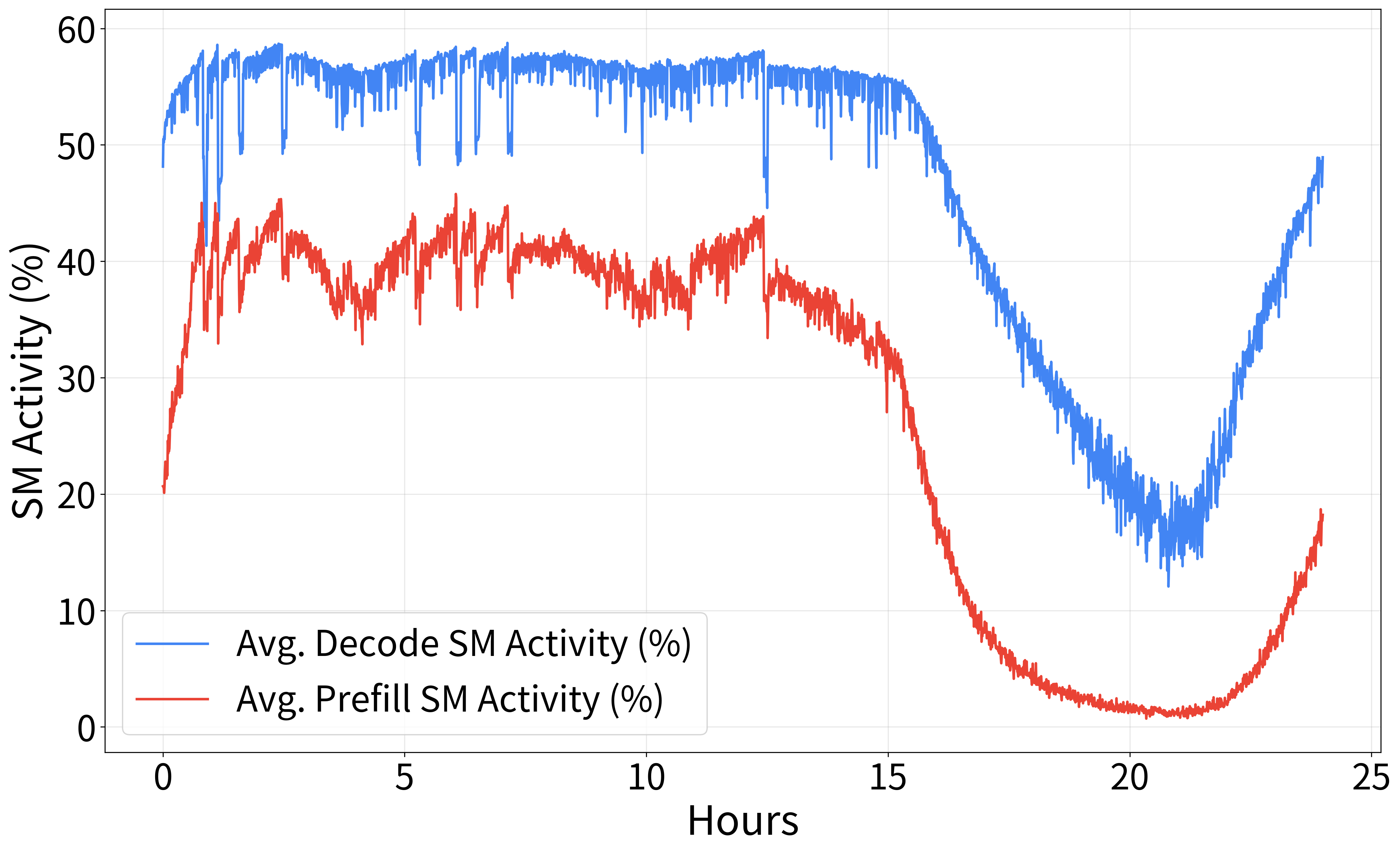}
        \caption{SM activity}
        \label{fig:sm-activity-vlm}
    \end{subfigure}

    % The MAIN caption for the entire figure, placed inside the environment
    \caption{Performance metrics of a \textbf{vision-language search} service with \ModelC \textbf{with TPS-based} autoscaling.}
    \label{fig:vlm-combined-with-autoscale}
\end{figure}

\clearpage        % finish any earlier floats
\onecolumn 
\section{Scaling Algorithms}
\label{sec:scaling_algorithms}
\vspace{0.5em}

This section provides the detailed pseudo-code for the scaling and scheduling algorithms implemented in HeteroScale.

\vspace{0.5em}

\begin{algorithm}[H]
\caption{Proportional Control Scaling}
\label{alg:proportional-scaling}
\SetAlgoLined
\KwIn{Current Instances $I_{curr}$, Observed Metric $M_{curr}$, Target Metric per Instance $M_{target}$, \\
\quad Scaling thresholds $\theta_{out}, \theta_{in}$, Cooling periods $C_{out}, C_{in}$, Last scaling timestamp $T_{last}$}

\KwOut{Scaling decision (ScaleOut, ScaleIn, NoChange), final instance count $I_{final}$}

$I_{expected} \gets I_{curr} \times \frac{M_{curr}}{M_{target}}$\;
$R \gets \frac{I_{expected}}{I_{curr}}$\;
$cooling \gets CurrentTime() - T_{last}$\;

\If{$R > 1 + \theta_{out}$ \textbf{and} $cooling \ge C_{out}$}{
    \Return{(ScaleOut, $I_{expected}$)}\;
}
\ElseIf{$R < 1 - \theta_{in}$ \textbf{and} $cooling \ge C_{in}$}{
    \Return{(ScaleIn, $I_{expected}$)}\;
}
\Else{
    \Return{(NoChange, $I_{curr}$)}\;
}
\end{algorithm}
\vspace{1.5em}
\begin{algorithm}[H]
\caption{Negative Feedback Control Scaling}
\label{alg:latency-scaling}
\SetAlgoLined
\KwIn{Current Instances $I_{curr}$, Current Latency $L_{curr}$, Target Latency $L_{target}$, Latency thresholds $\alpha_{out}, \beta_{out}, \gamma_{in}$, \\ 
Cooling periods $C_{out}, C_{in}$, Last scaling timestamp $T_{last}$}

\KwOut{Scaling decision (ScaleOut, ScaleIn, NoChange), final instance count $I_{final}$}

\If{$L_{curr} \ge L_{target} \times \alpha_{out}$}{
    $I_{expected} \gets I_{curr} \times 1.2$\;
}
\ElseIf{$L_{curr} \ge L_{target} \times \beta_{out}$}{
    $I_{expected} \gets I_{curr} \times 1.1$\;
}
\ElseIf{$L_{curr} \le L_{target} \times \gamma_{in}$}{
    $I_{expected} \gets I_{curr} \times 0.95$\;
}
\Else{
    \Return{(NoChange, $I_{curr}$)}\;
}

% $R \gets \frac{I_{expected}}{I_{curr}}$\;
$cooling \gets CurrentTime() - T_{last}$\;

\If{$cooling \ge C_{up}$}{
    \Return{(ScaleOut, $I_{expected}$)}\;
}
\ElseIf{$cooling \ge C_{in}$}{
    \Return{(ScaleIn, $I_{expected}$)}\;
}
\Else{
    \Return{(NoChange, $I_{curr}$)}\;
}
\end{algorithm}

\begin{algorithm}[h]
\caption{Network Affinity-aware Scheduling and Allocation}
\label{alg:hetero-resource-allocation}
\SetAlgoLined
\KwIn{Requests $R$, Resources $A$, Priorities $P$}
\KwOut{Allocation list $Allocations$}

$Allocations \gets [\ ]$\;
$TreeView \gets$ ConstructTreeView(A)\;
$SortedReqs \gets$ SortByPriority($R$, $P$)\;

\For{each $r \in SortedReqs$}{
    $C \gets$ GetAffinityConstraints($r$)\;
    $PodsDelta \gets$ CalcPodsDelta($r$)\;
\If{$Type(r) = ScaleOut$}{
        $SG{s} \gets$ FilterRDMASubGroups(C)\;
        $SG{s} \gets$ SortByGroupPriority(SG)\;
\For{$sg \in SG{s}$}{
            $PodsAlloc \gets 0$\;
\While{CanAssignOnePod($sg$) \textbf{and} $PodsDelta > 0$}{
                AssignOnePod(sg)\;
UpdateTreeView($TreeView$)\;
                $PodsDelta \gets PodsDelta - 1$\;
                $PodsAlloc \gets PodsAlloc + 1$\;
}
            $\text{append}(Allocations, (r, g, PodsAlloc))$\;
\If{$PodDelta = 0$}{
                \textbf{break}\;
}
        }
    }
    
    \ElseIf{$Type(r)  = ScaleIn$}{
        $DG{s} \gets$ GetDeploymentGroups($r$)\;
$DG{s} \gets SortByPriority(DG{s})$\;
        \For{each $dg \in DG{s}$}{
            $PodsDeduct \gets 0$\;
\While{CanDeductOnePod($dg$) \textbf{and} $PodDelta > 0$}{
                DeductOnePod($dg$)\;
$PodsDelta \gets PodsDelta - 1$\;
                $PodsDeduct \gets PodsDeduct + 1$\;
}
            $\text{append}(Allocations, (r, dg, PodsDeduct))$\;
\If{$PodDelta = 0$}{
                \textbf{break}\;
}
        }
    }
}
\Return{$Allocations$}\;
\end{algorithm}

\clearpage        % finish any earlier floats
\onecolumn        % full-width layout
\section{Experiment Results of Different Autoscaling Policies}
\label{sec:experiment-appendix}
\vspace{0.5em}

% ----- Group 1 -----
\begin{center}
  \includegraphics[width=0.3\textwidth]{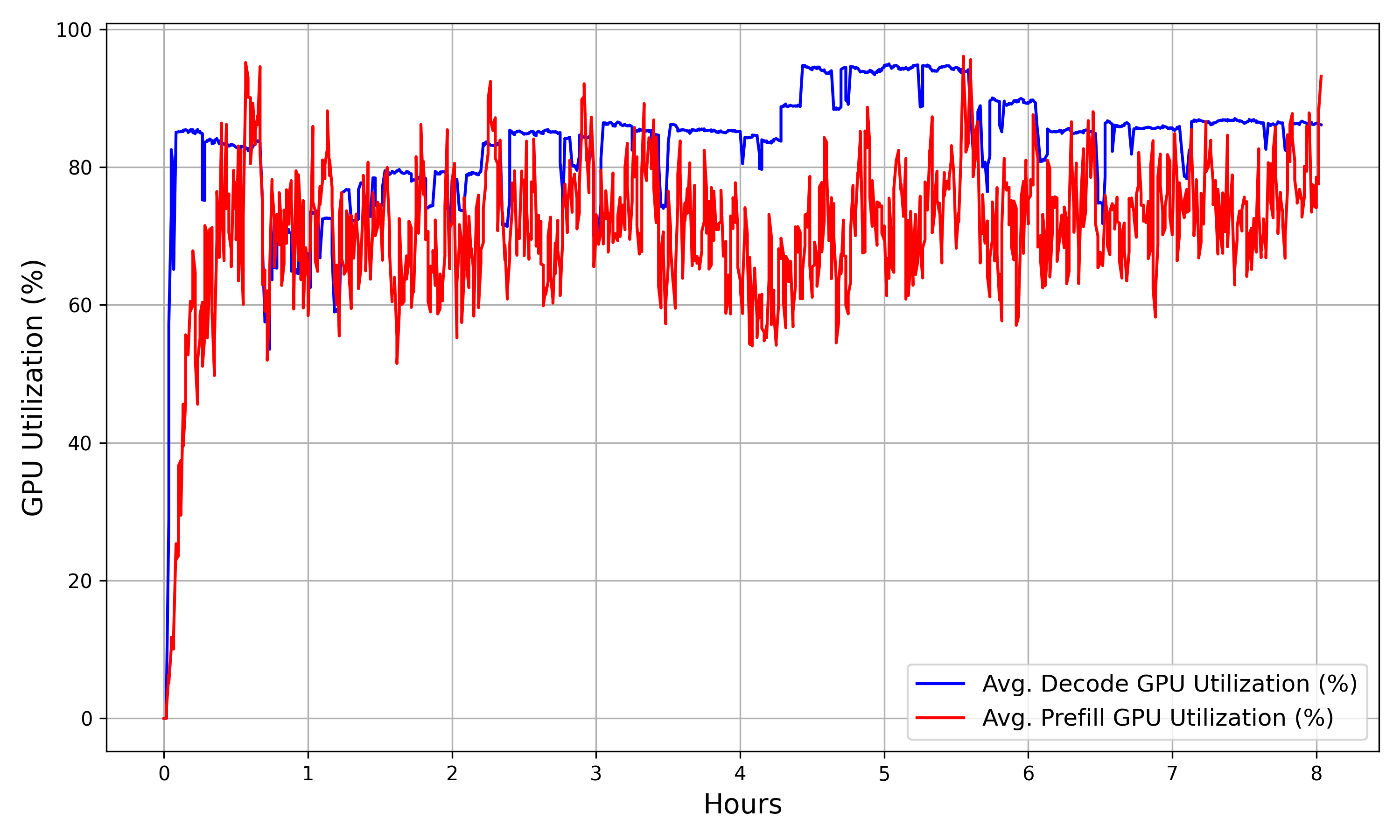}
  \includegraphics[width=0.3\textwidth]{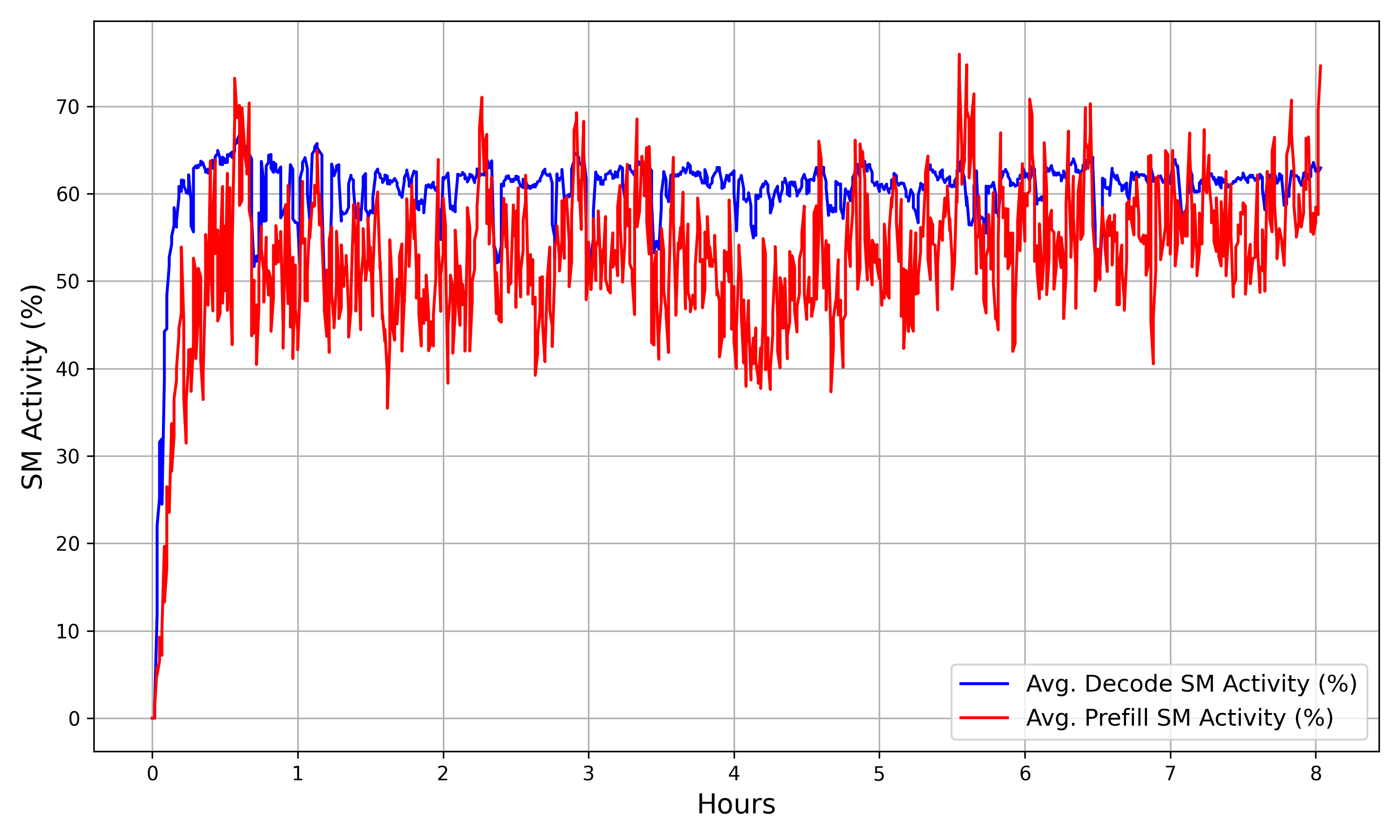}
  \includegraphics[width=0.3\textwidth]{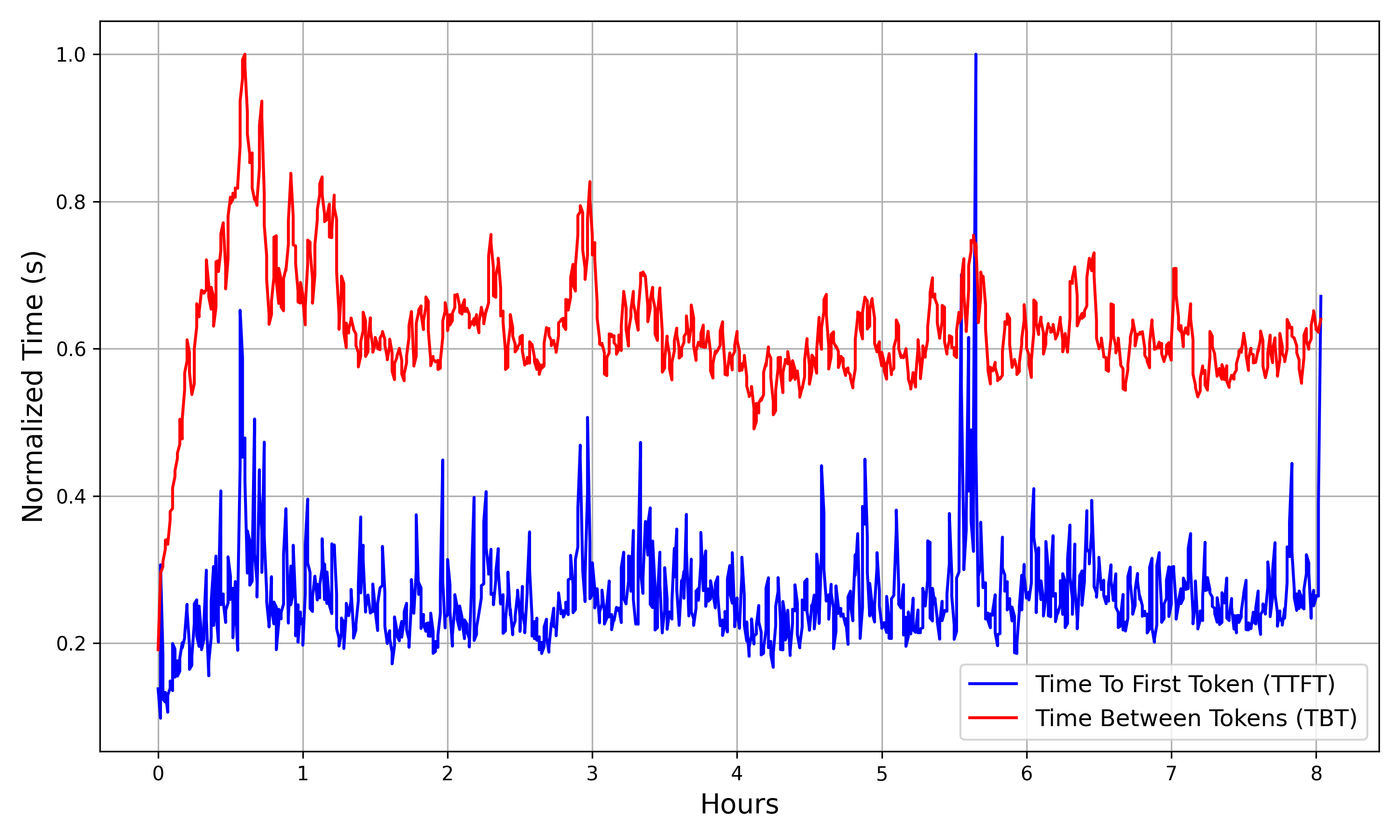}
  \captionof{figure}{Autoscaled by Prefill TPS}
\end{center}

% ----- Group 2 -----
\begin{center}
  \includegraphics[width=0.3\textwidth]{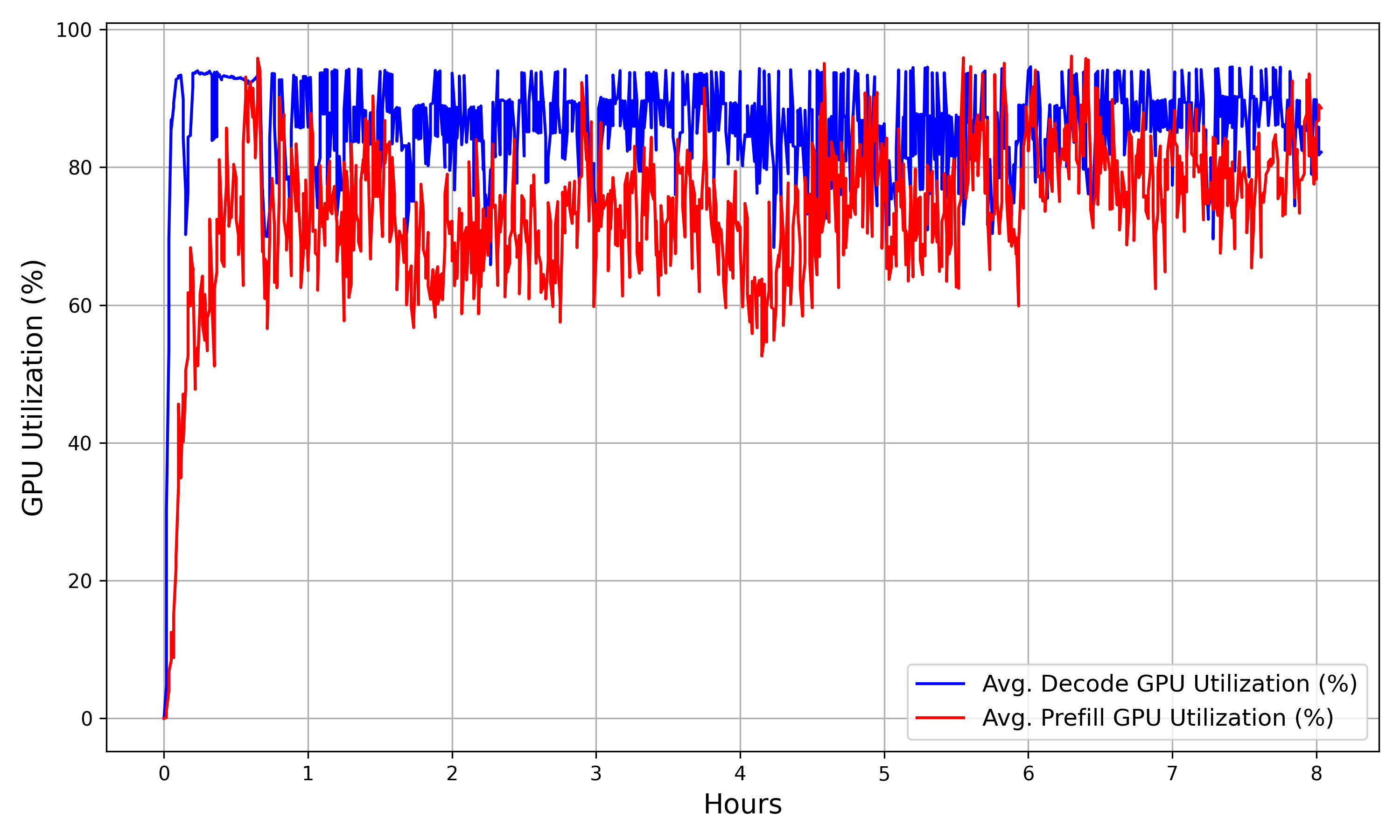}
  \includegraphics[width=0.3\textwidth]{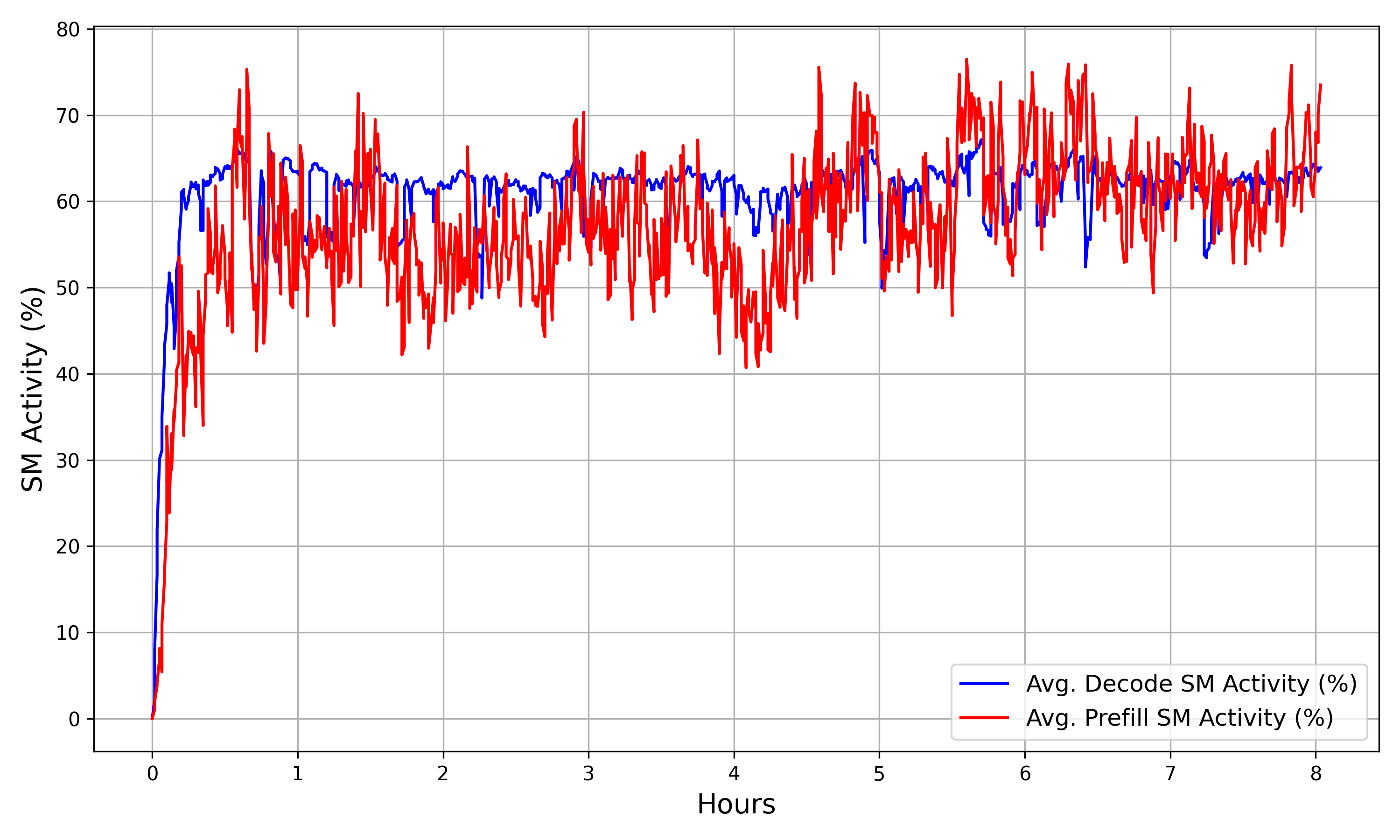}
  \includegraphics[width=0.3\textwidth]{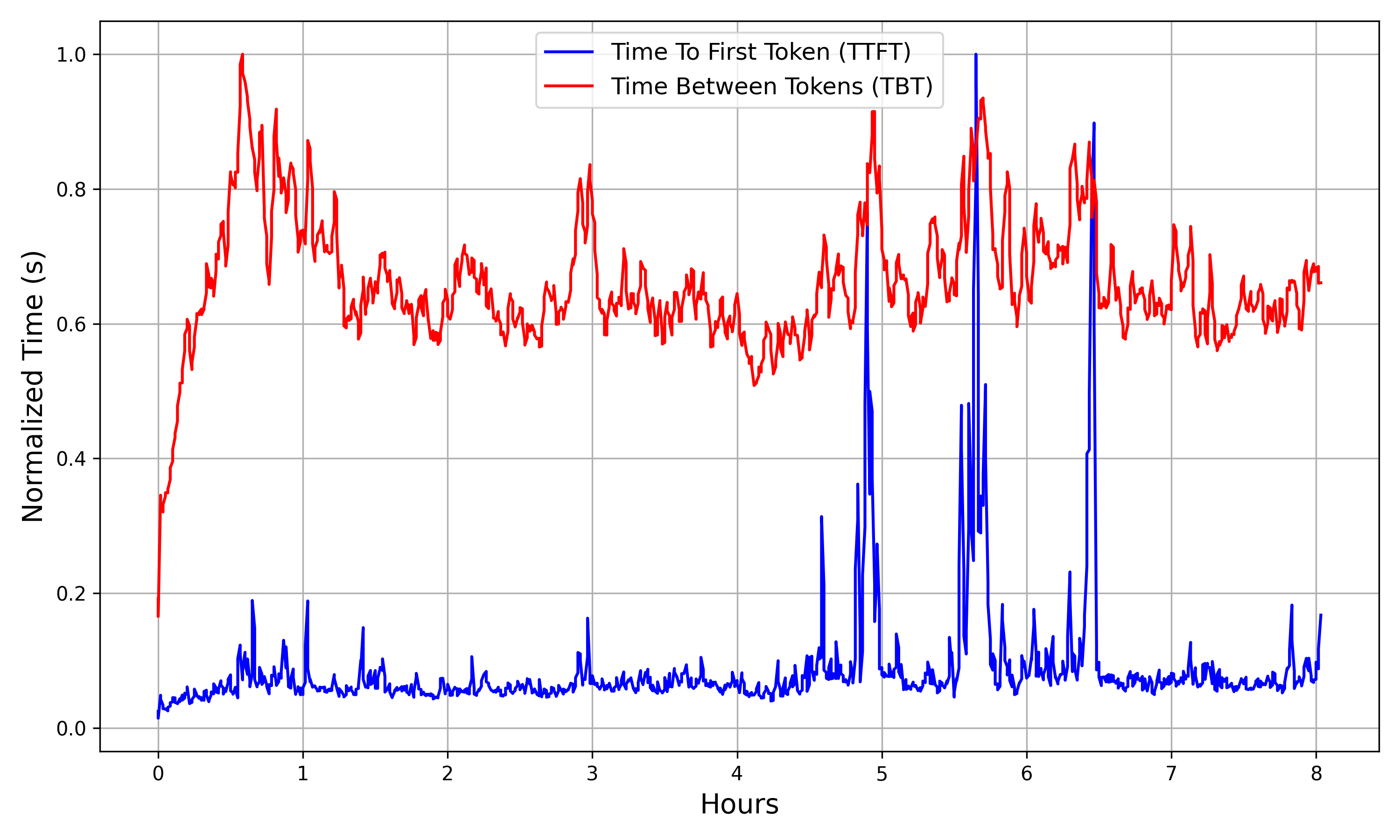}
  \captionof{figure}{Autoscaled by Decode TPS}
\end{center}

% ----- Group 3 -----
\begin{center}
  \includegraphics[width=0.3\textwidth]{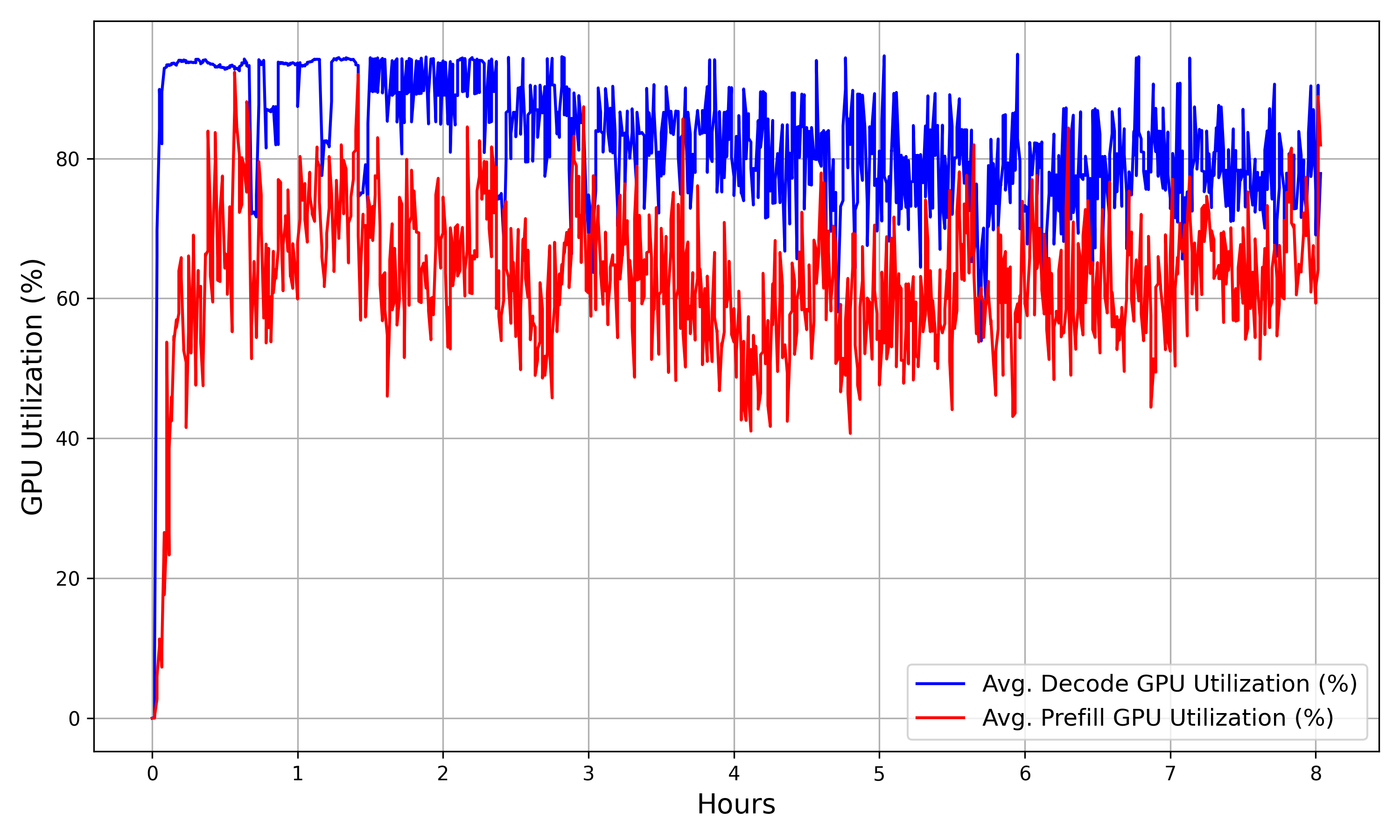}
  \includegraphics[width=0.3\textwidth]{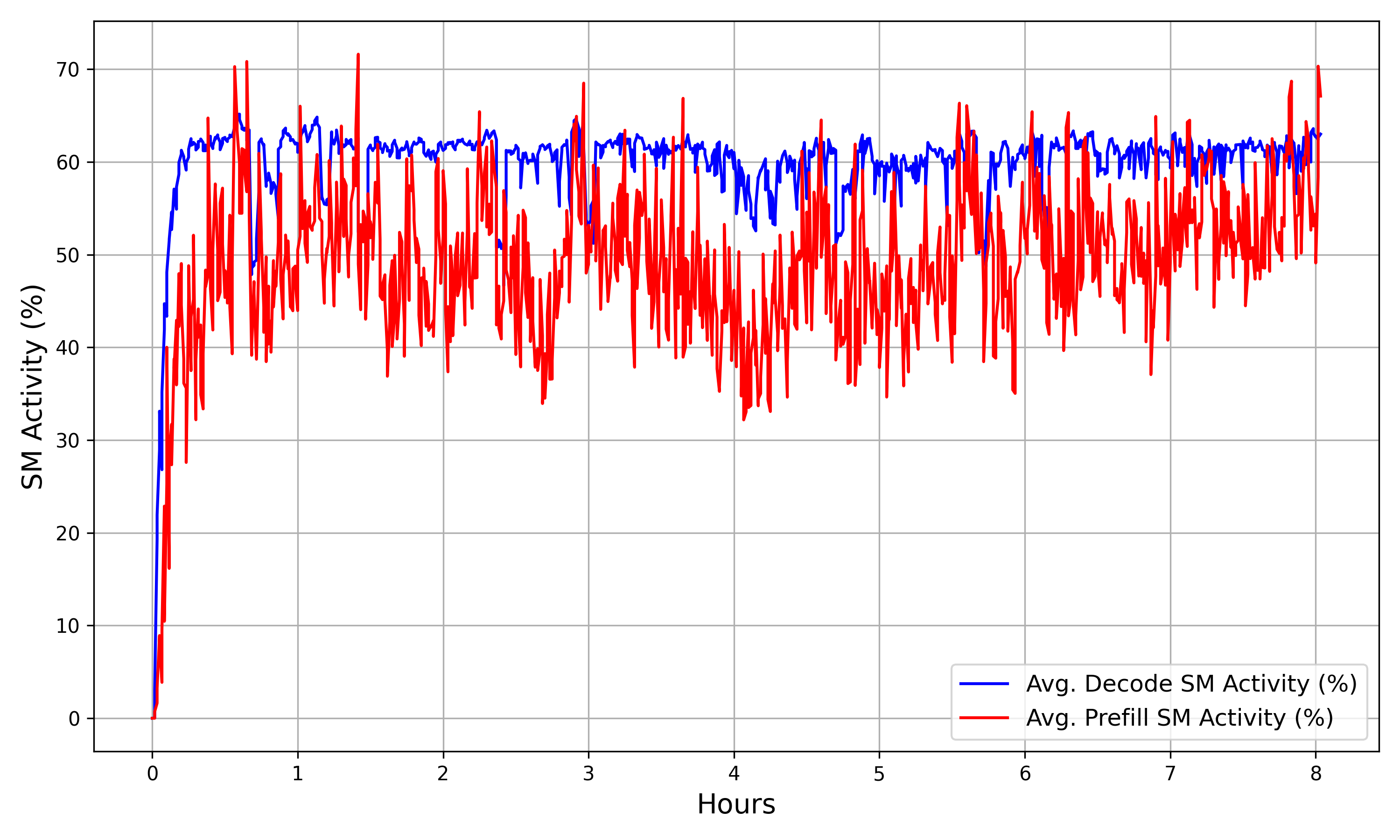}
  \includegraphics[width=0.3\textwidth]{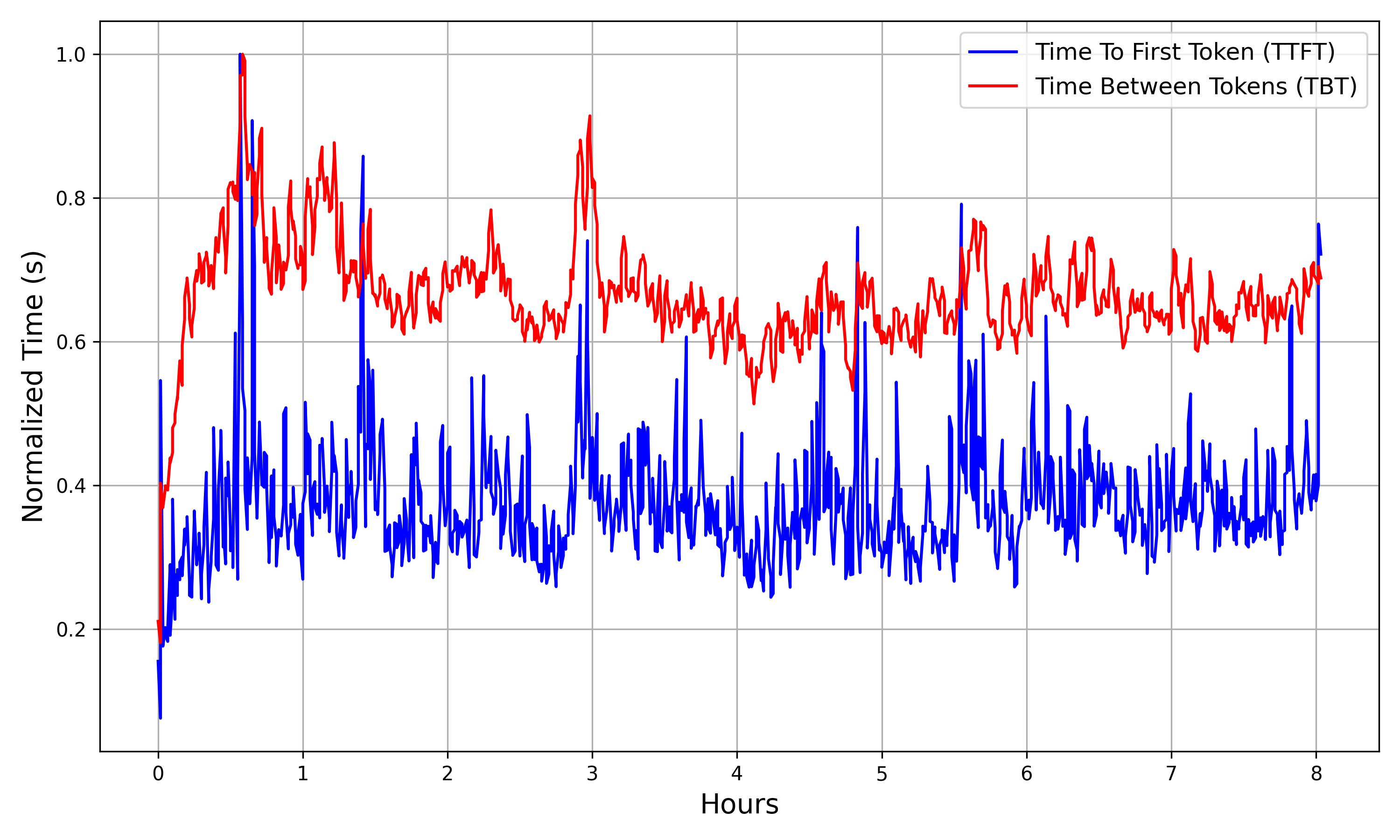}
  \captionof{figure}{Autoscaled by Prefill GPU Utilization}
\end{center}

% ----- Group 4 -----
\begin{center}
  \includegraphics[width=0.3\textwidth]{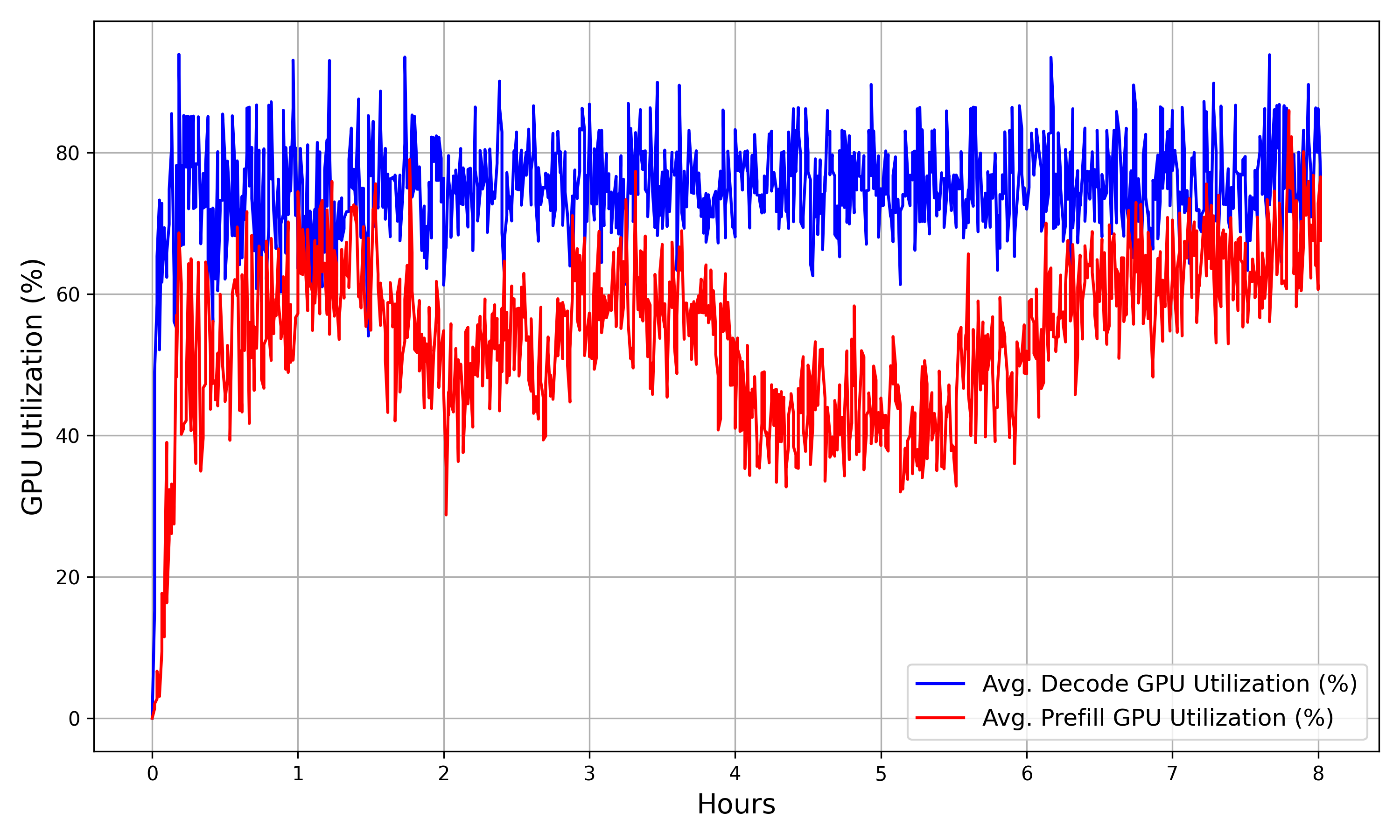}
  \includegraphics[width=0.3\textwidth]{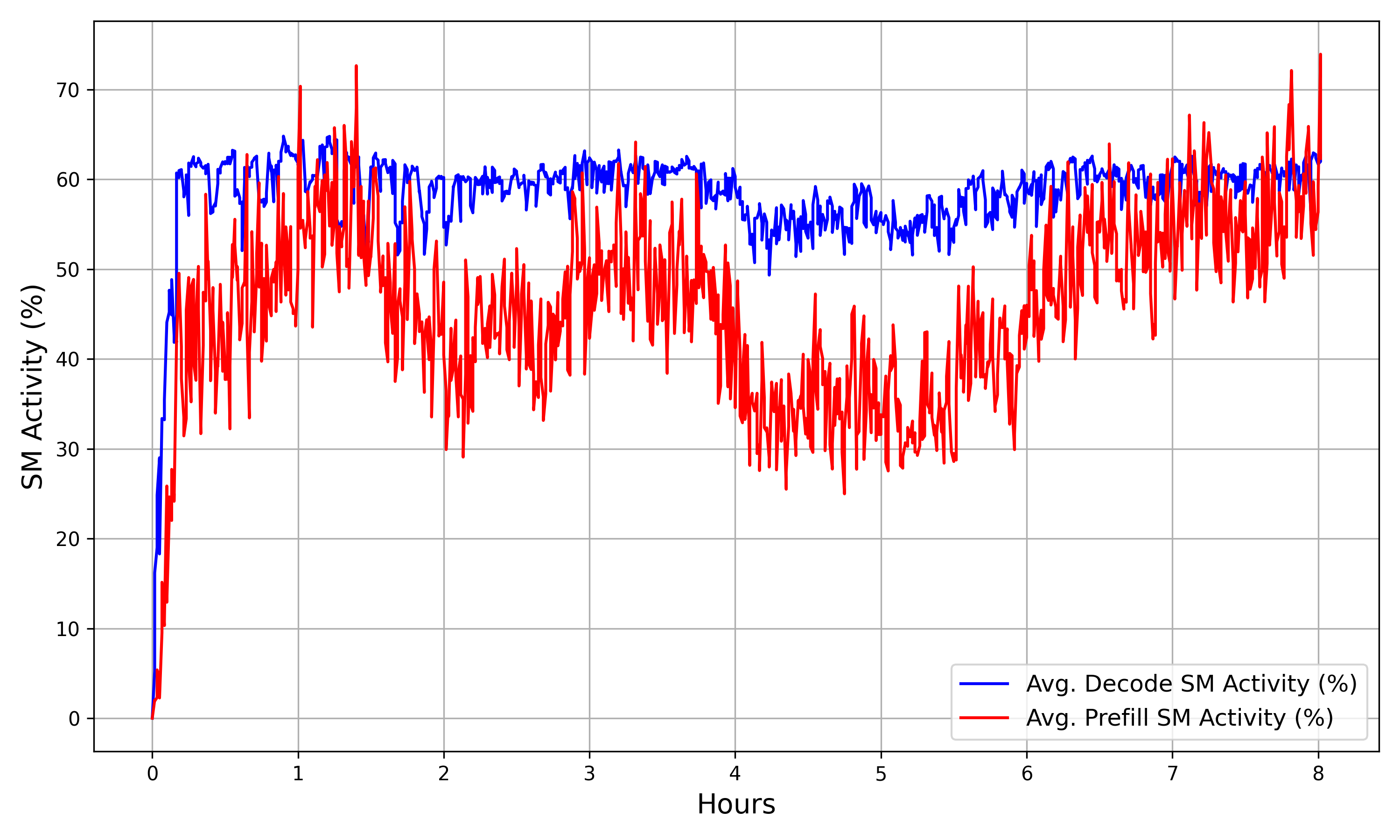}
  \includegraphics[width=0.3\textwidth]{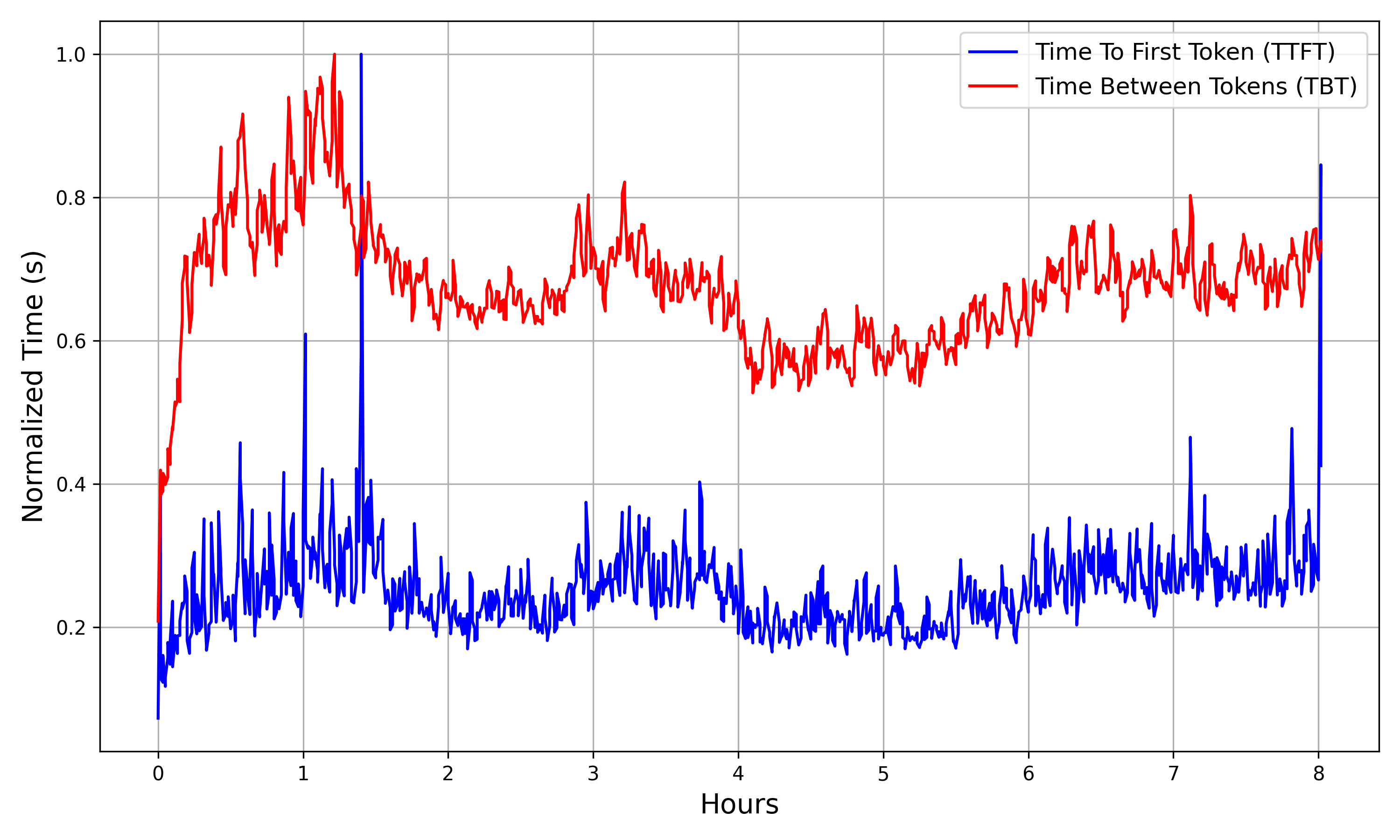}
  \captionof{figure}{Autoscaled by Decode GPU Utilization}
\end{center}

% ----- Group 5 -----
\begin{center}
  \includegraphics[width=0.3\textwidth]{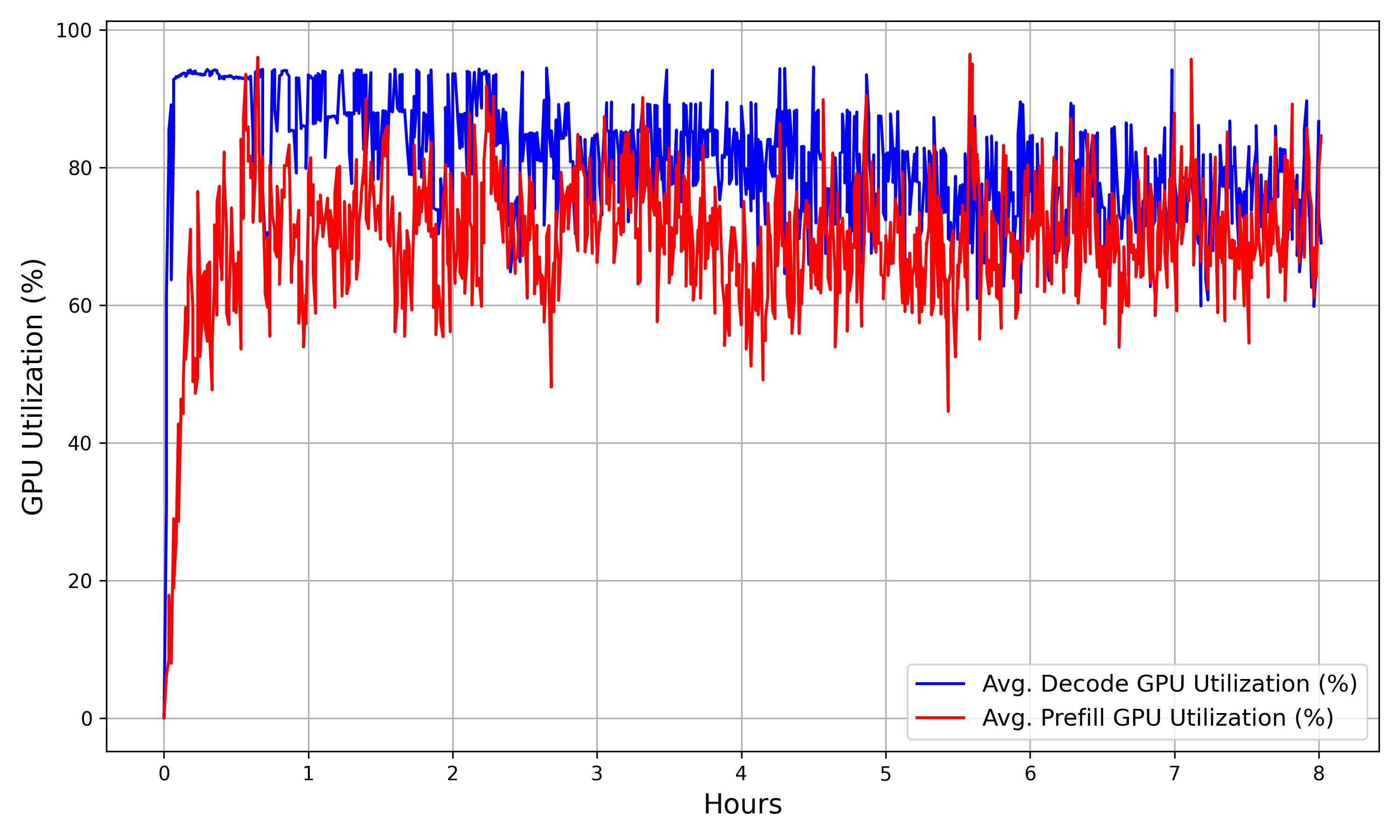}
  \includegraphics[width=0.3\textwidth]{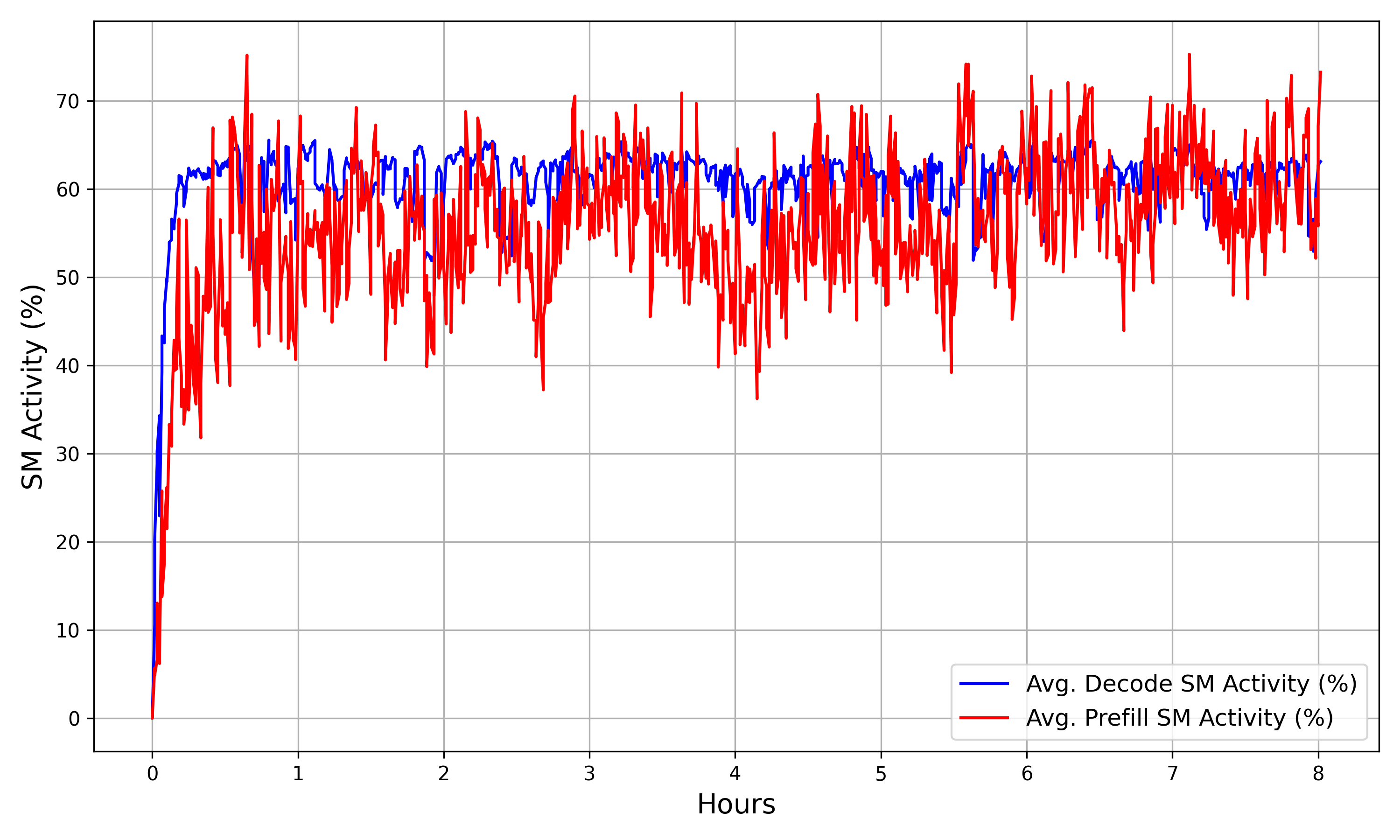}
  \includegraphics[width=0.3\textwidth]{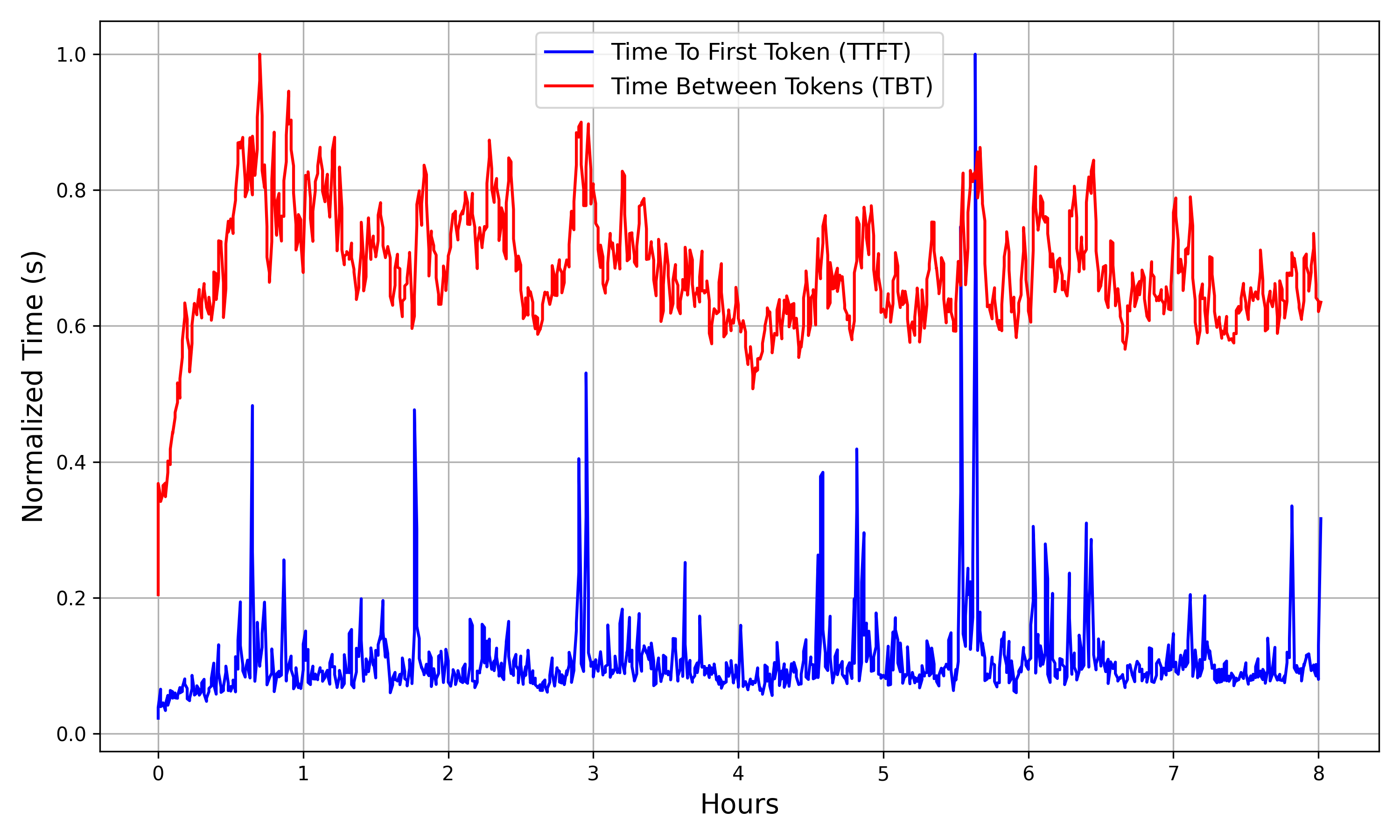}
  \captionof{figure}{Autoscaled by Prefill SM Activity}
\end{center}

% ----- Group 6 -----
\begin{center}
  \includegraphics[width=0.3\textwidth]{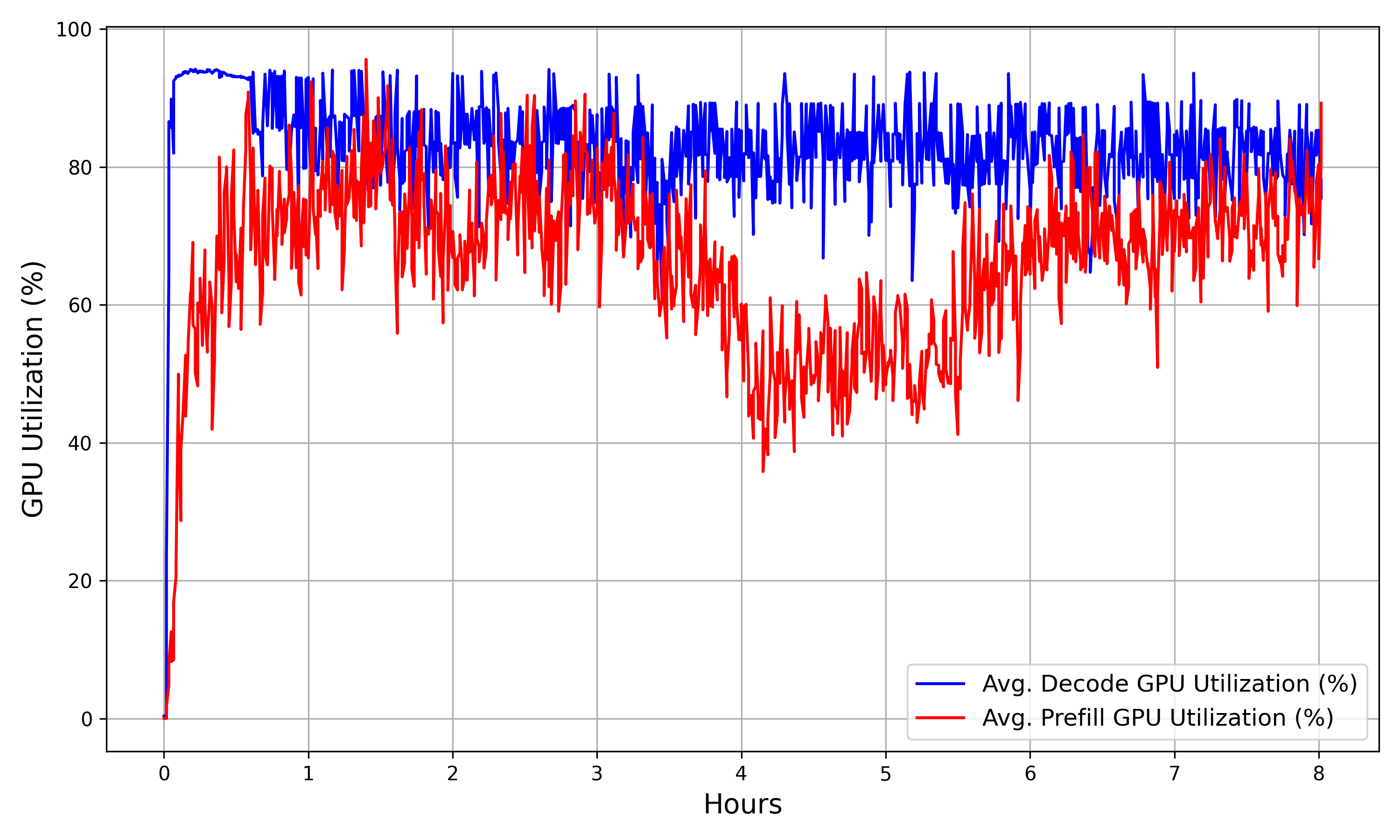}
  \includegraphics[width=0.3\textwidth]{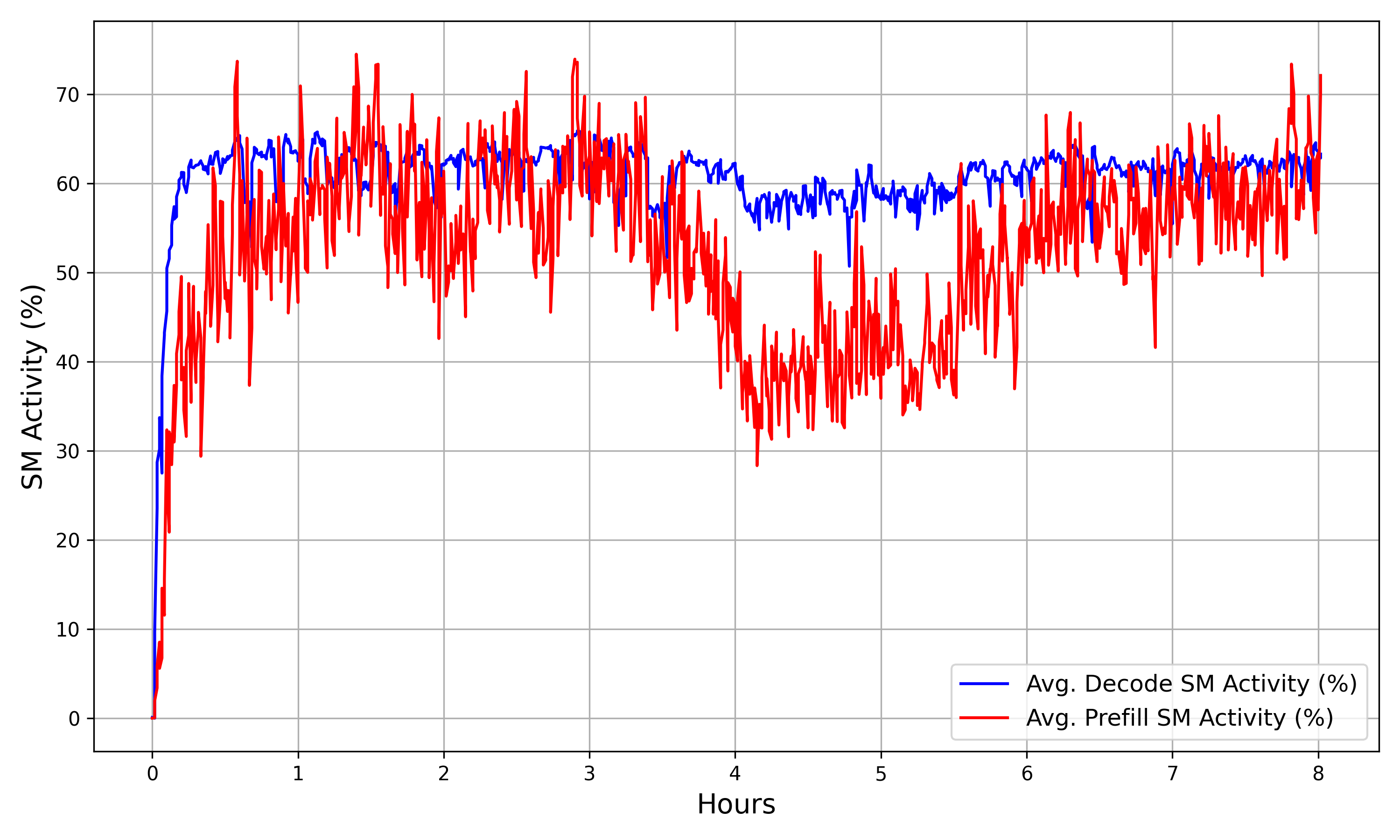}
  \includegraphics[width=0.3\textwidth]{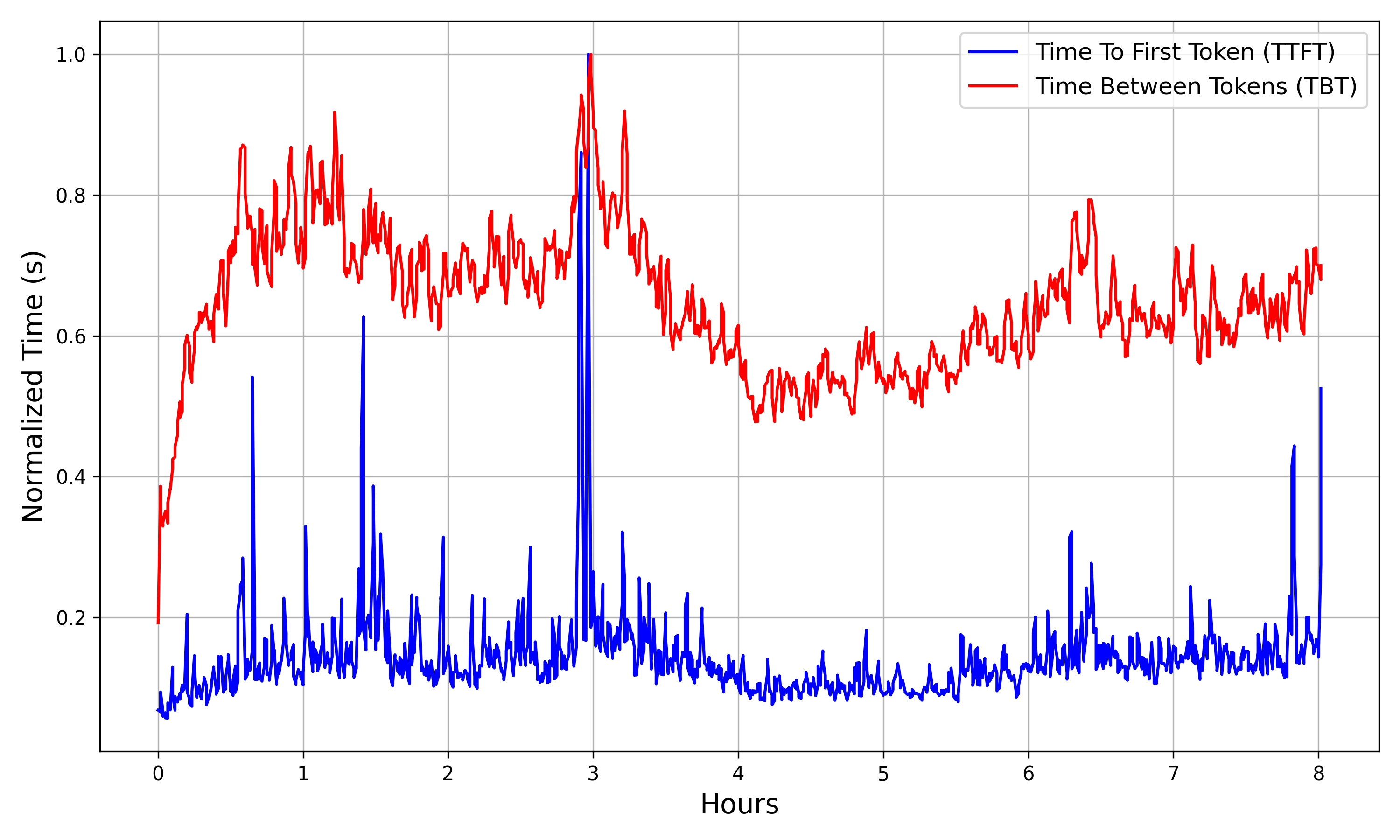}
  \captionof{figure}{Autoscaled by Decode SM Activity}
\end{center}

% ----- Group 7 -----
\begin{center}
  \includegraphics[width=0.3\textwidth]{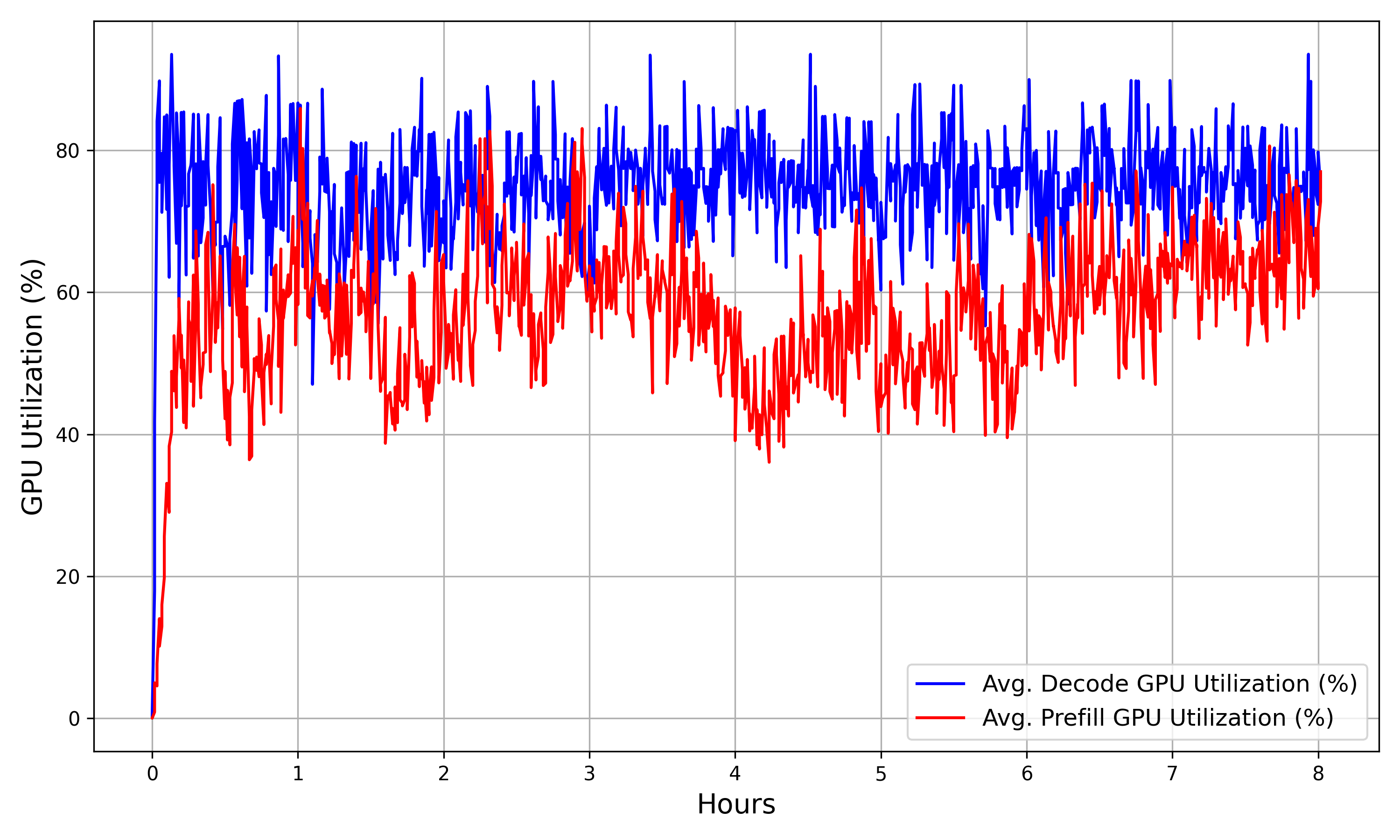}
  \includegraphics[width=0.3\textwidth]{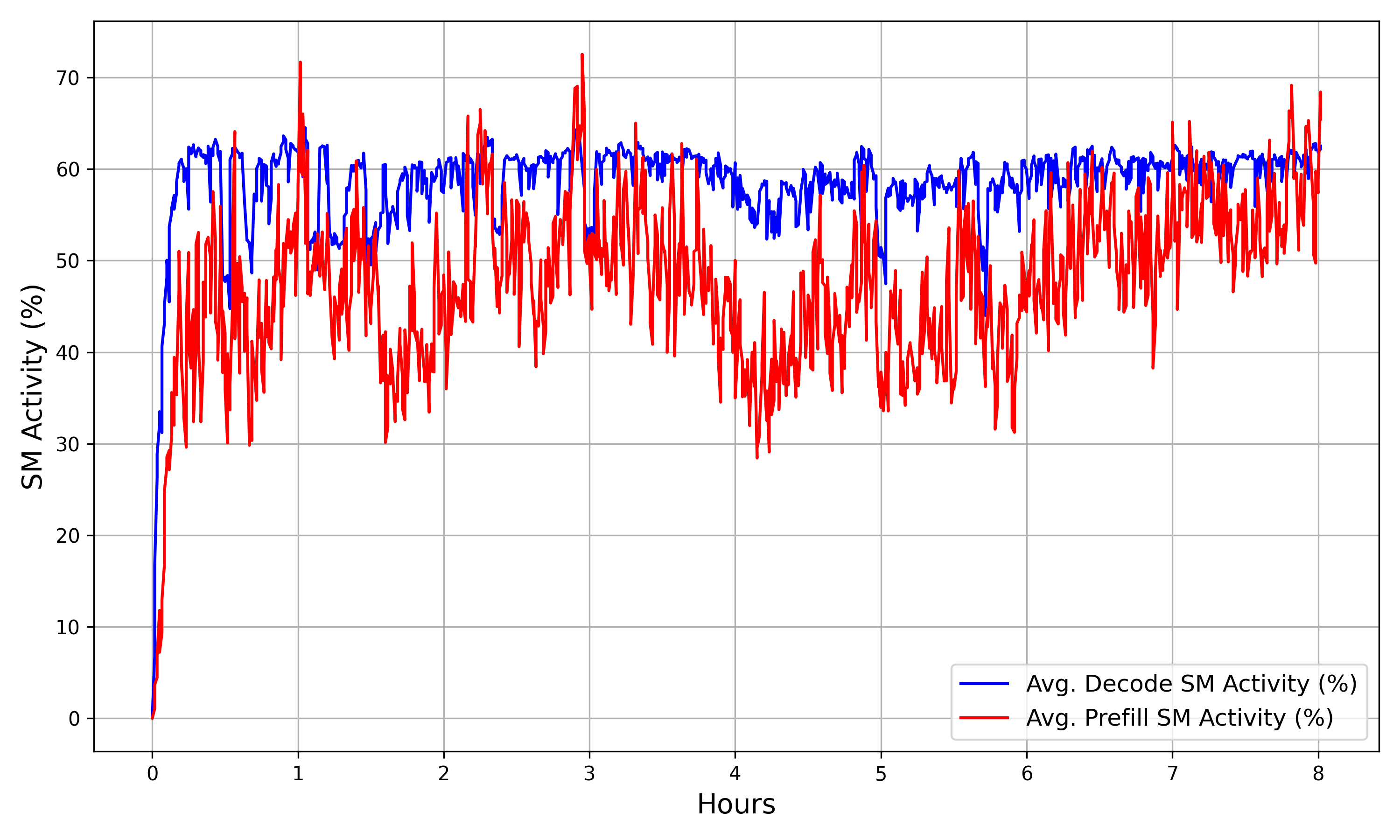}
  \includegraphics[width=0.3\textwidth]{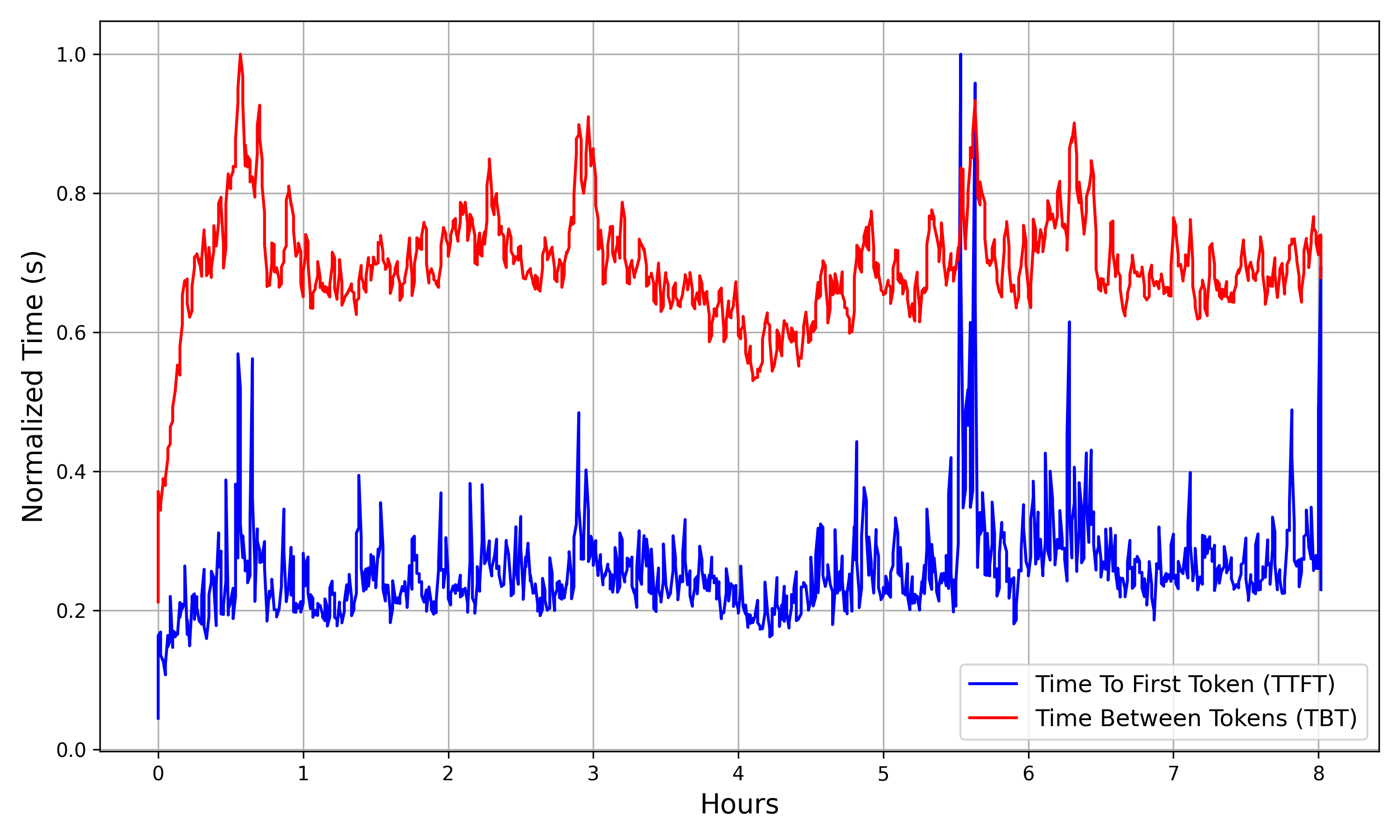}
  \captionof{figure}{Autoscaled by TTFT}
\end{center}

% ----- Group 8 -----
\begin{center}
  \includegraphics[width=0.3\textwidth]{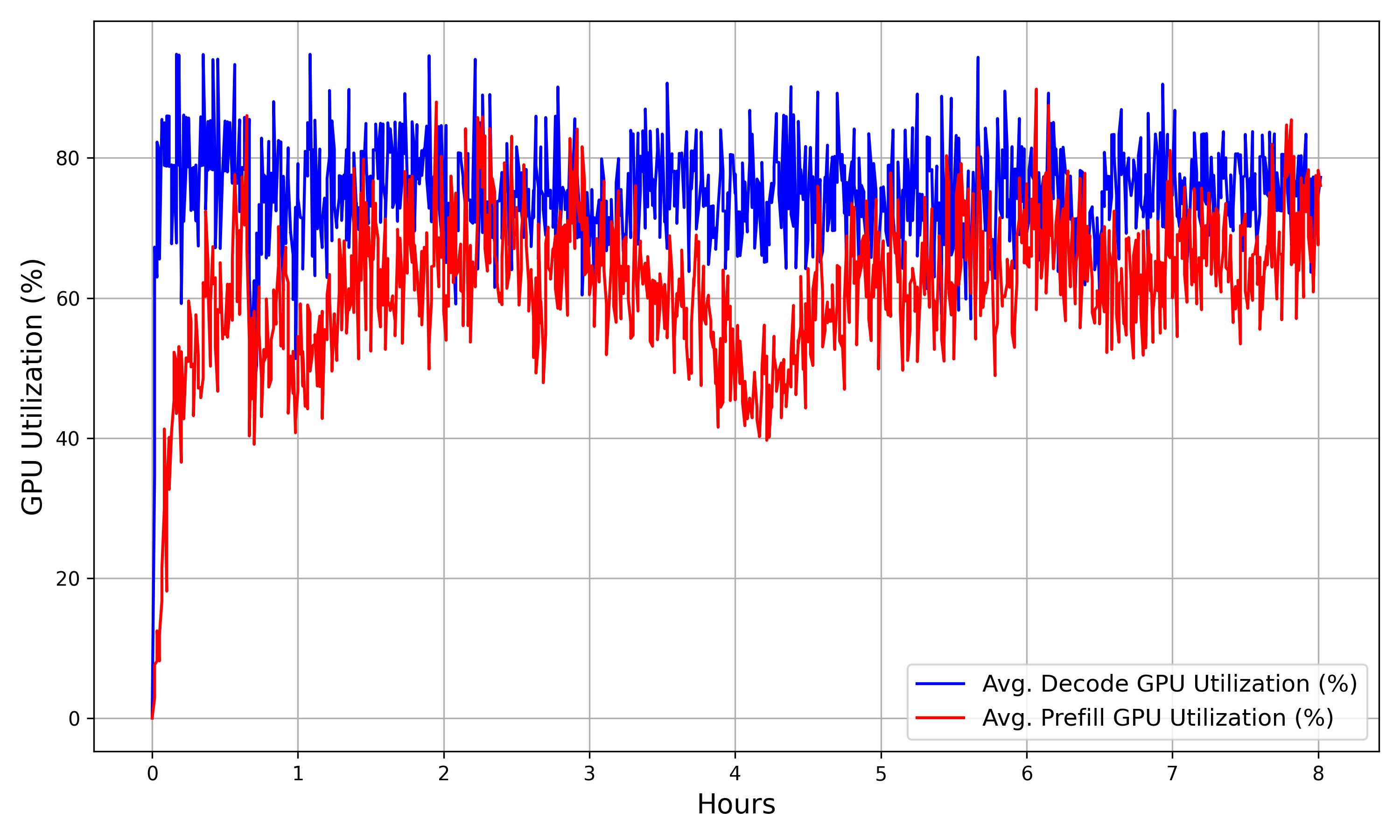}
  \includegraphics[width=0.3\textwidth]{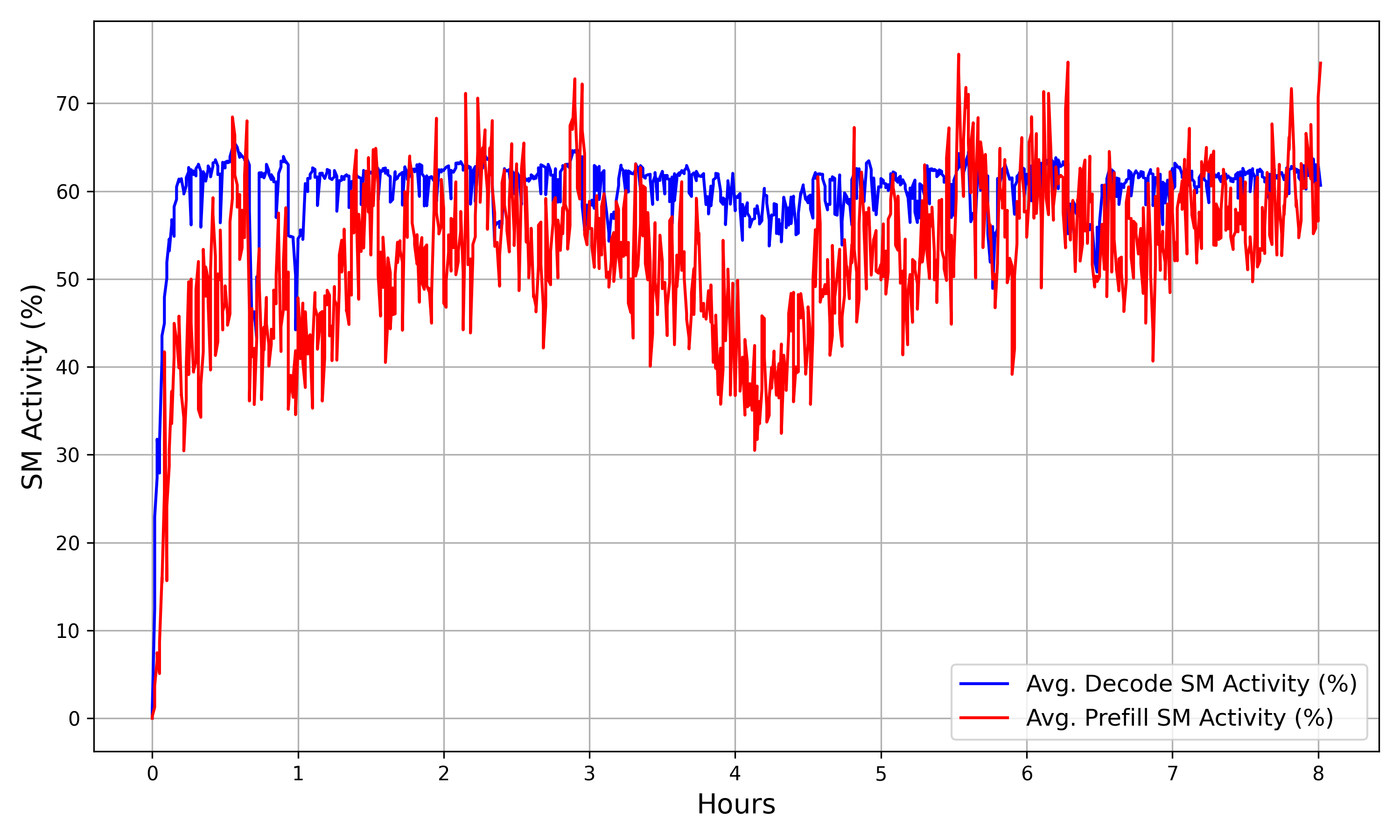}
  \includegraphics[width=0.3\textwidth]{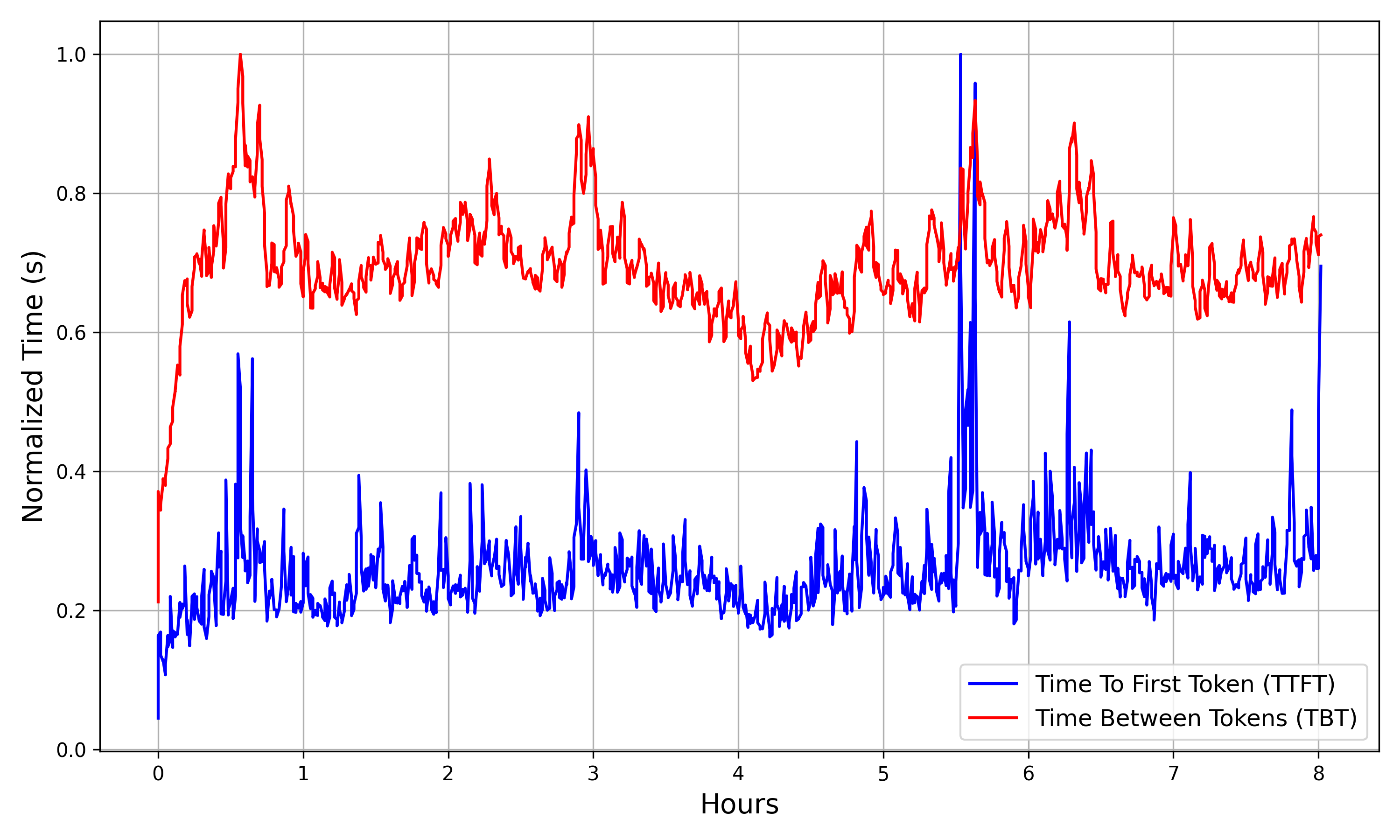}
  \captionof{figure}{Autoscaled by TBT}
\end{center}

% \clearpage
% \twocolumn % switch back to two-column mode if needed

\end{document}